\documentclass[a4paper,11pt]{article}
\usepackage{jcappub}

\title{Supersymmetric hybrid inflation with non-minimal coupling to
gravity}

\author{Umananda Dev Goswami}

\affiliation{Department of Physics, Dibrugarh University,
Dibrugarh 786004, Assam, India}

\emailAdd{umananda2@gmail.com}

\abstract{
We have studied the paradigm of cosmic inflation using the simplest model 
based on the idea of supersymmetric hybrid inflation with non-minimal coupling 
to gravity, specially under the slow-roll approximation following the 
superconformal approach to supergravity. It is found that 
within a range of values of the non-minimal coupling parameter $\xi$,
the model can accommodate the inflation data reported by the Planck ($n_s$ and
upper limit of $r$) and BICEP2/Keck (upper limit of $r$) collaborations. The 
study shows that the most probable value of $\xi$ should be 
$\sim 0.0134\pm0.0005$. That is
coupling is found to be very week. Within this range of $\xi$, the values of 
$r$ estimated from our model for $50 - 70$ e-foldings are found to be laying 
well below the upper limits set by the Planck and BICEP2/Keck collaborations.
Similarly, values of $n_s$ obtained for the said parameters are in good 
agreement with its latest data of the Planck collaboration. The constraint 
equations for the running of the scalar spectral index $n_{sk}$ 
and its running $n_{skk}$ are derived from the numerical solutions of our 
model for these parameters. These equations can be used to test our model from 
the data of future cosmological observations.}


\arxivnumber{1705.05540}

\begin{document}
\maketitle
\flushbottom

\section{Introduction}
The paradigm of cosmic inflation \cite{Starobinsky, Kazanas,Guth,Sato1,Sato2,Bassett,Martin} not only 
solved the outstanding problems of modern cosmology, such as the horizon and 
flatness problems \cite{Bassett,Martin}, but also it explains the origin of 
the temperature anisotropy of Cosmic Microwave Background (CMB) radiation as 
observed by the Cosmic Background Explorer (COBE) satellite \cite{COBE}. The 
recent observations of the Planck satellite \cite{PLANCK,PLANCK2}
on the CMB anisotropies has provided even more strong evidence of the 
existence of the epoch of inflation. The tempting evidence of the cosmic
inflation comes from the BICEP2 experiment \cite{BICEP2}, who recently 
claimed to has been detected the B-mode polarization of the CMB temperature 
anisotropy. These detected B-modes are supposed to be generated by primordial 
gravitational waves, which are due to quantum fluctuations of the graviton. 
So, if this result is confirmed by future checks then it will establish that 
the gravity is ultimately a quantum theory. This would imply that the physics 
of cosmic inflation, which we still do not understand in the fully fundamental 
sense, lies beyond the standard model. However, the latest results from
the analysis of data taken by the BICEP2 and Keck Array \cite{BICEP2Keck} set 
a (combined) upper limit for the B-mode polarization that is in consistent
with the upper limit set by the most recent results of Planck experiment 
\cite{PLANCK2} as well as set by the WMAP experiment \cite{WMAP} earlier.
                    
According to a large number of models of cosmic inflation, the inflation is 
generated by a homogeneous scalar field, called inflaton, which under suitable 
conditions may lead to an early-time accelerated expansion of the Universe. The
magnitude of the inflaton field is typically very large at the beginning of 
the inflation and then it rolls down towards a potential minimum where the 
inflation ends (as an example for chaotic inflation see Ref. \cite{Linde0}).
Another important but complex class of models which are based on a phase 
transition between two scalar fields are known as hybrid or double inflation 
models \cite{Linde1,Copeland}. In the hybrid inflation scenario one of the two 
interacting fields generates the inflation and is known as the inflaton field 
as usual. The second field which is dubbed as the waterfall field remains 
ineffective during inflation, but it ends the inflation by triggering an 
instability at a time when the inflaton field reaches the minimum of its 
potential. The waterfall field is a heavy field so that it can generate 
sufficient instability to stop the inflation.   

As the inflation takes place at a much higher energy scale (the Planck
scale) than the electroweak energy scale, a hierarchy problem exists between
these two energy scales \cite{Kawasaki}. One of the most effective natural
solution to this problem is the implementation of the theory of supersymmetry
(SUSY) \cite{Feng}, which ensures the stability of such a large hierarchy
against radiative corrections, to the theory of inflation. This stability is
very essential to keep the flatness of inflaton potential at the quantum level.
Thus, it is important to study inflation in the framework of the gravity
version of SUSY, i.e., in the framework of supergravity. The hybrid inflation
scenario has the natural structure that is suitable to incorporate the
framework of supergravity within it. The only new constraint brought to this
scenario by supergravity is that, the inflaton field values should be less
than the reduced Planck mass, otherwise flatness of the potential may be
destroyed making the onset of inflation more difficult \cite{Mendes,Lyth}.

Application of SUSY to the inflationary paradigm shows that the Higgs
doublet superfield has to play the role of inflaton field. This supersymmertic
Higgs model inflation requires only local SUSY, i.e. supergravity to solve the
Einstein equation. Again for the mass of the Higgs field to lie within the
bound set by current findings of LHC \cite{Carmi} and to get sufficiently flat
Higgs field potential, there must be a large non-minimal coupling of the Higgs
field to the curvature scalar, of the form $\xi \phi^2 R$. The Jordan frame
action of the inflaton field in such models contain a canonical kinetic term
of the Higgs field but the gravity part of the action changes due to the
non-minimal coupling of gravity with Higgs field. With this action it is not
possible to use familiar equation of general relativity, the inflationary
solutions and the standard slow-roll analysis. However, all these can be
achieved if one makes the transformation of the action from Jordan frame to the
Einstein frame where variables show minimal coupling between the Higgs field
and the gravity \cite{Ferrara0}. This transformation can be done by the
superconformal approach to supergravity developed in \cite{Ferrara}, so
that the original superconformal symmetry can be recovered.
Most recently there is a wave of interest to implement the supergravity in
the inflationary paradigm using non-minimal gravity coupling with the Higgs
inflation field via superconformal approach to supergravity (e.g. see
\cite{Kawasaki,Lee,Salopek,Futamase,Fakir,Mukaigawa,Bezrukov,Barvinsky,Shaposhnikov,Gorbunov,Bellido,Simone,Barvinsky1,Kamenshchik,Magnin,Choudhury,Noorbala,Elizalde,Inagaki,Odintsov,Myrzakulov,Elizalde1,Kallosh} and 
references therein). This wave is mainly to study 
the behaviour of various models derived in the past in the context of general 
relativity coupled to scalar fields with or without supersymmetry. 
The flow in this 
context is mainly concentrated towards the direction of the chaotic inflation 
scenario only. 
However, in Refs.\cite{David,Kaiser,Greenwood,Sfakianakis,Schutz} different 
aspects of inflation including primordial perturbations
and isocurvature modes were studied using the multifield model with 
non-minimal coupling to gravity within the domain of standard theories.
Similarly, the connection between hybrid inflation and the models of Higgs
inflation in supergravity with non-minimal coupling was analysed in
the Refs.\cite{Pallis1,Pallis2}, wherein it was shown that the observable
gravitational waves can be achieved by avoiding transplanckian excursions of
the initial (non-canonically normalized) inflaton and without provoking any
problem with perturbative unitarity. 
Thus study of hybrid inflationary models in supergravity with non-minimal 
coupling to gravity would be interesting and hence it invites our attention in 
this direction. 

There are two types of hybrid inflation models derived from SUSY:
F-term and D-term hybrid inflation \cite{Mazumdar}. As the F-term type
naturally fits with the Higgs mechanism so most of the studies are done on 
this type of hybrid inflation models only. As such, in this work we study 
the simplest F-term SUSY hybrid inflation model \cite{Dvali,Orani} with a 
non-minimal coupling to gravity to see the consequences of this coupling in
a simplest possible mean. In this context, it should be mentioned that in the
Ref.\cite{Koushik} a  systematic study was done on the suitability of inflation 
model building in Jordan frame supergravity, where it shows that the
inflationary models have strong dependency on non-minimal coupling to
gravity and in many cases the supergravity contributes to the flattening of
the potential in the Einstein frame, that would be suitable for slow-roll 
inflation. Here, we 
will basically investigate the above cited model in the Einstein frame under 
the slow-roll inflation scenario, in the light of recent cosmological 
observations. The rest of the paper is organized as follows: For 
clarity of the theoretical basis of the work we introduce the idea of the
superconformal approach to supergravity in the next section. In the section 3
we develop the cited hybrid inflation model with a non-minimal coupling to 
gravity. Features of the developed model and their possible consequent 
implications following the slow-roll approximation scenario are discussed in 
the section 4 and 5. We conclude our work finally in 
the section 6.  

\section{Superconformal theory in supergravity} \label{sec2}
Until recently it was not possible to implement the idea of Higgs inflation in
SUSY models because the superconformal theory brings the supergravity action
directly to the Einstein frame, after which the original superconformal 
symmetry can not be recovered \cite{Kallosh}. With the new approach developed in
\cite{Ferrara}, it is now possible to formulate supergravity action in the 
Jordan frame by setting the superconformal compensator in a more general and
flexible way. The complete $\mathcal{N} = 1$, $4\mathcal{D}$ supergravity 
action in Jordan frame derived in \cite{Ferrara} is
\begin{multline}
\mathcal{S}_{J} = \int d^4x\sqrt{-g_J}\bigg[-\frac{\mathcal{F}(\mathcal{X},\bar{\mathcal{X}})}{6}\{R_J-\bar{\Psi}_\mu R^\mu\}            -\frac{1}{6}(\partial_\mu\Phi)(\bar{\Psi}\cdot\gamma\Psi^\mu) +\mathcal{L}_0+\mathcal{L}_{1/2} + \mathcal{L}_{1} \\ + \mathcal{L}_m + \mathcal{L}_{mix} +\mathcal{L}_\mathcal{D} + \mathcal{L}_{\mathcal{D}f} - V_J\bigg].
\label{eq1}
\end{multline} 
In this action, $\mathcal{F}(\mathcal{X},\bar{\mathcal{X}})$ is the general 
frame function, which is an arbitrary function of complex scalar superfields 
$\mathcal{X}$ and $\bar{\mathcal{X}}$. $\Psi_\mu$ is 
the gravitino field with its kinetic term $R^\mu$. The kinetic terms of spin 
$0$, $\frac{1}{2}$ and $1$ fields are represented by $\mathcal{L}_0$, 
$\mathcal{L}_{1/2}$ and $\mathcal{L}_{1}$ respectively. $\mathcal{L}_m$, 
$\mathcal{L}_{mix}$, $\mathcal{L}_\mathcal{D}$ and $\mathcal{L}_{\mathcal{D}f}$
represent the fermion mass terms, mixing terms, kinetic $\mathcal{D}$ terms
and $4$-fermion terms respectively. The details about all these terms can be
found in \cite{Ferrara}. $V_J$ is the Jordan frame potential given by 
\cite{Ferrara0,Ferrara}
\begin{multline}
V_J = \frac{\mathcal{F}^2(\mathcal{X},\bar{\mathcal{X}})}{9}\bigg[e^{\mathcal{K}(\mathcal{X},\bar{\mathcal{X}})}\bigg\{\nabla_\alpha\mathcal{W}(\mathcal{X})g^{\alpha\bar{\beta}}\nabla_{\bar{\beta}}\mathcal{\bar{W}}(\mathcal{X})
 - 3\mathcal{W}(\mathcal{X})\mathcal{\bar{W}}(\mathcal{X})\bigg\} \\+ \frac{1}{2}\left(\mbox{Re}\;f(\mathcal{X})\right)^{-1\;AB}\mathcal{P}_A\mathcal{P}_B\bigg],
\label{eq2}
\end{multline} 
where $\mathcal{K}(\mathcal{X},\bar{\mathcal{X}})$ is K\"{a}hler potential, 
$\mathcal{W}(\mathcal{X})$ is the holomorphic superpotential, 
$f_{AB}(\mathcal{X})$ is the kinetic gauge matrix, $\mathcal{P}_A$ is the 
momentum map or Killing potential (it encodes the Yang-Mills transformation of 
the scalars), $g_{\alpha\bar{\beta}}$ is the K\"{a}hler metric, which is given 
by the second derivative of the K\"{a}hler potential
\begin{equation}
g_{\alpha\bar{\beta}}(\mathcal{X},\bar{\mathcal{X}}) = \frac{\partial^2\mathcal{K}(\mathcal{X},\bar{\mathcal{X}})}{\partial \mathcal{X}^\alpha \partial\bar{\mathcal{X}}^{\bar{\beta}}} \equiv \mathcal{K}_{\alpha \bar{\beta}} (\mathcal{X},\bar{\mathcal{X}}),
\label{eq3}
\end{equation}
and $\nabla_\alpha\mathcal{W}(\mathcal{X})$ denotes the K\"{a}hler-covariant 
derivative of the superpotential which has the form,
\begin{equation}
\nabla_\alpha\mathcal{W}(\mathcal{X}) \equiv \mathcal{W}_{\alpha}(\mathcal{X}) + \mathcal{K}_{\alpha}(\mathcal{X},\bar{\mathcal{X}})\mathcal{W}(\mathcal{X}).
\label{eq4}
\end{equation}    
Here the subscript $\alpha$ denotes derivative with respect to complex field 
$\mathcal{X}^{\alpha}$. 

If we set $\mathcal{F} = -\;3$ in the action (\ref{eq1}), the general 
$\mathcal{N} = 1$, $4\mathcal{D}$ supergravity action in an arbitrary Jordan 
frame reduces to the well known action of $\mathcal{N} = 1$, $4\mathcal{D}$ 
supergravity in the Einstein frame, where the curvature $R$ appears in the 
action only through the Einstein-Hilbert term $\frac{1}{2}\sqrt{-g_E}R(g_E)$, 
with $g^{\mu\nu}_E$ as the space-time metric in the Einstein frame. It should 
be noted that the bosonic sector action of $\mathcal{N} = 1$ supergravity 
coupled to chiral and vector matter multiplets is usually given in the 
Einstein frame \cite{Ferrara}. The corresponding potential in the Einstein 
frame takes the form:
\begin{multline}
V_E = e^{\mathcal{K}(\mathcal{X},\bar{\mathcal{X}})}\left[\nabla_\alpha\mathcal{W}(\mathcal{X})g^{\alpha\bar{\beta}}\nabla_{\bar{\beta}}\mathcal{\bar{W}}(\mathcal{X}) - 3\mathcal{W}(\mathcal{X})\mathcal{\bar{W}}(\mathcal{X})\right]\\ + \frac{1}{2}\left(\mbox{Re}\;f(\mathcal{X})\right)^{-1\;AB}\mathcal{P}_A\mathcal{P}_B = V^F_E + V^D_E,
\label{eq5}
\end{multline}   
where
\begin{equation}
V^F_E = e^{\mathcal{K}(\mathcal{X},\bar{\mathcal{X}})}\left[\nabla_\alpha\mathcal{W}(\mathcal{X})g^{\alpha\bar{\beta}}\nabla_{\bar{\beta}}\mathcal{\bar{W}}(\mathcal{X}) - 3\mathcal{W}(\mathcal{X})\mathcal{\bar{W}}(\mathcal{X})\right]
\label{eq6}
\end{equation}
is the F-term potential and
\begin{equation}
V^D_E = \frac{1}{2}\left(\mbox{Re}\;f(\mathcal{X})\right)^{-1\;AB}\mathcal{P}_A\mathcal{P}_B
\label{eq7}
\end{equation}
is the D-term potential. Consequently the potential (\ref{eq2}) in the Jordan
frame is related to the potential (\ref{eq5}) in the Einstein frame by
\begin{equation}
V_J = \frac{\mathcal{F}^2(\mathcal{X},\bar{\mathcal{X}})}{9}V_E.
\label{eq8}
\end{equation} 
The action in the Jordan frame can be obtained from the Einstein frame action
by rescaling only the metric and the farmions. In \cite{Ferrara} this is done
by using an $\mathcal{N} = 1$, $4\mathcal{D}$ superconformal theory 
\cite{Kallosh1} with local $SU(2,2|1)$ symmetry. Under this superconformal 
transformation the metric in a Jordan frame is related to the metric in the 
Einstein frame by
\begin{equation}
g^{\mu\nu}_J = \Omega^2 g^{\mu\nu}_E,
\label{eq9}
\end{equation}  
where the superconformal factor $\Omega^2$ can be identified as
\begin{equation}
\Omega^2 = - \frac{\mathcal{F}(\mathcal{X},\bar{\mathcal{X}})}{3} = e^{-\frac{\mathcal{K}(\mathcal{X},\bar{\mathcal{X}})}{3}}
> 0.
\label{eq10}
\end{equation}
The second choice of $\Omega^2$ in this equation yields a purely bosonic 
action in $\mathcal{N} = 1$, $4\mathcal{D}$ supergravity in a particular 
Jordan frame considered in \cite{Cremmer,Proeyen} for the superHiggs Effect. 
Moreover,
in order to have canonical kinetic terms in a purely bosonic action (which has 
special interest in cosmology), the frame function $\mathcal{F}$ should have
the form \cite{Ferrara}:
\begin{equation}
\mathcal{F}(\mathcal{X},\bar{\mathcal{X}}) = - 3 + \delta_{\alpha \bar{\beta}}\mathcal{X}^{\alpha}\bar{\mathcal{X}}^{\bar{\beta}} + \mathcal{J}(\mathcal{X}) + \bar{\mathcal{J}}(\bar{\mathcal{X}}),
\label{eq11}
\end{equation}
i.e. for this choice
\begin{equation}
\Omega^2 = 1 - \frac{1}{3}\left(\delta_{\alpha \bar{\beta}}\mathcal{X}^{\alpha}\bar{\mathcal{X}}^{\bar{\beta}} + \mathcal{J}(\mathcal{X}) + \bar{\mathcal{J}}(\bar{\mathcal{X}})\right),
\label{eq12}
\end{equation}
where $\mathcal{J}(\mathcal{X})$ and $\bar{\mathcal{J}}(\bar{\mathcal{X}})$ are
the holomorphic functions of scalars. 

The supergravity theories in which 
scalars are conformally coupled to gravity, kinetic terms are canonical, and 
the supergravity potential coincides with the global theory potential are 
known as Canonical Superconformal Supergravity (CSS) models \cite{Ferrara0}. 
The supergravity potential in CSS Jordan frame for generic cubic superpotential
is same as the global SUSY scalar potential \cite{Ferrara0,Kallosh,Mazumdar}. 
The F-term of this potential is
\begin{equation}
V_J = \sum\limits_\alpha\left|\frac{\partial \mathcal{W}(\mathcal{X^\alpha})}{\partial\mathcal{X^\alpha}}\right|^2,
\label{eq13}
\end{equation}
where sum is taken over all the superfields $\mathcal{X^\alpha}$. The 
corresponding F-term supergravity potential in the Einstein frame is related 
with the frame function (\ref{eq11}) or with the conformal factor (\ref{eq12}) 
via the relation (\ref{eq8}) \cite{Ferrara0}. That is in the CSS models, for 
a cubic superpotential the F-term Einstein frame potential can be obtained 
from the F-term Jordan frame potential itself.
But in general, where the superpotential is not cubic, the Einstein frame 
F-term potential in the generic supergravity theory is quite complicated as 
given in the equation (\ref{eq6}). In such a case the F-term Jordan frame 
potential has to be calculated from the corresponding Einstein frame potential
using the relation (\ref{eq8}).   
 
We can preserved the superconformal symmetry of matter multiplets by setting
$\mathcal{J}(\mathcal{X}) = 0$. In this case, the non-minimal coupling of the 
scalar fields become fixed and hence the scalar fields are conformally coupled 
to gravity. However, we can break the superconformal symmetry explicitly by 
setting an appropriate holomorphic function $\mathcal{J}(\mathcal{X})$, which
will include the nontrivial non-minimal coupling to gravity \cite{Ferrara0,Lee,Kallosh}. An important possibility, in our context, to break the 
superconformal symmetry of the matter multipletes in the supergravity action is
to introduce a dimensionless parameter in the holomorphic function 
$\mathcal{J}(\mathcal{X})$. As a simple illustration, if we denote this 
parameter by $\chi$ and introduced two complex scalar fields, viz., 
$X$ and $\Phi$, then after the gauge fixing, the holomorphic function 
$\mathcal{J}(\mathcal{X})$ can be expressed as $-\frac{3\chi}{4}\Phi^2$ 
\cite{Ferrara0,Kallosh}, which is independent of the field $X$. 
Under this situation the frame function (\ref{eq11}) and the conformal factor 
(\ref{eq12}) turn out to be \cite{Ferrara0}
\begin{multline}
\mathcal{F}(X,\Phi,\bar{X},\bar{\Phi}) = -3 + (|X|^2 + |\Phi|^2) - \frac{3}{4}\chi \left(\Phi^2+\bar{\Phi}^2\right)\\
= -3 + |X|^2 - \frac{3}{2}\xi\left(\Phi+\bar{\Phi}\right)^2 - \frac{3}{2}\left(\frac{1}{3}+\xi\right)\left(\Phi-\bar{\Phi}\right)^2
\label{eq14}
\end{multline}
and
\begin{equation}
\Omega^2 (X,\Phi,\bar{X},\bar{\Phi}) = 1 - \frac{1}{3}|X|^2 + \frac{1}{2}\xi\left(\Phi+\bar{\Phi}\right)^2
+ \frac{1}{2}\left(\frac{1}{3}+\xi\right)\left(\Phi-\bar{\Phi}\right)^2,
\label{eq15}
\end{equation}  
where $\bar{X}$ and $\bar{\Phi}$ are complex conjugate fields of $X$ and $\Phi$
respectively, and $\xi = -\frac{1}{6} +\frac{\chi}{4}$. It is now obvious that 
when $\xi = -\frac{1}{6}$ or $\chi = 0$, all scalar fields in Jordan frame are 
conformally coupled to gravity. However, when $\xi = -\frac{1}{3}$ or 
$\chi = -\frac{2}{3}$ the imaginary part of the scalar field $\Phi$ is 
decoupled from the curvature scalar $R$, i.e. it is minimally coupled to 
gravity. Similarly, when $\xi = 0$ or $\chi = \frac{2}{3}$, the real part of 
the scalar field $\Phi$ is minimally coupled to gravity \cite{Ferrara0,Kallosh}.
As we are interested in the 
non-minimal coupling of the real part of the scalar field $\Phi$ with gravity,
so we must have $\xi > 0$. This will introduce the non-minimal coupling term 
in the action (\ref{eq1}) for the real part of the scalar field $\Phi$ as 
$\frac{\xi}{2}\phi^2 R$, where $\phi/\sqrt{2} = \left(\Phi+\bar{\Phi}\right)/2$ 
is the real part of $\Phi$. For such non-minimal coupling, the F-term 
supergravity potential in the Einstein frame in general, with our present 
scalar fields can be calculated from the formula in equation (\ref{eq6}) and 
this potential in the Jordan frame can be found as 
\begin{multline}
V_J (X,\Phi,\bar{X},\bar{\Phi})
= V_E(X,\Phi,\bar{X},\bar{\Phi})\left[1 - \frac{1}{3}|X|^2 + \frac{1}{2}\xi\left(\Phi+\bar{\Phi}\right)^2 + \frac{1}{2}\left(\frac{1}{3}+\xi\right)\left(\Phi-\bar{\Phi}\right)^2\right]^2.
\label{eq16}
\end{multline}                        
Thus, the above discussion showed that it is possible to get inflation models 
in supergravity with the non-minimal coupling to gravity. In the following
we will make a study of the inflationary dynamics arising out of
the generic F-term supergravity potential in the Einstein frame for a given 
superpotential discussed in the next section. Throughout our above and further 
discussion we used the unit with $M_P = 1$, which can be recovered wherever is 
required.    

\section{The Model} \label{sec3}
We consider the simplest F-term SUSY hybrid inflation model, which is 
given by the most general form of superpotential \cite{Dvali,Orani}
\begin{equation}
\mathcal{W}(\Phi, \mathcal{S},\mathcal{\tilde{S}}) = \lambda\Phi\left(\mathcal{S}\mathcal{\tilde{S}} - \frac{v^2}{2}\right),  
\label{eq17}
\end{equation}
where $\lambda$ is a dimensionless coupling parameter, $v$ is another constant
parameter, $\Phi$ is a gauge singlet superfield containing the 
inflaton and $\mathcal{S}$, $\mathcal{\tilde{S}}$ is a conjugate pair of 
supperfields play the role of waterfall field. The superpotential (\ref{eq17})
has the $R$ symmetry under which $\mathcal{W}\rightarrow 
e^{i\gamma}\mathcal{W}$, $\Phi \rightarrow e^{i\gamma}\Phi$ and 
$\mathcal{S}\mathcal{\tilde{S}}$ is invariant \cite{Orani}. Because of this
symmetry this model does not require fine tuned parameters, which can be 
naturally obtained by the breaking of GUT gauge symmetry. E. g., such GUT gauge 
symmetry groups are $G_{B-L} (\equiv SU(3)_C\times SU(2)_L\times U(1)_Y\times U(1)_{B-L})$, $G_{LR} (\equiv SU(3)_C\times SU(2)_L\times SU(2)_R\times U(1)_{B-L})$ and $G_{5_X} (\equiv SU(5)\times U(1)_X)$ (for details see 
Refs. \cite{Matthew,Tatsuo} and related references therein). This model is
attractive mainly because it is possible to embed it in the particle physics
framework. Moreover, it is simple as it has only two parameters $\lambda$ and
$v$.    

The Jordan frame action of our model (i. e. for the scalar fields 
$z^\alpha$ = $\Phi$, $\mathcal{S}$, $\mathcal{\tilde{S}}$) can be written as 
\cite{Ferrara,Lee,Pallis1}  
\begin{equation}
S_{J} = \int d^4x\sqrt{-g_J}\bigg[-\frac{\mathcal{F}}{6}R_J- \mathcal{F}_{\alpha \bar{\beta}}\mathcal{D}_\mu z^\alpha \mathcal{D}^\mu \bar{z}^{\bar{\beta}} + \mathcal{F} \mathcal{A}_\mu \mathcal{A}^\mu - V_J \bigg],
\label{eq17a}
\end{equation}
where $\mathcal{F}$ is the frame function as given by equation (\ref{eq14}),
$\mathcal{F}_{\alpha \bar{\beta}} = \frac{\partial^2\mathcal{F}}{\partial z^\alpha \partial\bar{z}^{\bar{\beta}}}$ and $\mathcal{D}_\mu = 
\partial_\mu - \mathbf{A}^A_\mu \kappa^\alpha_A$ are the covariant 
derivatives of scalar fields $z^\alpha$. Here $\mathbf{A}^A_\mu$ represents
the vector gauge fields and $\kappa^\alpha_A$ for the Killing vector. 
$\mathcal{A}_\mu$ is the purely bosonic part of the on-shell value of the
auxiliary field $A_\mu$ of supergravity, which is given by\footnote{Depending
on a gauge field, the auxiliary pseudovector has an additional contribution,
$+ \frac{1}{6}i\mathbf{A}_\mu^A(r_A - \bar{r}_A)$ if the K\"ahler potential is
not gauge-invariant in direction A. Here $r_A$ is the holomorphic part of the 
transformation of the K\"ahler potential under gauge symmetry \cite{Ferrara,Lee}.}
\begin{equation}
\mathcal{A}_\mu = - \frac{i}{2 \mathcal{F}} 
\left(\mathcal{F}_{\alpha} \mathcal{D}_{\mu} z^\alpha - 
\mathcal{F}_{\bar{\alpha}} \mathcal{D}_{\mu} \bar{z}^{\bar{\alpha}} \right).  
\end{equation}
Here $\mathcal{F}_\alpha = \frac{\partial \mathcal{F}}{\partial z^\alpha}$. 
Thus the kinetic terms of the scalar fields are partly determined by the 
value of $\mathcal{A}_\mu$. However, to achieve the canonical kinetic terms of
scalar fields in the Jordan frame within the relation between $\mathcal{K}$ and
$\mathcal{F}$ given by the equation (\ref{eq10}), it is required that 
$\mathcal{F}_{\alpha \bar{\beta}} = \delta_{\alpha \bar{\beta}}$ and 
$\mathcal{A}_\mu = 0$ \cite{Ferrara,Lee}. The first requirement can be 
achieved with our choice of the frame function (\ref{eq11}) (whose explicit
form is given by the equation (\ref{eq14})). But second requirement can not be 
achieved in general due to the angular modes of complex scalar fields, except
in the case when the moduli $\left|z^\alpha \right|$ dominate the dynamics
during the cosmological evolution, the scalar kinetic terms can be of 
canonical form \cite{Ferrara,Lee}. 

Now using the superconformal transformation relation (\ref{eq9}) via relation
(\ref{eq10}), the Einstein frame action of our model can be expressed as
\cite{Ferrara,Lee,Pallis1}

\begin{equation}
S_{E} = \int d^4x\sqrt{-g_E}\left[\frac{1}{2}R_E- \mathcal{K}_{\alpha \bar{\beta}}g_E^{\mu\nu}\mathcal{D}_\mu z^\alpha \mathcal{D}_\nu \bar{z}^{\bar{\beta}} - V_E \right],
\label{eq17b}
\end{equation}
where the property (\ref{eq3}) is used in this expression and 
$V_E$ is the Einstein frame potential which is 
given by the equation (\ref{eq6}) in general. As discussed in the previous 
section, it is clear from the first term of the r.h.s. of this expression that 
action in the Einstein frame $S_E$ exhibits minimal couplings of the scalar 
fields $z^\alpha$ to $R_E$.
As already mentioned in the previous section, due
to our choice of holomorphic function $\mathcal{J}$, our model will give 
supergravity inflation with non-conformal but non-minimal coupling to gravity.

For the superfields $\Phi$, $\mathcal{S}$ and $\tilde{\mathcal{S}}$, the
superconformal factor (\ref{eq15}) can be written in a convenient form as
\begin{equation}
\Omega^2 = 1 - \frac{1}{3}(\mathcal{S}\bar{\mathcal{S}} + 
\tilde{\mathcal{S}}\bar{\tilde{\mathcal{S}}} + \Phi\bar{\Phi}) + 
\frac{1}{4}\chi(\Phi^2 + \bar{\Phi}^2) + \frac{1}{3}\zeta ( \Phi\bar{\Phi})^2,
\label{eqa}
\end{equation}
where $\chi = 4(\xi + \frac{1}{6})$ and the last term with the second order
coupling parameter $\zeta$ is introduced to stabilize the inflation by
avoiding negative value of the potential $V_E$ during inflation.

Again from the relation (\ref{eq10}), we may get the relation between the 
K\"ahler potential $\mathcal{K}$ and $\Omega^2$ as
\begin{equation}
\mathcal{K} = -\;3\ln\Omega^2.
\label{eqb}
\end{equation}
This relation leads another relation between $\mathcal{K}$ and $\Omega^2$, 
which is
\begin{equation}
e^{\mathcal{K}} = \Omega^{-6}.
\label{eqc}
\end{equation}
Using equations (\ref{eq4}) and (\ref{eqc}), the supergravity F-term scalar 
potential in the Einstein frame $V_E$ can be obtained from the equation 
(\ref{eq6}) for the superfields $\Phi$, $\mathcal{S}$ and 
$\tilde{\mathcal{S}}$ in the form:
\begin{equation}
V_E = \Omega^{-6}g^{\alpha\bar{\alpha}}\Big[\mathcal{W}_{\alpha}\bar{\mathcal{W}}_{\bar{\alpha}} + \mathcal{W}_{\alpha}\mathcal{K}_{\bar{\alpha}}\bar{\mathcal{W}} +
\mathcal{K}_{\alpha}\bar{\mathcal{W}}_{\bar{\alpha}}\mathcal{W} + \mathcal{K}_{\alpha}\mathcal{K}_{\bar{\alpha}}|\mathcal{W}|^2\Big] - 3\Omega^{-6}|\mathcal{W}|^2,
\label{eqd}
\end{equation}
where subscripts $\alpha$ and $\bar{\alpha}$ denote derivatives with respect to
our superfields and their complex conjugates respectively, and 
$g^{\alpha\bar{\alpha}}$ are the diagonal elements of conjugate K\"ahler 
metric for these superfileds. Here, off-diagonal elements of
$g^{\alpha\bar{\beta}}$ are omitted as they will contribute nothing to the
inflationary dynamics to be discussed here.     

Now, using the the superconformal factor (\ref{eqa}) in the definition 
(\ref{eqb}) of the K\"ahler potential, we may obtain the following derivatives
of K\"ahler potential with respect to the superfields $\Phi$, $\mathcal{S}$,
$\tilde{\mathcal{S}}$ and to their respective complex conjugates as
\begin{eqnarray}\nonumber
\mathcal{K}_{\mathcal{S}} = \Omega^{-2}\bar{\mathcal{S}},\; 
\mathcal{K}_{\bar{\mathcal{S}}} = \Omega^{-2}\mathcal{S},\;
\mathcal{K}_{\tilde{\mathcal{S}}} = \Omega^{-2}\bar{\tilde{\mathcal{S}}},\;
\mathcal{K}_{\bar{\tilde{\mathcal{S}}}} = \Omega^{-2}\tilde{\mathcal{S}},\\
\mathcal{K}_{\Phi} = \Omega^{-2}\Big(\bar{\Phi} - \frac{3}{2}\chi\Phi -2\zeta\Phi\bar{\Phi}^2\Big),\;
\mathcal{K}_{\bar{\Phi}} = \Omega^{-2}\Big(\Phi - \frac{3}{2}\chi\bar{\Phi} - 2\zeta\Phi^2\bar{\Phi}\Big).
\label{eqe}
\end{eqnarray}
According to the definition of the  K\"ahler metric given in the 
equation (\ref{eq3}), its diagonal components for our model can 
be obtained from these results as given by
\begin{eqnarray}
g_{\mathcal{S}\bar{\mathcal{S}}} = \Omega^{-2}\Big[1+\frac{1}{3}\Omega^{-2}\mathcal{S}\bar{\mathcal{S}}\Big],
\label{eqf}\\
g_{\tilde{\mathcal{S}}\bar{\tilde{\mathcal{S}}}} = \Omega^{-2}\Big[1+\frac{1}{3}\Omega^{-2}\tilde{\mathcal{S}}\bar{\tilde{\mathcal{S}}}\Big], \label{eqg}\\
\qquad g_{\Phi\bar{\Phi}} = \Omega^{-2}\Big[(1-4\zeta\Phi\bar{\Phi}) + \frac{1}{3}\Omega^{-2}G(\Phi\bar{\Phi})\Big],
\label{eqh}
\end{eqnarray}
where
\begin{equation}
G(\Phi\bar{\Phi}) = f(\Phi\bar{\Phi})f(\bar{\Phi}\Phi)
\end{equation}
with
\begin{eqnarray}
f(\Phi\bar{\Phi}) = \Phi-\frac{3}{2}\chi\bar{\Phi} - 2\zeta\Phi^2\bar{\Phi},\\
f(\bar{\Phi}\Phi) = \bar{\Phi}-\frac{3}{2}\chi\Phi - 2\zeta\Phi\bar{\Phi}^2.
\end{eqnarray}
The corresponding components of the conjugate  K\"ahler metric are
\begin{eqnarray}
g^{\mathcal{S}\bar{\mathcal{S}}} = \frac{\Omega^4}{\Omega^2+\frac{1}{3}\mathcal{S}\bar{\mathcal{S}}},
\label{eqi}\\
g^{\tilde{\mathcal{S}}\bar{\tilde{\mathcal{S}}}} = \frac{\Omega^4}{\Omega^2+\frac{1}{3}\tilde{\mathcal{S}}\bar{\tilde{\mathcal{S}}}},
\label{eqj}\\
g^{\Phi\bar{\Phi}} = \frac{\Omega^4}{\Omega^2(1-4\zeta\Phi\bar{\Phi})+\frac{1}{3}G(\Phi\bar{\Phi})}.
\label{eqk}
\end{eqnarray} 
Similarly, derivatives of the potential (\ref{eq17}) and its conjugate with 
respect to the superfields and to their respective conjugate fields are
\begin{eqnarray}\nonumber
\mathcal{W}_{\mathcal{S}} = \lambda\tilde{\mathcal{S}}\Phi,\;
\mathcal{W}_{\tilde{\mathcal{S}}} = \lambda\mathcal{S}\Phi,\;
\mathcal{W}_{\Phi} = \lambda\Big(\mathcal{S}\tilde{\mathcal{S}} -\frac{v^2}{2}\Big),\\
\bar{\mathcal{W}}_{\bar{\mathcal{S}}} = \lambda\bar{\tilde{\mathcal{S}}}\bar{\Phi},\;
\bar{\mathcal{W}}_{\bar{\tilde{\mathcal{S}}}} = \lambda\bar{\mathcal{S}}\bar{\Phi},\;
\bar{\mathcal{W}}_{\bar{\Phi}} = \lambda\Big(\bar{\mathcal{S}}\bar{\tilde{\mathcal{S}}} -\frac{v^2}{2}\Big).
\label{eql}
\end{eqnarray}    

Using results (\ref{eqe}), (\ref{eqi}), (\ref{eqj}), (\ref{eqk}) and 
(\ref{eql}), the supergravity F-term potential (\ref{eqd}) in the Einstein
frame for the superpotential (\ref{eq17}) can be expressed explicitly in terms
of superfields as
\begin{eqnarray}\nonumber
V_E = \frac{\Omega^{-2}}{\Omega^{2} + \frac{1}{3}|\mathcal{S}|^2}\lambda^2|\Phi \tilde{\mathcal{S}}|^2 + \frac{\Omega^{-2}}{\Omega^{2} + \frac{1}{3}|\mathcal{\tilde{S}}|^2}\lambda^2|\Phi \mathcal{S}|^2\\\nonumber 
+ \frac{\Omega^{-2}}{\Omega^2(1-4\zeta\Phi\bar{\Phi})+\frac{1}{3}G(\Phi\bar{\Phi})}\lambda^2(\mathcal{S}\mathcal{\tilde{S}} - \frac{v^2}{2})(\bar{\mathcal{S}}\bar{\tilde{\mathcal{S}}} - \frac{v^2}{2})\\\nonumber
+ \frac{\Omega^{-4}}{\Omega^{2} + \frac{1}{3}|\mathcal{S}|^2}\lambda^2|\Phi|^2\left[\tilde{\mathcal{S}}\mathcal{S}(\bar{\mathcal{S}}\bar{\tilde{\mathcal{S}}} - \frac{v^2}{2}) + \bar{\mathcal{S}}\bar{\tilde{\mathcal{S}}}(\mathcal{S}\tilde{\mathcal{S}} - \frac{v^2}{2}) \right] \\\nonumber
+ \frac{\Omega^{-4}}{\Omega^{2} + \frac{1}{3}|\tilde{\mathcal{S}}|^2}\lambda^2|\Phi|^2\left[\mathcal{S}\tilde{\mathcal{S}}(\bar{\mathcal{S}}\bar{\tilde{\mathcal{S}}} - \frac{v^2}{2}) + \bar{\mathcal{S}}\bar{\tilde{\mathcal{S}}}(\mathcal{S}\tilde{\mathcal{S}} - \frac{v^2}{2})\right] + \nonumber
\end{eqnarray}
\begin{multline} 
\frac{\Omega^{-4}}{\Omega^2(1-4\zeta\Phi\bar{\Phi})+\frac{1}{3}G(\Phi\bar{\Phi})}\lambda^2\Big[\bar{\Phi}f(\Phi\bar{\Phi}) + \Phi f(\bar{\Phi}\Phi)\Big]
\times (\mathcal{S}\tilde{\mathcal{S}} - \frac{v^2}{2})(\bar{\mathcal{S}}\bar{\tilde{\mathcal{S}}} - \frac{v^2}{2})\\ 
+ \frac{\Omega^{-6}}{\Omega^{2} + \frac{1}{3}|\mathcal{S}|^2}\lambda^2|\Phi \mathcal{S}|^2\left|\mathcal{S}\tilde{\mathcal{S}} - \frac{v^2}{2}\right|^2
+ \frac{\Omega^{-6}}{\Omega^{2} + \frac{1}{3}|\tilde{\mathcal{S}}|^2}\lambda^2|\Phi \tilde{\mathcal{S}}|^2\left|\mathcal{S}\tilde{\mathcal{S}} - \frac{v^2}{2}\right|^2\\ 
+  \frac{\Omega^{-6}}{\Omega^{2}(1-4\zeta\Phi\bar{\Phi}) + \frac{1}{3}G(\Phi\bar{\Phi})}\lambda^2|\Phi|^2G(\Phi\bar{\Phi})\left|\mathcal{S}\tilde{\mathcal{S}} - \frac{v^2}{2}\right|^2 - 3\Omega^{-6}\lambda^2|\Phi|^2\left|\mathcal{S}\tilde{\mathcal{S}} - \frac{v^2}{2}\right|^2.
\label{eq18}
\end{multline}
From this potential we find that SUSY global minimum lies at,
\begin{equation}
\langle\mathcal{S}\rangle\; =\; \langle\mathcal{\tilde{S}}\rangle\; =\; \pm\frac{v}{\sqrt{2}}, \;\;\; \langle\Phi\rangle = 0,
\label{eq19}
\end{equation}
which preserves SUSY. However, at tree level there is no term in 
equation (\ref{eq18}) which may drive $\Phi$ to zero. This problem will 
eliminate if
radiative corrections are taken into account because, these corrections drive 
the system towards the global minimum by lifting the potential in the $\Phi$ 
direction \cite{Dvali,Orani}. Moreover, with these vacuum expectation of values of $\mathcal{S}$ and $\mathcal{\tilde{S}}$, $\mathcal{W}$ leads to the 
spontaneous breaking of GUT gauge symmetry to SM gauge symmetry, which may 
develop topological defects in the end of inflation depending on the group 
which is broken. This serious problem will not arise if the GUT gauge 
groups, such as $G_{LR}$ and $G_{5_X}$ are broken to respective SM gauge 
groups. In both these cases no cosmic defects are formed during the breaking.
In contrast, the breaking of GUT gauge groups, such as $SU(4)_C\times SU(2)_L \times SU(2)_R$, $SU(5)$ or $SO(10)$ to SM groups lead to the generation of 
magnetic monopoles (for details see the Ref. \cite{Matthew} and related 
references therein).    

Let us define various superfields as given by
\begin{equation}
\mathcal{S} = (S_1 + iS_2)/\sqrt{2},\;\; \tilde{\mathcal{S}} = (\tilde{S}_1 + i\tilde{S}_2)/\sqrt{2}\;\; \mbox{and}\;\; \Phi = (\phi + i\theta)/\sqrt{2}, 
\label{eq20}
\end{equation}
where $S_1$, $S_2$, $\tilde{S}_1$, $\tilde{S}_2$, $\phi$ and $\theta$ are all
real scalar fields. However, for simplification we consider that 
$S_1 = \tilde{S}_1 = S$ and $S_2 = \tilde{S}_2 = \theta = 0$. This simplified
consideration leads the potential (\ref{eq18}) to the following convenient 
form:
\begin{equation}
V_E = \frac{3\,S^{2}\lambda^2\phi^2\big(S^2-v^2+2\,\Omega^2_s\big)^2}{4\,\Omega^6_s\big(S^2+6\,\Omega^2_s\big)} + \frac{3\,\big(S^2-v^2\big)^2\lambda^2\big(-3\,(1+4\,\xi)\phi^2+ 4\,\zeta\phi^4 +2\,\Omega^2_s\big)}{4\,\Omega^4_s\big(6\,\Omega^2_s\,(1-2\zeta\phi^2) + \phi^2(6\xi + \zeta\phi^2)^2 \big)},
\label{eq23}
\end{equation}
where
$\Omega^2_s = 1 - \frac{1}{3}S^2 +\xi\phi^2 + \frac{1}{12}\zeta\phi^4.$
This consideration is justified by our premise that only the real part of the 
superfield $\Phi$ is coupled the gravity. We may now refer this real part 
$\phi$ as the inflaton field. Moreover, as $\mathcal{S}$ and 
$\tilde{\mathcal{S}}$ are members the conjugate pair of superfields play the
role of waterfall field, so magnitudes of corresponding parts of these fields 
must be same. In homotopy of the inflaton field we may take the real parts of
these fields together as the waterfall field. Again, if we consider that
$S_2 = \tilde{S}_2 = S'$ and $S_1 = \tilde{S}_1 = \theta = 0$, then we 
will get a potential similar to (\ref{eq23}) but with a replacement of 
$S\rightarrow S'$ and $(S^2-v^2)\rightarrow (S'^2+v^2)$.        

As we are interested in inflationary trajectories with non-minimally coupled 
real part of scalar field to gravity, so afterwards we will focus our 
attention only on potential (\ref{eq23}). It should be noted that this 
potential has two (gauge) Symmetry Breaking Vacua (SBV) at 
\begin{equation}
\phi = 0, \;\;\; S = \pm\, v,
\label{eq24}
\end{equation}
where the inflation ends and SUSY is preserved with a fast phase transition 
driven by the waterfall field $S$.

At this point, it is important to
remember that the field $\phi$ in the Einstein frame is not canonically
normalized due to non-minimal coupling parameters $\xi,\,\zeta \ne 0$. The 
canonically normalized inflaton field $\varphi$ in the 
Einstein frame is related to the field $\phi$ for only $\xi\ne0$ as follows 
\cite{Ferrara0,Kallosh}:

\begin{equation}
\frac{d\phi}{d\varphi} = \frac{\omega_0^2}{\sqrt{\omega_0^2 + 6\,\xi^2\phi^2}},
\label{eq24a}
\end{equation}
where $\omega_0^2 = 1 + \xi \phi^2$. This sort of normalization is already 
included in our analysis as under above mentioned considerations related to the 
superfields, $g^{\Phi\bar{\Phi}}$ component given in the equation (\ref{eqk}) 
can expressed along $S=0$ as 
\begin{equation}
g^{\Phi\bar{\Phi}} = \frac{\Omega_0^4}{\Omega_0^2(1-2\,\zeta\phi^2) + \frac{1}{6}\phi^2(6\,\xi + \zeta\phi^2)^2},
\label{eq24b}
\end{equation} 
where $\Omega_0^2 = 1 +\xi\phi^2 + \frac{1}{12}\zeta\phi^4$.
This equation will take the same form of the equation (\ref{eq24a}) in the
situation of $\zeta = 0.$ Moreover, the asymptotic form of solution of the 
equation (\ref{eq24a}) shows 
that in the interval $0<\phi<< \frac{1}{\xi}$, $\varphi \approx \phi$ 
(for details see  \cite{Ferrara0}). Since in our work, the effective range of 
$\phi$ belongs approximately to this interval due to small values of $\xi$ as
well as $\zeta$ (see below), so we may consider $\phi$ itself as the 
canonically normalized inflaton field in the Einstein frame for our further 
analysis below.

\begin{figure*}[hbt]
\centerline{
\includegraphics[scale=0.47]{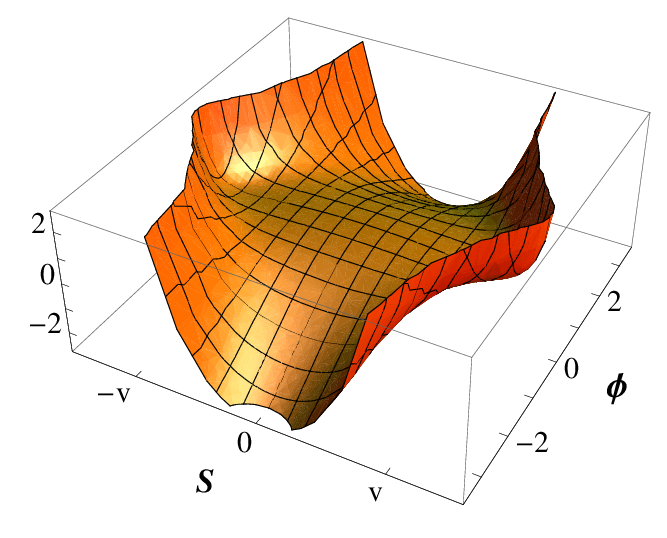}
\hspace{0.2cm}
\includegraphics[scale=0.47]{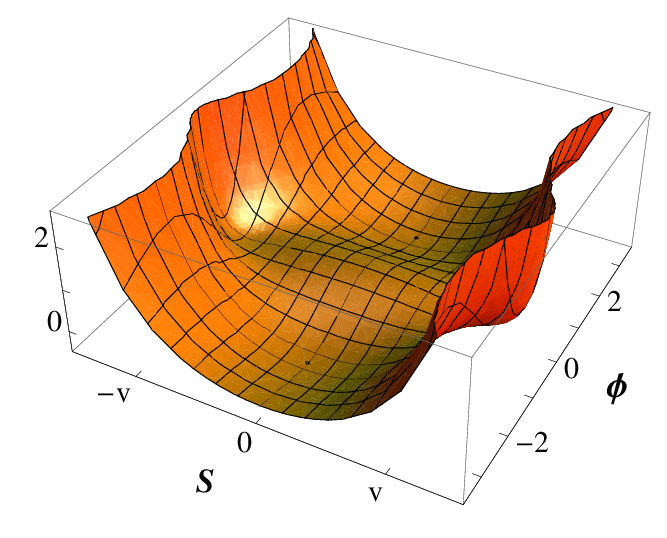}
\hspace{0.2cm}
\includegraphics[scale=0.47]{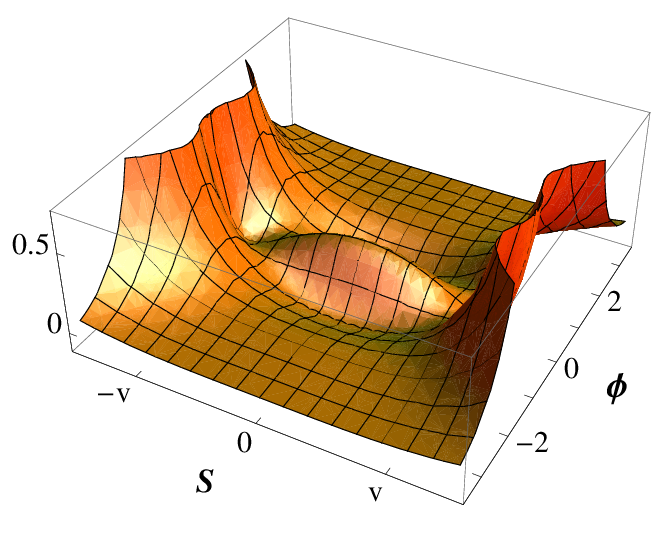}
\vspace{0.7cm}}
\centerline{
\includegraphics[angle=-90,origin=c,scale=0.33]{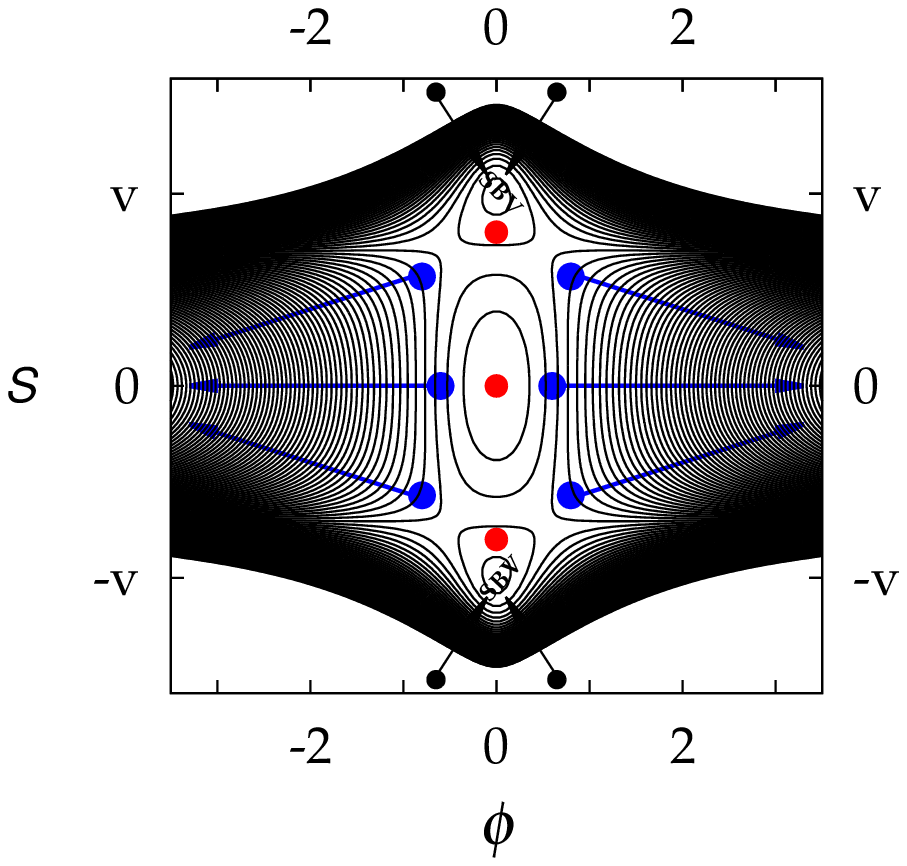}
\includegraphics[angle=-90,origin=c,scale=0.33]{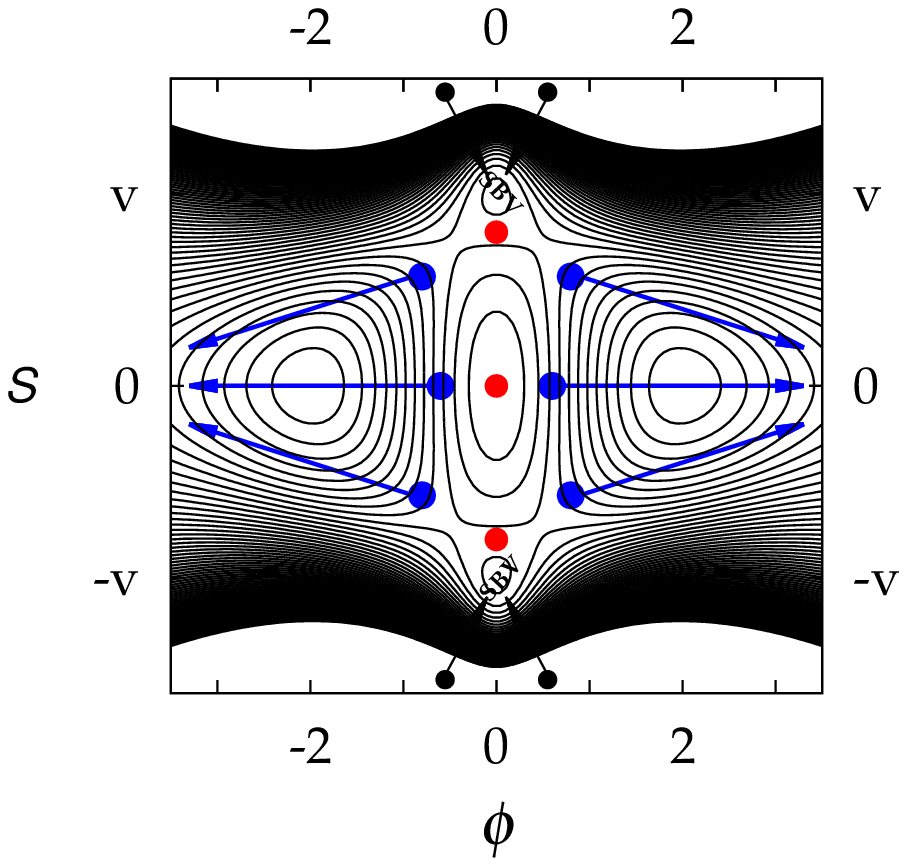}
\includegraphics[angle=-90,origin=c,scale=0.33]{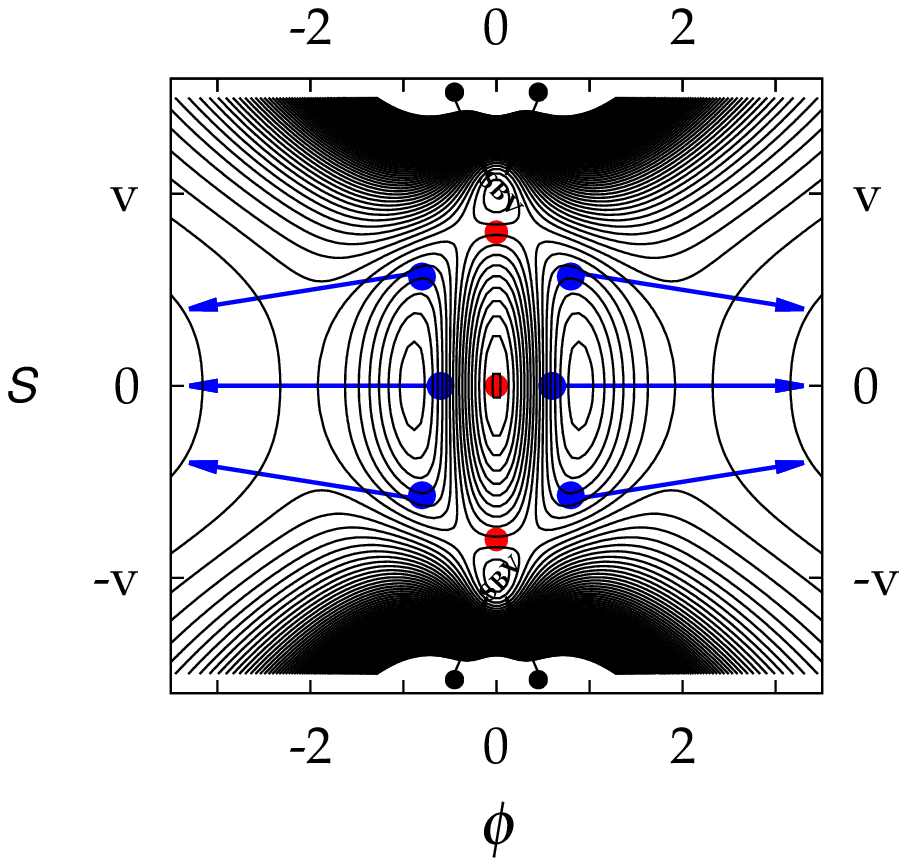}}
\caption{3D plots of the potential (\ref{eq23}) (upper panel). Plots from
left to right are for the values of non-minimal coupling parameter $\xi$
equal to $0.01$, $0.10$ and $0.50$ respectively. The corresponding contour
plots are shown in the lower panel. For these plots the relation between
parameters $\xi$ and $\zeta$ is taken as $\zeta = 3\,\xi^2$.
In contour plots,
the black points with arrows indicate the possible range of inflationary
trajectories that may end in the SBV, whereas the blue points with arrows
are used to show possible range of normal inflationary trajectories. The red
points in these plots indicate the critical points. For the purpose of the
plotting, we have taken $v = M_{pl}$ and $\lambda = 1$. Fields $\phi$ and
$S$ are in unit of $M_{pl}$.
This is applicable to all other plots related to the fields.}
\label{fig1}
\end{figure*}
From the numerical calculations we found that the sensible inflations can be
obtained for different small values (e.g. $\sim$ 10$^{-3}$) of the parameter 
$\zeta$ with corresponding small values (e.g. $\sim$ 10$^{-2}$) of the 
parameter $\xi$. Similarly, same amount of sensible inflation can be obtained 
for different small values of $\zeta$ by adjusting the corresponding small 
values of $\xi$ so that the ratio of their values remains constant. These 
behavours
of the parameters $\xi$ and $\zeta$ develop an ambiguous situation to 
constraint their values. It is also seen that the value of $\xi$ should be 
many times greater than the value of $\zeta$ for a reasonable inflation. To
avoid such equivocal situation as well as to simplify the formulations and 
calculations, and also keeping the numerically found relation between $\xi$ and 
$\zeta$ in mind as mentioned above, we consider an analytical relation between 
these two parameters as 
\begin{equation}
\zeta = 3\,\xi^2
\label{zxi}
\end{equation} 
for our present work. So, this relation is applicable to all values of $\xi$
mentioned below to find the related values of $\zeta$ and hence we will not
mentioned the value of $\zeta$ explicitly for any value of $\xi$ used in
our analysis below. We have derived this 
relation from the third and fourth terms of the equation (\ref{eqa}) on the 
basis of our consideration that $\xi$ and $\zeta$ are the respective first and 
second order non-minimal coupling strengths to gravity. It is to be noted that
there may be some other possible relations of type (\ref{zxi}) between these 
two parameters for reasonable inflations.                  
 
\begin{figure*}[hbt]
\centerline{
\includegraphics[scale=0.32]{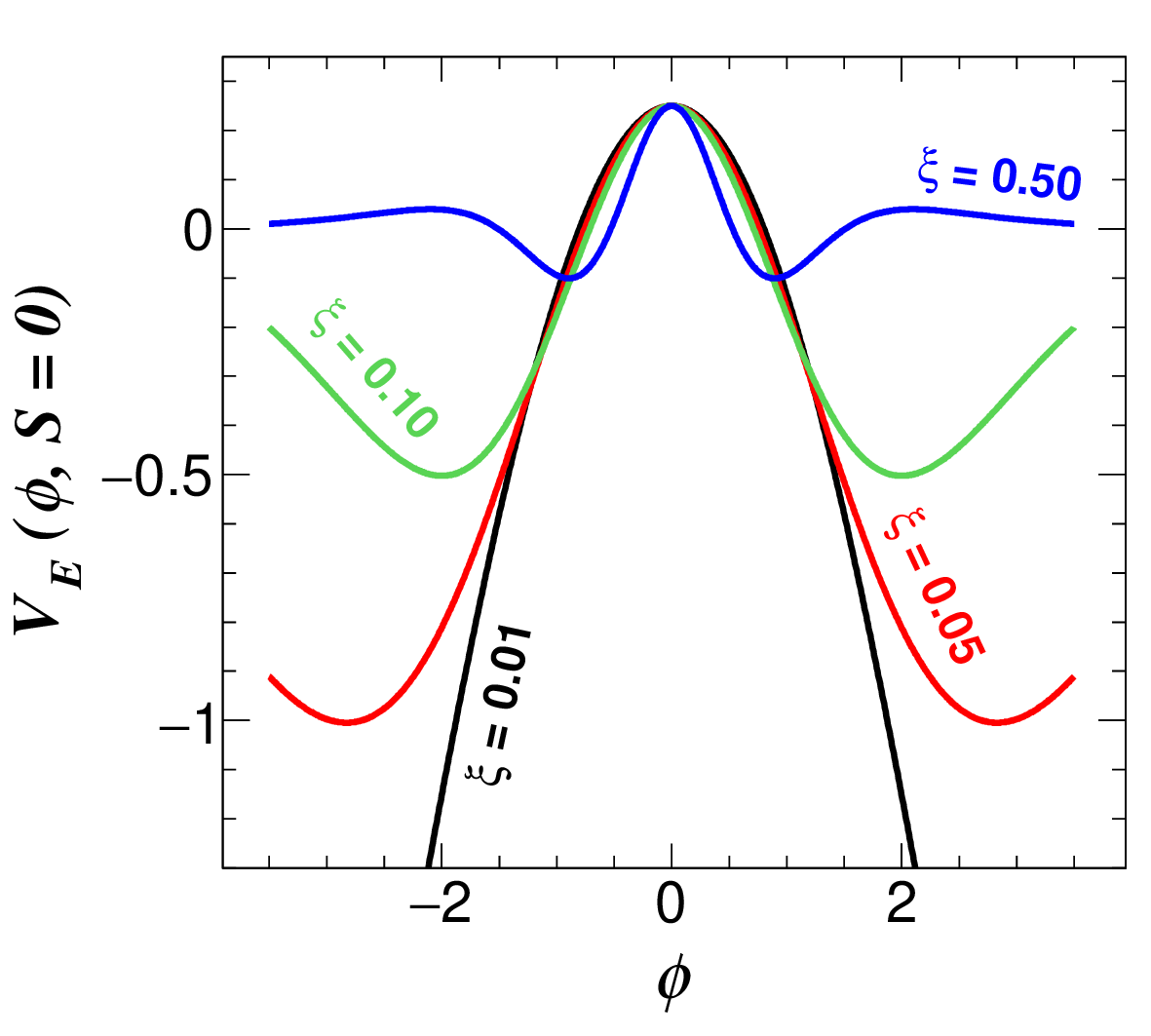}
}
\centerline{
\includegraphics[scale=0.32]{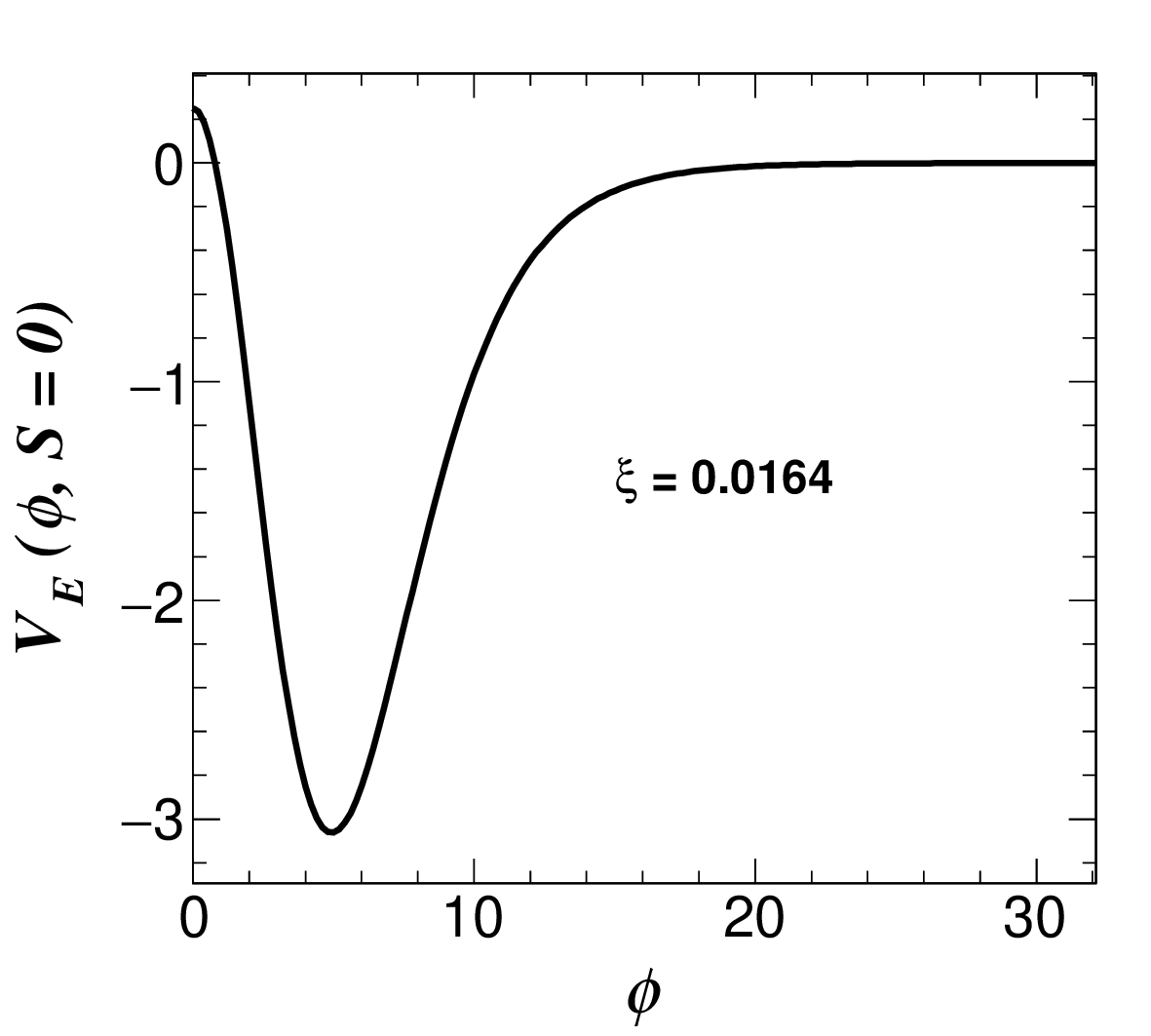}\hspace{-0.2cm}
\includegraphics[scale=0.32]{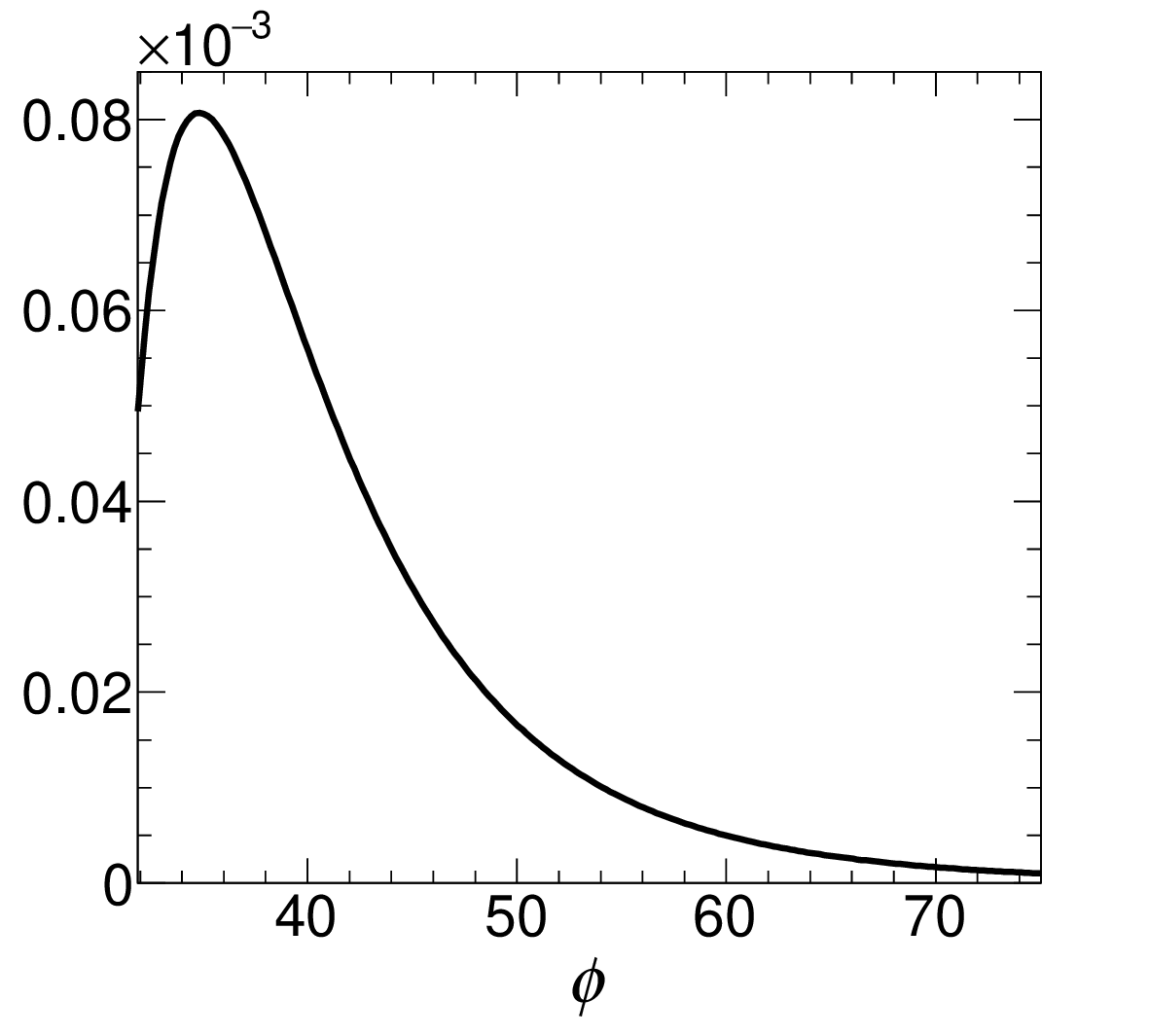}
}
\caption{Behaviour of the potential (\ref{eq23}) along $S = 0$ (i.e. potential
(\ref{eq25a})) with respect to
the inflaton field $\phi$ for different values of non-minimal coupling
parameter $\xi$. After certain initial values of $\phi$ depending on the value
of $\xi$ the potential becomes sufficiently flat along the direction of $S = 0$.}
\label{fig2}
\end{figure*}

Fig.\ref{fig1} shows the 3D and contour plots of the potential 
(\ref{eq23}) for three different values, viz., $0.01$, $0.10$ and $0.50$ 
respectively of the non-minimal coupling parameter $\xi$. It is observed that 
the field system or the inflationary dynamics becomes unstable with a higher 
value of the parameter $\xi$, because with a increasing value of $\xi$, the 
range of the inflationary trajectories to be end at SBV squeezes, whereas the 
range for the inflationary trajectories along direction of field $S$ increases. 
The $\phi = S = 0$ point is one of the unstable or critical points and hence 
from this point with $S = 0$, $\phi$ tends to roll to its higher value from 
both sides to the minimum of its potential. This rolling is faster for smaller 
value of the coupling parameter $\xi$. That is, for higher values of $\xi$ the 
potential becomes sufficiently flat along the normal inflationary trajectory 
(i.e. along $S = 0$) in comparison to its flatness for the smaller values of 
$\xi$ (also see the Fig.\ref{fig2}).

The potential along the normal inflationary trajectory is found from the 
potential (\ref{eq23}) as given by
\begin{equation}
V_E(S = 0) = \frac{\lambda^2v^4\left[2 - (3 + 10\xi)\phi^2 + \frac{25}{6}\zeta\phi^4\right]}{8\,\Omega_0^4\big[\Omega_0^2 + 6\phi^2(\xi^2-\frac{1}{3}\zeta)\big]}.
\label{eq25}
\end{equation}
Now using the relation (\ref{zxi}) between $\xi$ and $\zeta$, this potential
can be simplified further as
\begin{equation}
V_E(S = 0) = \frac{4\,\lambda^2\,v^4\,Q}{f_c^6},
\label{eq25a}
\end{equation}
where $Q = 4 - 2(3 + 10\xi)\phi^2 + 25\xi^2\phi^4$ and $f_c = 
2 + \xi\phi^2.$
It is seen that this potential is not minimum at $\phi = 0$, instead it becomes
negative for $0< |\phi| < \sqrt{\frac{2(3 + 10\xi)}{25\xi^2}}$. It 
attains its lowest 
negative (minimum) value at $\phi = \pm\, \sqrt{2/5\xi}$, depending on the 
value of $\xi$. After this value as $\phi$ increases, the potential starts to 
increase towards positive side and then remains positive for 
$|\phi|> \sqrt{\frac{2(3 + 10\xi)}{25\xi^2}}$ (see the bottom plot of the 
Fig.\ref{fig2}). As the negative potential can not be used to define the 
conventional inflation \cite{Felder2002} due to associated tachyonic 
instabilities, we will use this potential for 
$|\phi|> \sqrt{\frac{2(3 + 10\xi)}{25\xi^2}}$ to study its inflation 
dynamics depending on the value of $\xi$.            

\begin{figure*}[hbt]
\centerline{
\includegraphics[scale=0.3]{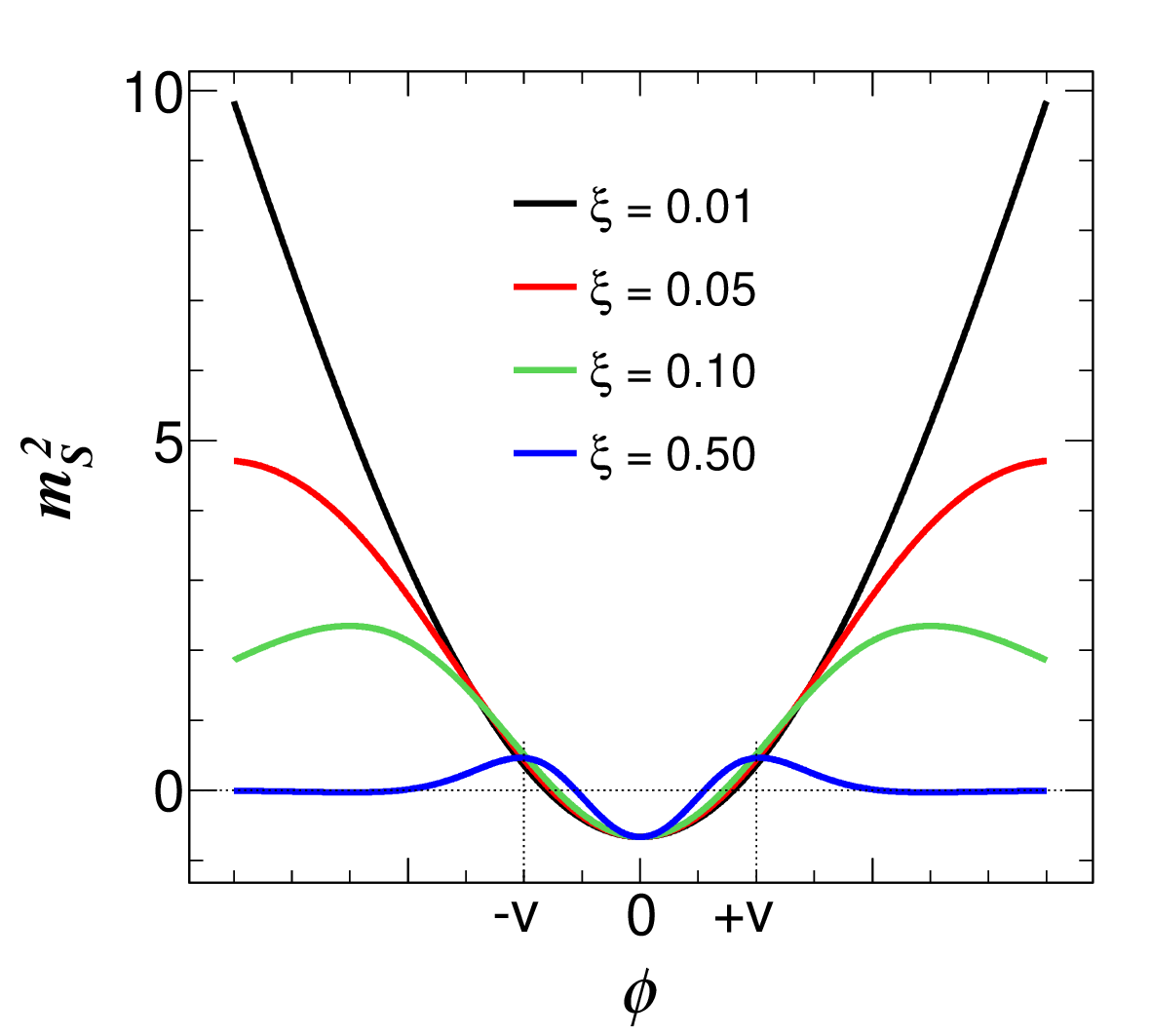}
\hspace{0.5cm}
\includegraphics[scale=0.3]{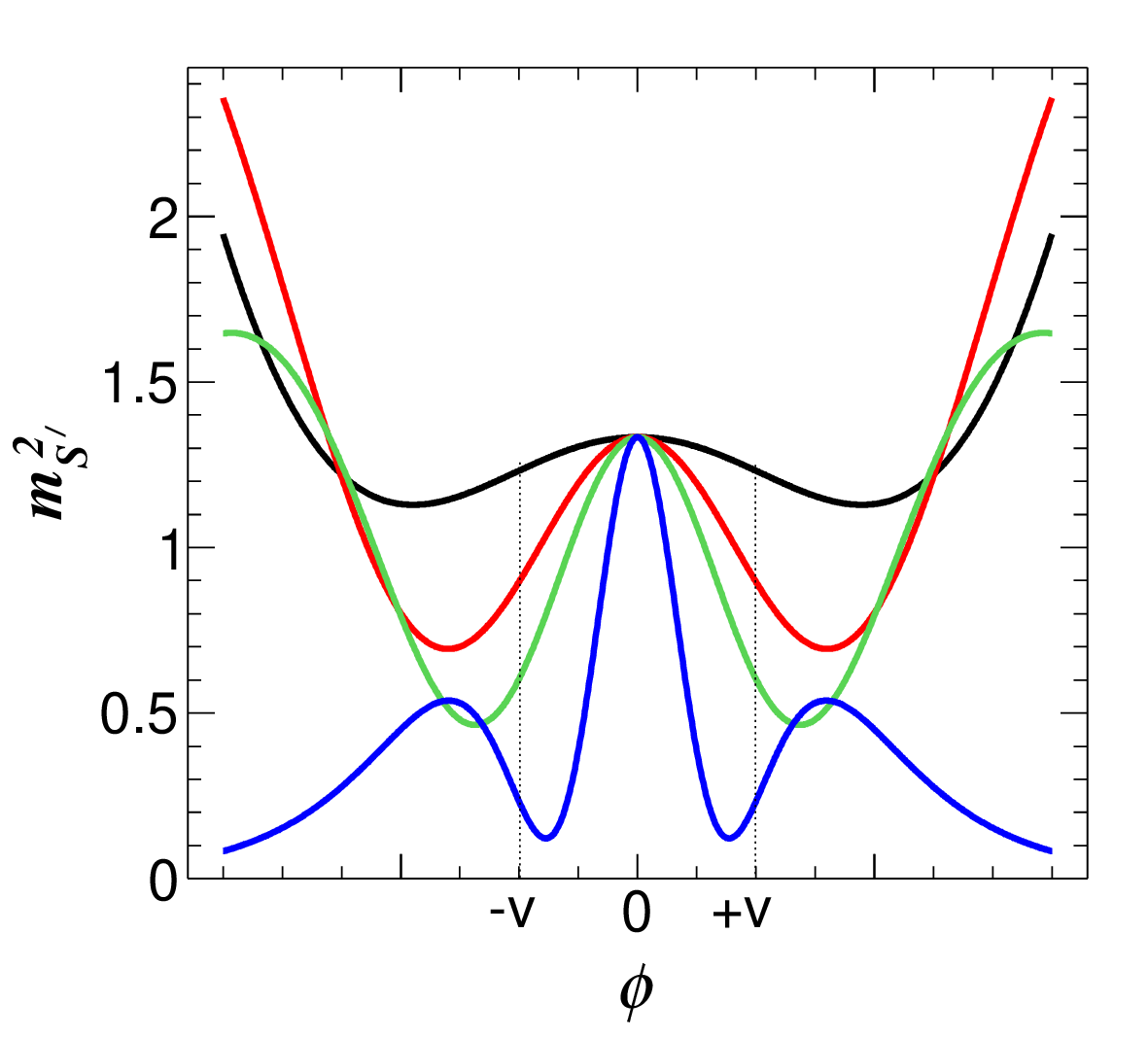}}
\centerline{
\includegraphics[scale=0.3]{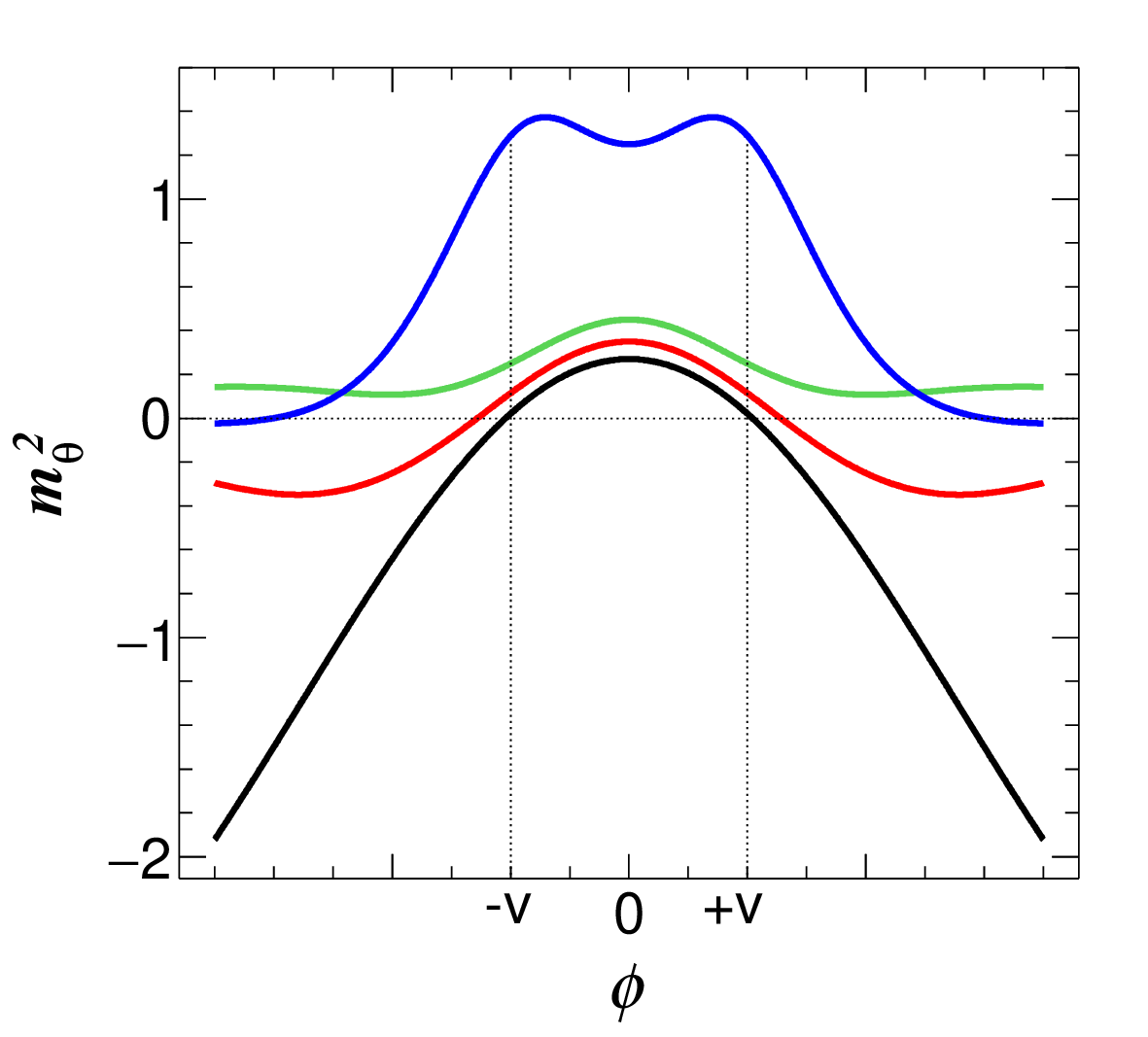}
\hspace{0.5cm}
\includegraphics[scale=0.3]{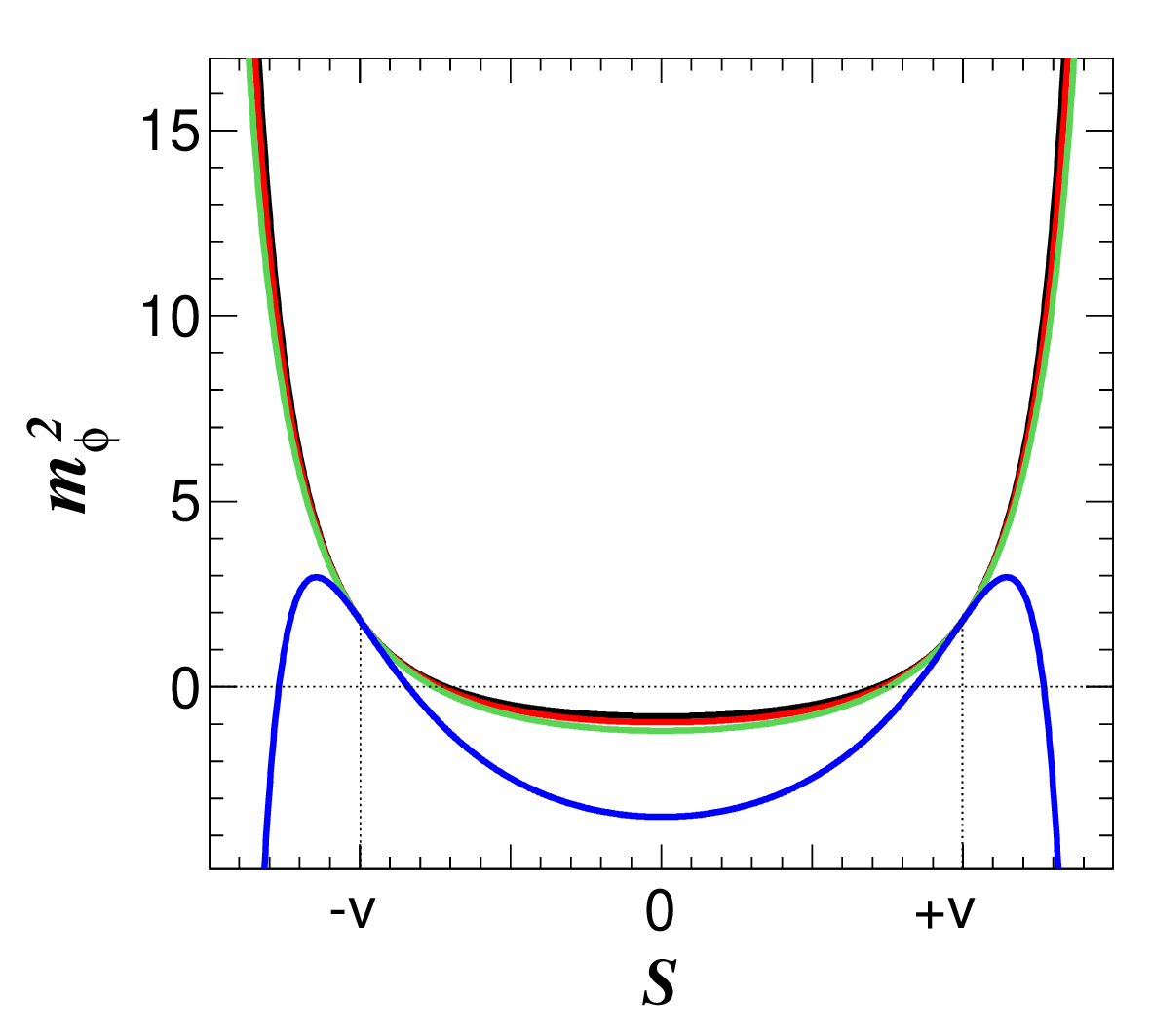}
}
\centerline{
\includegraphics[scale=0.27]{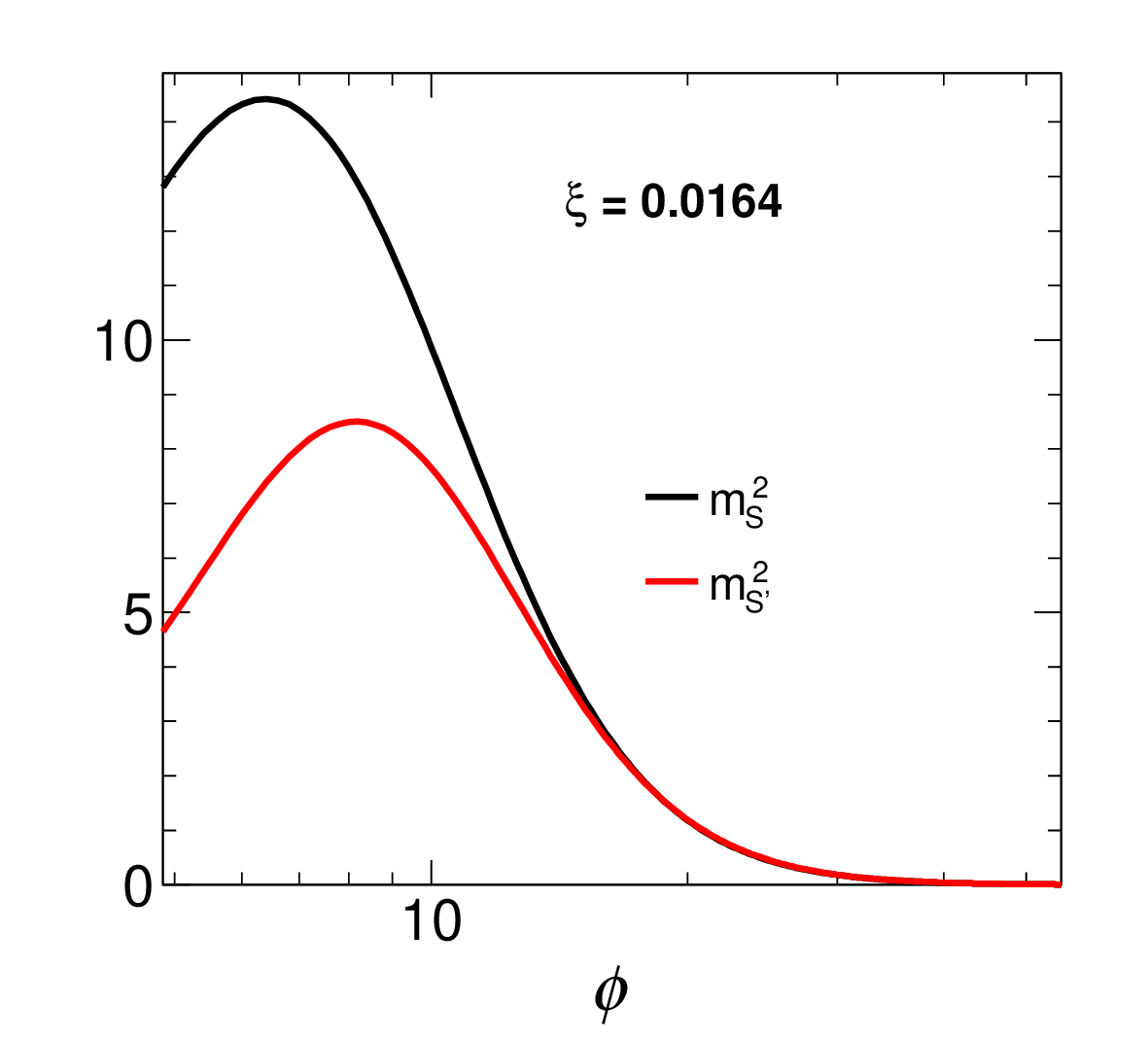}\hspace{-0.4cm}
\includegraphics[scale=0.27]{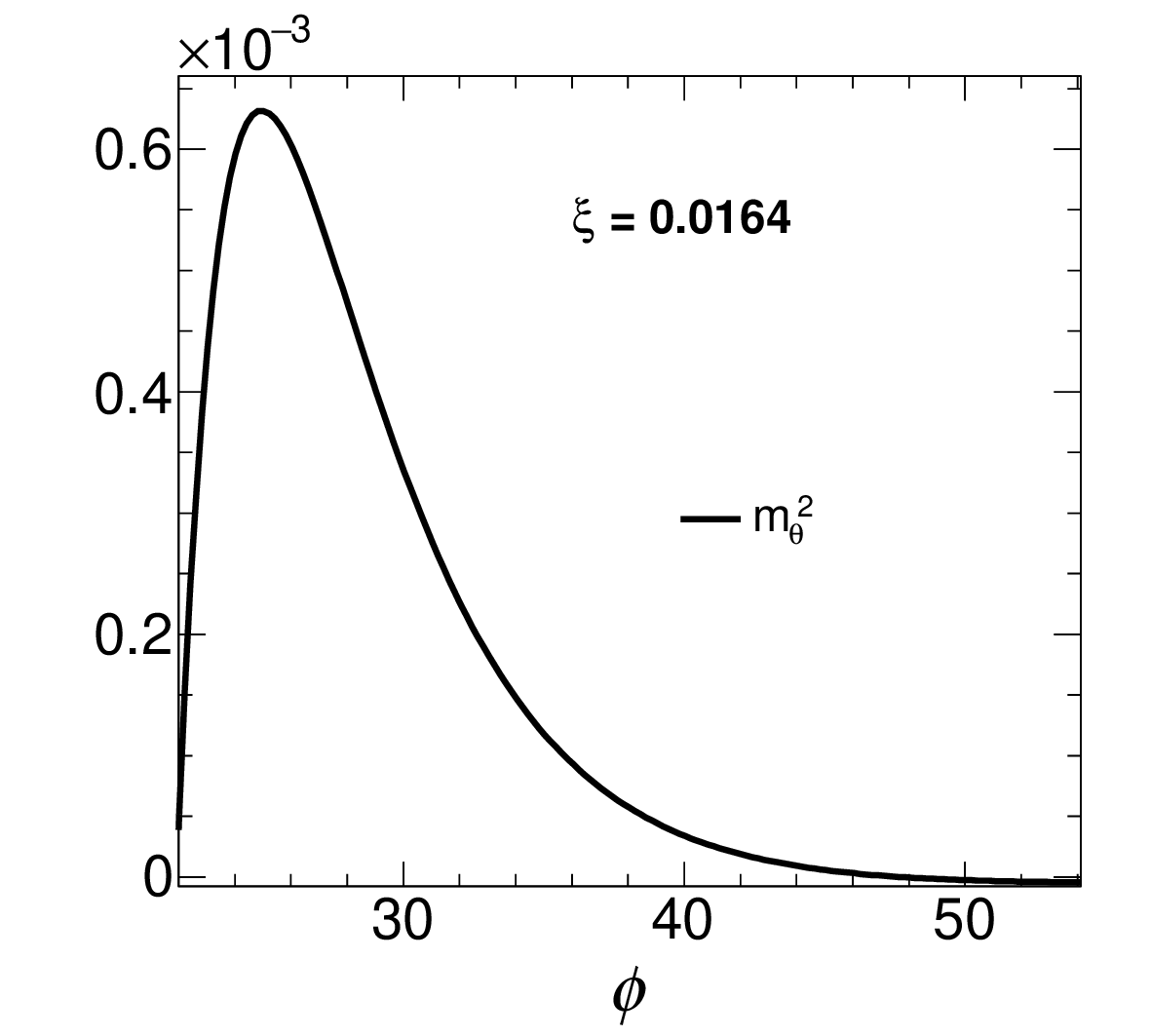}\hspace{-0.4cm}
\includegraphics[scale=0.27]{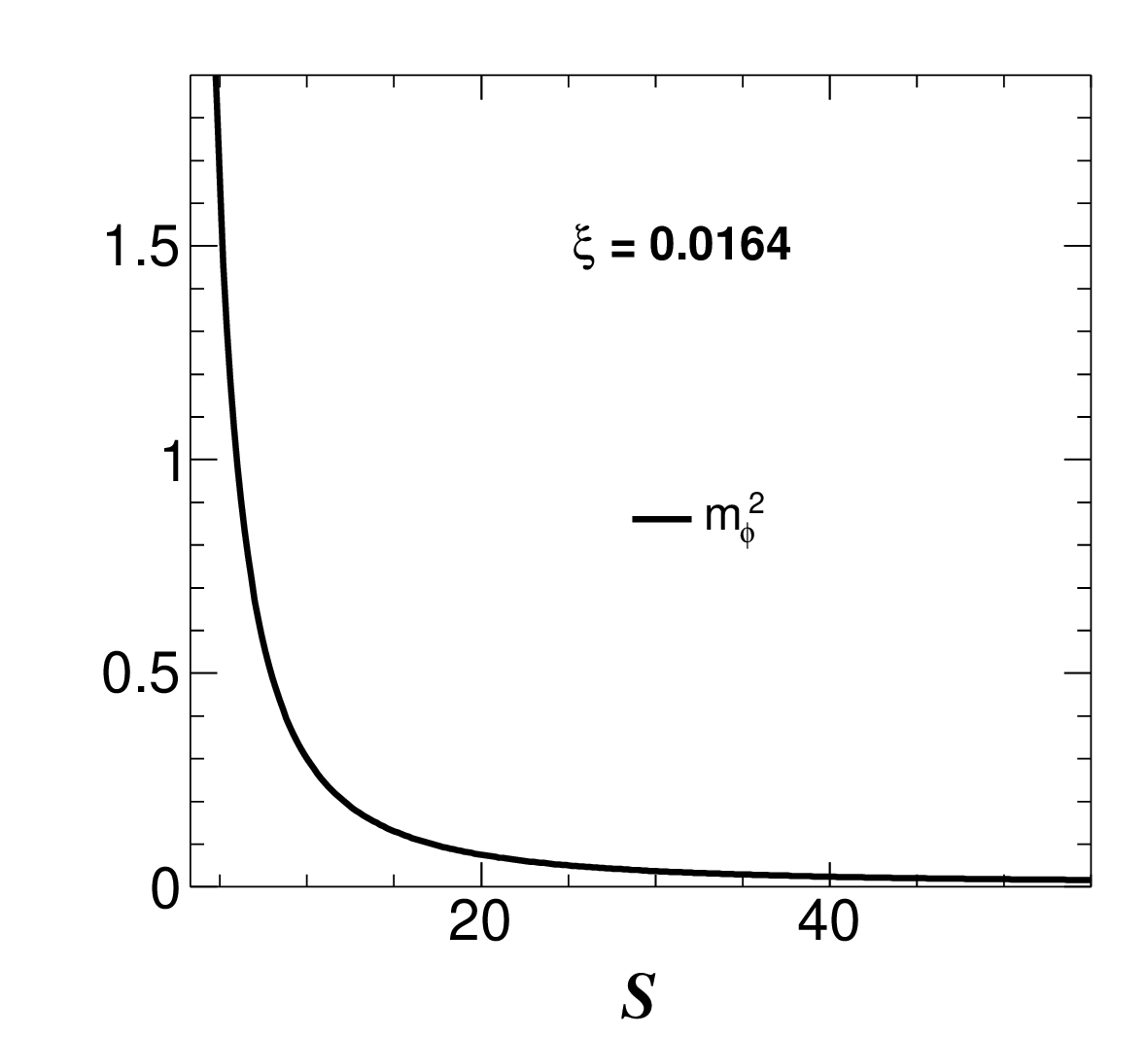}
}
\caption{Variation of mass square of the scalar fields $S$ (top left plot),
$S'$ (top right plot), $\theta$ (middle left plot) and $\phi$
(middle right plot) obtained from the equations (\ref{eq26}), (\ref{eq27}),
(\ref{eq28}) and (\ref{eq29}) respectively for different values of coupling 
parameter $\xi$. Bottom plots are to show the variation patterns of mass 
square of these fields with respect to the higher values of respective 
governing fields for $\xi = 0.0164$.}
\label{fig3}
\end{figure*}

To check the stability of the trajectory of inflation we calculate the 
masses squared of all scalar fields, viz., $S$, $S'$, 
$\theta$ and $\phi$. The mass squared of the field $S$ obtained from the 
potential (\ref{eq23}) is
\begin{multline}
m^2_S = \frac{16\lambda^2}{3f_c^8}\Big[3 \phi^2f_c^4 - 
   3 v^2 f_c^2 \big(4 - 2 (1 + 10 \xi) \phi^2\, + 25 \xi^2 \phi^4\big) - 
   4 v^4 \big(-4 + 2 (3 + 16 \xi + 6 \xi^2) \phi^2\\ - 
      5\, \xi^2 (11 + 12\, \xi) \phi^4 + 75\, \xi^4 \phi^6\big)\Big].
\label{eq26}
\end{multline}
From this equation we found that irrespective of the value of $\xi$, the mass 
of the field $S$ remains positive when $\phi \ge \pm \;v$, but may become 
tachyonic only for the values of $\phi < \pm\; v$ (in the case when
$v = M_{pl}$) with a slight dependence on the value of $\xi$. This tachyonic 
state, however, depends on the value of $v$. 
This is clearly indicated by the fact that at $\phi = 0$, $m^2_S =  
\lambda^2v^2(v^2 -3)/3$, which shows that the tachyonic state is possible only 
when $v<\sqrt{3} M_{pl}$. This means that within this range of values of $\phi$ 
there is an instability and hence the inflation is not well defined there. The 
top left plot of the Fig.\ref{fig3} shows these features of the $m_{S}$ with 
respect to field $\phi$ at different values of the $\xi$ as obtained from the 
equations (\ref{eq26}). It is seen that for $\phi >  |v|$, 
$m_{S}^2$ is highly dependent on the value of the $\xi$, as within this range 
of $\phi$, $m_{S}^2$ decreases very fast with increasing value of the $\xi$.
That is, the field $S$ becomes more heavier with the increasing $\phi$ only for
the smaller value of the parameter $\xi$.\\    
\indent The mass squared of the field $S'$ can be derived from the potential
for ${S'}$ corresponding to the potential (\ref{eq23}) as discussed above.
The expression for the $m^2_{S'}$ can be obtained as
\begin{multline}
m^2_{S'} = \frac{16\lambda^2}{3f_c^8}\Big[3 \phi^2f_c^4 + 
   3 v^2 f_c^2 \big(4 - 2 (1 + 10 \xi) \phi^2\, + 25 \xi^2 \phi^4\big) - 
   4 v^4 \big(-4 + 2 (3 + 16 \xi + 6 \xi^2) \phi^2 \\- 
      5\, \xi^2 (11 + 12\, \xi) \phi^4 + 75\, \xi^4 \phi^6\big)\Big].
\label{eq27}
\end{multline}
It is seen that $m^2_{S'}$ remains positive for all values of $\phi$.
At $\phi = 0$, $m^2_{S'} = \lambda^2v^2(v^2 +3)/3$. So there is no possible 
instability region to be associated with the field $S'$. Just like the field 
$S$, within the initial period of the field $\phi$ the mass of the field $S'$ 
is also highly coupling parameter $\xi$ 
dependent with a similar tendency, but with more prominence for almost all 
values of $\phi$ within the said period. That is, for this field also mass 
decreases with increasing value of $\xi$ and becomes more heavier with 
increasing $\phi$ for smaller value of $\xi$ only. However, the pattern of 
variation of 
$m^2_{S'}$ with respect to $\phi$ is appeared to be get inverted for almost 
for all values of $\xi$ in contrast to the patterns of $m^2_S$ for all values 
$\xi$  (see the top right plot of Fig.\ref{fig3}). It is also found that mass 
of the field $S$ is heavier than the field $S'$ within the initial period of 
the field $\phi$, but in the latter stage masses of these two fields become 
almost equal and also in this stage masses of both these fields decrease 
rapidly towards zero even for smaller value of $\xi$ (see the bottom left plot 
of Fig.\ref{fig3})  

The mass squared of the field $\theta$ as a function of the field 
$\phi$ can be found from the potential (\ref{eq18}) by assuming that 
$\mathcal{S} = \tilde{\mathcal{S}} = 0$ in the potential. Under this 
assumption the expression of the mass squared of the field $\theta$ is 
obtained from the said potential as
\begin{multline}
m^2_{\theta} = -\frac{4v^4\lambda^2}{f_c^6}\Big[-32 \xi - 120 \xi^3 \phi^4 + 50 \xi^4 \phi^6 + 4 (-1 + \phi^2) - 7 \xi^2 \phi^2 (24 + 5 \phi^2)\Big].
\label{eq28}
\end{multline}
It is obvious from this equation that the field $\theta$ becomes tachyonic
for $|\phi|>1$ (i.e. for $|\phi|>v$) when $\xi\rightarrow 0$. But in the case 
of non-zero $\xi$ the tachyonic state of the field $\theta$ depends highly on
the parameter $\xi$ and the stage of the inflaton field $\phi$. It is seen
that in the initial stage of $\phi$, the field $\theta$ gradually comes out
of the tachyanic state and at sufficiently high value of $\xi$ the field 
remains effectively almost out of this state (see for 
example the cases of $\xi \ge 0.1$ in the middle left plot of Fig.\ref{fig3}). 
Thus for the case of smaller values of $\xi<0.1$, there are some instabilities 
along the inflationary trajectory with $S=0$ for the field $\theta$ within 
certain initial range of values of $\phi$. However, beyond this range, i.e. in 
the latter stage with $|\phi|> \sqrt{\frac{2(3 + 10\xi)}{25\xi^2}}$ as mention
earlier, the mass of the field $\theta$ becomes positive and gradually decreases
towards zero with further increasing $\phi$, and finally its mass becomes 
almost zero even for very small value of $\xi$ (see the bottom middle plot of 
Fig.\ref{fig3}). Thus within this range of values of $\phi$ depending on
the value of $\xi$ the inflation could be well defined without any 
instability associated with $\theta$.  Again at $\phi = 0$, $m^2_{\theta} = 
\frac{1}{4}\lambda^2v^4(1 + 8\xi)$. This shows that the mass of the 
field $\theta$ is highly $\xi$ parameter dependent, specially for smaller 
values of inflaton field $\phi$ apart from its dependence on parameters 
$\lambda$ and $v$ (see the middle left plot of Fig.\ref{fig3}). 

From the Fig.\ref{fig3} it is seen that among the fields $S$, $S'$ and 
${\theta}$, the field $S$ appears to be more massive, whereas the field 
${\theta}$ appears to be least massive for the value of $\phi$
rolling away from origin with a smaller value of coupling parameter $\xi$, 
effectively upto certain period of inflation. As the field ${\theta}$ remains
non-tachyonic within the effective range of inflation as well as for the 
values of $\xi$ of our interest (see the section \ref{mp}), so it will not
generate any instability along the inflationary track ($S=0$) as already 
mentioned above.    

Finally, the mass squared of the inflaton field $\phi$ given by the 
potential (\ref{eq23}) is
\begin{multline}
m^2_\phi = \frac{9\lambda^2}{4\,(S^2 -3)^3(S^2-6)}
\Big[12 S^8 \xi^2 - 18 v^4 (3 + 16 \xi) + S^6 \big(11 + 48 \xi - 24 (3 + v^2) \xi^2\big)\\ + S^2 \big(72 + 36 v^2 (1 + 16 \xi) + v^4 (27 + 48 \xi - 72 \xi^2)\big) \\+ 6 S^4 \big(-5 - 48 \xi + 2 v^4 \xi^2 + v^2 (-5 - 16 \xi + 24 \xi^2)\big)\Big].
\label{eq29}
\end{multline}
It is found from this equations that at $S = v$, 
$m_{\phi} = \lambda v\sqrt{18/(v^4 -9v^2 +18)}$. Thus, 
it is interesting to note that the inflaton field acquired the constant mass 
independent of the parameter $\xi$, when waterfall field takes the constant 
parameter value $v$. This is a distinguishable behaviour of 
$m_{\phi}$ from the masses of other three fields. The middle right plot of 
Fig.\ref{fig3} shows the variation pattern of $m_{\phi}^2$ with respect to the 
waterfall field $S$ for different values of the coupling parameter $\xi$.
From the plot it is seen that the minimum value of the field $S$ above which 
the inflaton field $\phi$ remains as a real field, increases gradually with 
the value of the parameter $\xi$. At $S = 0$, 
$m^2_{\phi} = - \frac{1}{4}\lambda^2v^4(3 + 16\xi)$. Thus along the 
inflationary path ($S=0$), the field $\phi$ remains in tachyonic state.  
Moreover, we have seen that the mass of the inflaton field $\phi$ is more 
sensitive to the higher values ($\ge 0.5$) of the parameter $\xi$, as it 
should be. For the sake of completeness it should be mention that for smaller 
values of $\xi < 0.5$ the mass of the field $\phi$ increases vary rapidly with 
the field $S$ initially, but the mass decreases effectively to zero as $S$ 
increases to its higher value (see the bottom right plot of the Fig.\ref{fig3}).             
\section{Slow-roll inflation parameters and related observables}
As mentioned in the previous section, in this section we will use the 
potential (\ref{eq23}) only along the normal inflationary trajectory $S=0$, i.e.
the potential (\ref{eq25a}) to study the inflationary observables under
the slow-roll approximation. Also we have seen from the previous section that
the flatness of the potential (\ref{eq23}) or (\ref{eq25a}) is the coupling 
parameter dependent, such that the potential becomes sufficiently flat after 
some initial values of the inflation field $\phi$ depending upon the values 
$\xi$. So, within a reasonable 
degree of approximation it would be possible to apply the slow-roll 
approximation to the inflation driven by the potential (\ref{eq25a}).      
In the slow-roll inflation scenario, the slow-roll parameters are usually used 
to set constraints on the observable parameters produced by a given 
inflationary potential for the requirement of acceptable inflation. For a 
given potential $V(\varphi)$, the usual slow-roll parameters are expressed as
\begin{equation}
\epsilon \equiv \frac{1}{2}\left|\frac{V'}{V}\right|^2,\;\; \eta \equiv \frac{V''}{V},\;\; \xi_2 \equiv \frac{V'V'''}{V^2},\;\; \xi_3 \equiv \frac{V'^2V''''}{V^3},
\label{eq30}
\end{equation} 
in the unit of $M_p = 1$, where primes denote the derivatives with respect to
$\varphi$. These parameters are used to express the observable in the slow-roll
approximation of inflation, which are
\begin{eqnarray}
n_t \;=\; -2\epsilon = -\frac{r}{8}, \label{eq31}\\
n_s \;=\; 1+2\eta-6\epsilon, \label{eq32}\\
n_{tk} \;=\; \frac{dn_t}{d\;ln\ \kappa} = 4\epsilon(\eta - 2\epsilon), \label{eq33}\\
n_{sk} \;=\; \frac{dn_s}{d\;ln\; \kappa} = 16\epsilon\eta - 24\epsilon^2 - 2\xi_2, \label{eq34}\\\nonumber
n_{skk} \;=\; \frac{d^2n_s}{d\; ln\; \kappa^2} = - 192\epsilon^3 + 192\epsilon^2\eta - 32\epsilon\eta^2 \nonumber\\
 - 24\epsilon\xi_2 + 2\eta\xi_2 + 2\xi_3, \label{eq35}\\
\delta_H^2(\kappa)\;=\; \frac{1}{150\pi^2}\frac{\Lambda^4}{\epsilon_H},
\label{eq36} 
\end{eqnarray} 
where $n_t$ is the tensor spectral index, $n_{tk}$ is its running, $n_s$ is 
the scalar spectral index, $n_{sk}$ is its running and
$n_{skk}$ is the running of the running. $\delta_H^2(\kappa)$ is the density
perturbation at wave number $\kappa$ and $\Lambda \equiv V_H^{1/4}$ is the
scale of inflation. $r$ is the ratio of tensor to scalar
perturbations, defined by
\begin{equation}
r \equiv \frac{P_t}{P_s} = 16\epsilon,
\label{eq37}
\end{equation}
\begin{equation}
\mbox{where}\;\; P_s = \frac{1}{2\epsilon}\left(\frac{H}{2\pi}\right)^2,\;\; 
\mbox{and}\;\; P_t = 8\left(\frac{H}{2\pi}\right)^2
\label{eq38} 
\end{equation}
are the power spectrum of scalar and tensor perturbations respectively. It 
should be noted that all the quantities with a subscript $H$ are evaluated 
at the inflation scale $\phi_H$, at which some 50 -- 60 e-folds are produced 
before the end of inflation. If we define a quantity $\delta_{ns}$ by 
\begin{equation}
\delta_{ns} \equiv 1 - n_s,
\label{eq39} 
\end{equation}
then using the equation (\ref{eq32}) we may
rewrite the equation (\ref{eq33}) as a constraint equation among the
observables as \cite{Mariana} 
\begin{equation}
n_{tk} = \frac{r}{64}(r-8\delta_{ns}).
\label{eq40}
\end{equation}    
This constraint equation should be satisfied by any model of inflation that is 
based on the slow-roll approximation as it is model independent. Hence it
would indicate a departure from the slow-roll approximation if the values
obtained for the observables fail to satisfy this equation. 

The reported value of the tensor-to-scalar ratio from the BICEP2 \cite{BICEP2} 
experiment is $r = 0.20^{+0.07}_{-0.05}$. After the foreground subtraction
based on dust model its value is reported as $r = 0.16^{+0.07}_{-0.05}$. It is
a significantly higher value than the upper bound set by the Planck 
Collaboration ($r < 0.11$ at 95\% c.l.) \cite{PLANCK,PLANCK2}, indicating a large scale 
inflation. From this reported data of the BICEP2 experiment it can be inferred 
that the scale of inflation to lie in the range $2.03\times 10^{16}$ GeV 
$<\Delta<$ $2.36\times 10^{16}$ GeV. But, most recent combined analysis of the
data of BICEP2 and Keck Array experiments \cite{BICEP2Keck} sets the upper 
limit of $r$ as $< 0.07$ at $95$\% c.l., which is in consistent with the 
Planck's upper bound. Similarly, the latest data of Planck \cite{PLANCK2} give
the constraint on the scalar spectral index with $n_s$ = $0.968\pm 0.006$ at
$68$\% c.l.. As the BICEP2 experiment data on $r$ reported in the 
Ref. \cite{BICEP2} is overestimated and also there is(are) no conclusive 
observation(s) yet on the expected value $r$, we will use upper limits of 
Planck's and BICEP2/Keck's data on $r$ to analyze our
model. Hence, in the light of these data, if we take $\delta_{ns} = 0.032$, a 
simple calculation from the equation (\ref{eq40}) gives that $n_{tk}$ will 
take values within the ranges of $-2.51\times 10^{-4} < n_{tk} < 0$ and 
$-2.03\times 10^{-4} < n_{tk} < 0$ for the Planck's and BICEP2/Keck's upper 
limits on $r$ respectively. Thus these will additionally set constraints to 
test our model.

\section{The model's predictions}
\label{mp}
Substituting the scalar potential (\ref{eq25a}) in expressions for the slow-roll
parameters in the equation (\ref{eq30}), we obtain the explicit expressions 
for these parameters in terms of the inflaton field $\phi$ and the non-minimal
coupling parameters $\xi$ along the inflation direction $S=0$ as
\begin{eqnarray}
\epsilon = \frac{8 \phi^2 \big(6 - 100 \xi^2 \phi^2 + 50 \xi^3 \phi^4 + \xi (32 - 15 \phi^2)\big)^2}{f_c^2Q^2},\;\;\;\;\;\;\;
\label{eq41}\\
\eta = \frac{4}{f_c^2Q}\Big[-12 - 1600 \xi^3 \phi^4 + 450 \xi^4 \phi^6 + 8 \xi (-\,8 +\;\;\;\;\;\;\nonumber\\ 21 \phi^2) + \xi^2 (1016 \phi^2 - 165 \phi^4)\Big], \;\;\;\;\;\;\;\label{eq42}\\
\xi_2 = \frac{1}{f_c^4Q^2}\Big[192 \xi \phi^2 \big(6 - 100 \xi^2 \phi^2 + 
50 \xi^3 \phi^4 + \xi (32 -\;\;\;\;\;\nonumber\\ 15 \phi^2)\big) \big(-72 - 2050 \xi^3 \phi^4 + 375 \xi^4 \phi^6 +\;\;\;\;\;\;\nonumber\\ \xi (-424 + 306 \phi^2) + \xi^2 (2252 \phi^2 - 165 \phi^4)\big)\Big],\;\;\;\;\;\;\;\label{eq43}\\
\xi_3 = \frac{1}{f_c^6Q^3}\Big[768 \xi \phi^2 \big(6 - 100 \xi^2 \phi^2 + 50 \xi^3 \phi^4 + \xi (32 -\;\;\;\;\;\nonumber\\ 15 \phi^2)\big)^2
\big(144 - 31900 \xi^4 \phi^6 + 4125 \xi^5 \phi^8 +\;\;\;\;\;\;\nonumber
\end{eqnarray}
\begin{eqnarray}
 \xi (848 - 3060 \phi^2) + 80 \xi^2 \phi^2 (-259 + 78 \phi^2)\,+\;\;\;\;\;\; \nonumber\\ \xi^3 (54280 \phi^4 - 2145 \phi^6)\big)\Big].\;\;\;\;\;\;\;
\label{eq44} 
\end{eqnarray}  

These equations, viz., (\ref{eq41}), (\ref{eq42}), (\ref{eq43}) and 
(\ref{eq44}) for the parameters $\epsilon$, $\eta$, $\xi_2$ and $\xi_3$ 
respectively are very complex in form. So, it is not easy to find any 
analytical relation among these parameters and hence we need to calculate
these parameters from these generic equations to study the slow-roll 
inflation to be obtained for the potential (\ref{eq25a}).  
 
\begin{figure*}[hbt]
\centerline{
\includegraphics[scale=0.32]{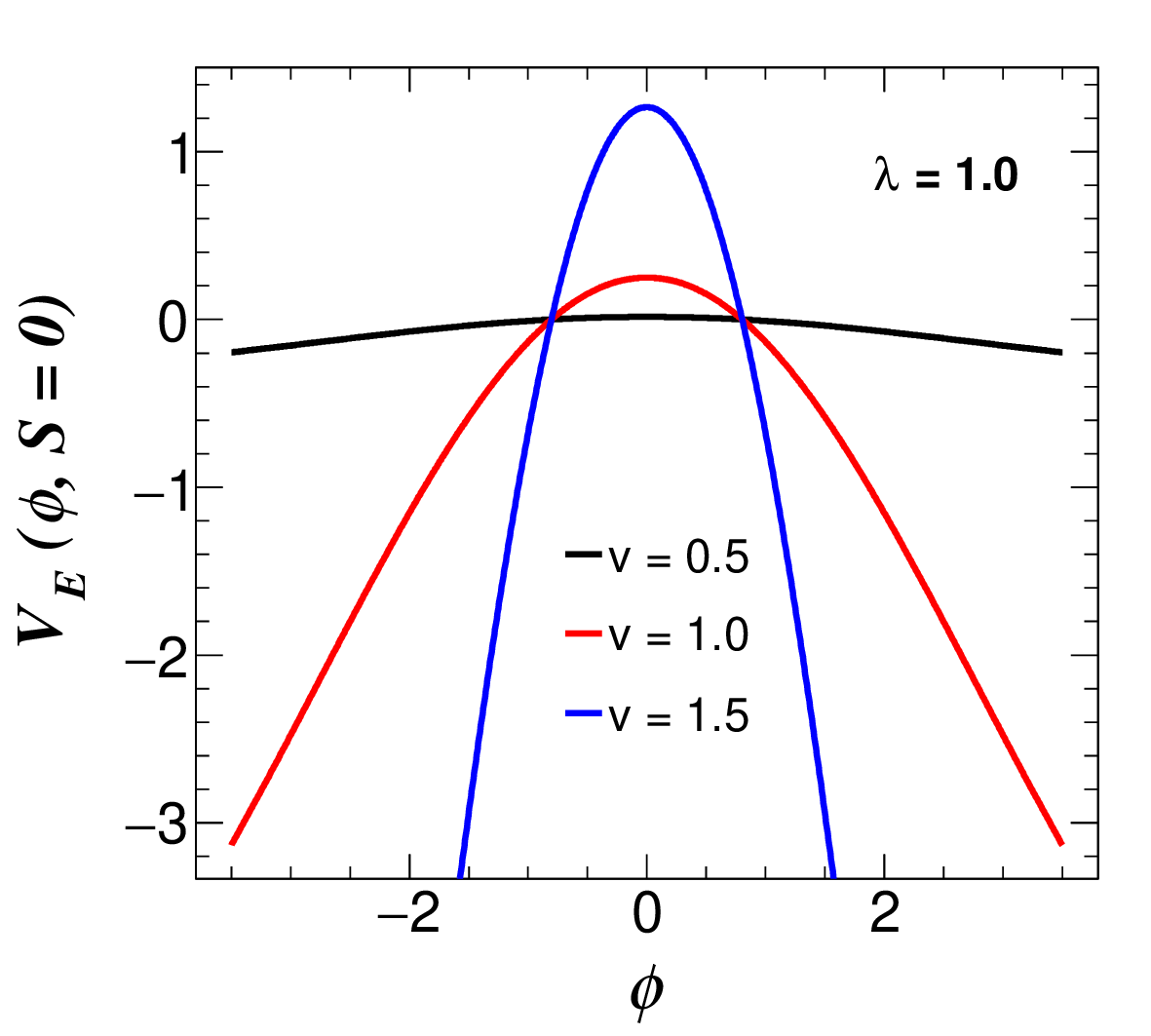}\hspace{0.5cm}
\includegraphics[scale=0.32]{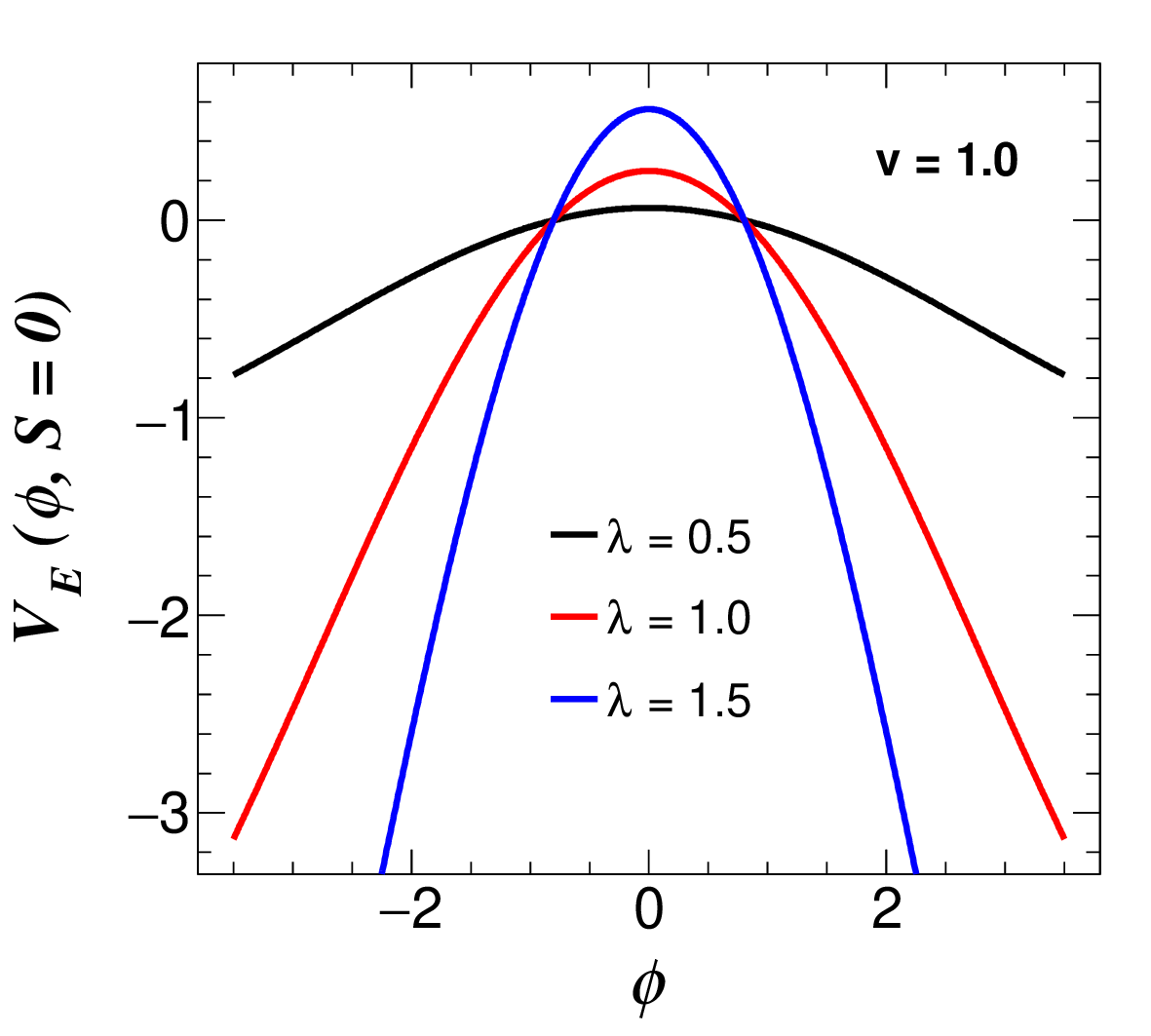}}
\caption{Effect of $\lambda$ and $v$ on the potential (\ref{eq25a}). $v$ is 
taken in unit of $M_{pl}$. The potential depends heavily on $v$ and $\lambda$,
and is more sensitive to $v$. Here the value of $\xi$ is fixed at 0.01.} 
\label{fig4}
\end{figure*}

It is interesting to see that the slow-roll parameters $\epsilon$, $\eta$, 
$\xi_2$ and 
$\xi_3$ are independent of potential parameters $\lambda$ and $v$. 
Consequently $\lambda$ and $v$ do not have any direct influence on observables
of the slow-roll inflation. However, as the potential (\ref{eq25a}), and the
masses of the scalar fields $S$, $S'$, $\theta$ and $\phi$ depend 
heavily on parameters $\lambda$ and $v$ (see equations 
(\ref{eq26}) -- (\ref{eq29})), the flatness of the potential and hence 
stability range of inflation should be dependent on these two parameters. In 
this context, we have studied the behaviour of the potential (\ref{eq25a}) for 
two different sets of combinations of values of $\lambda$ and $v$ as shown in 
the Fig.\ref{fig4}. From the figure it is clear that the shape of the 
potential depends very much on the values of $\lambda$ and $v$, and it is more
sensitive to $v$. With decreasing values of $\lambda$ and $v$, the
potential becomes sufficiently flat. Consequently it implies that with 
decreasing values of $\lambda$ and $v$, the value of the term $\lambda^2v^2$
becomes very small rapidly and hence in particular the tachyonic range of 
$m_S$ or the instability range of 
the inflaton field $\phi$ will reduce considerably (see the Fig.\ref{fig3}).   
      
\subsection{Relations between $r$, $n_s$  and $\phi$}
\begin{figure*}[hbt]
\centerline{
\includegraphics[scale=0.26]{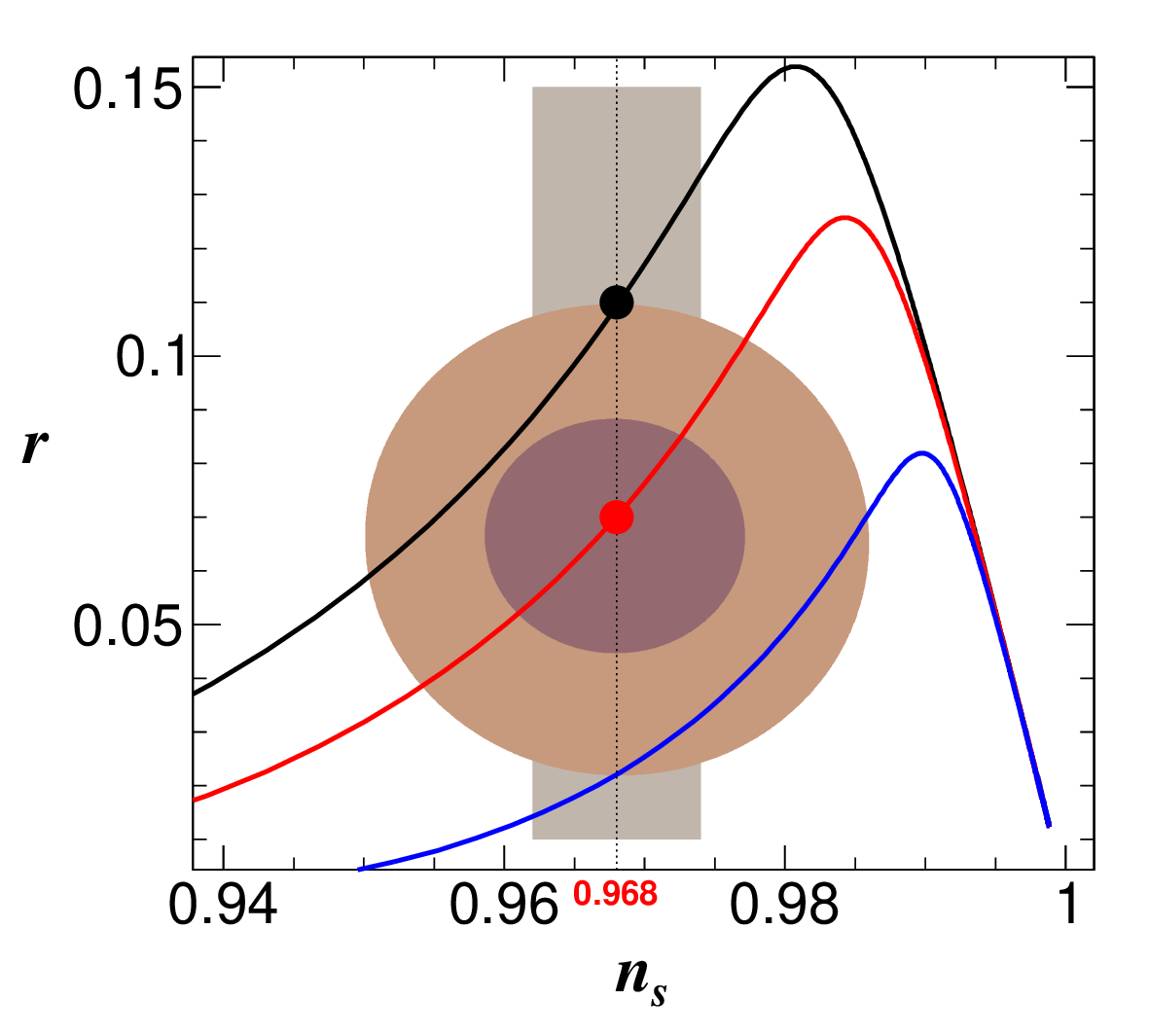}
\hspace{0.5cm}
\includegraphics[scale=0.26]{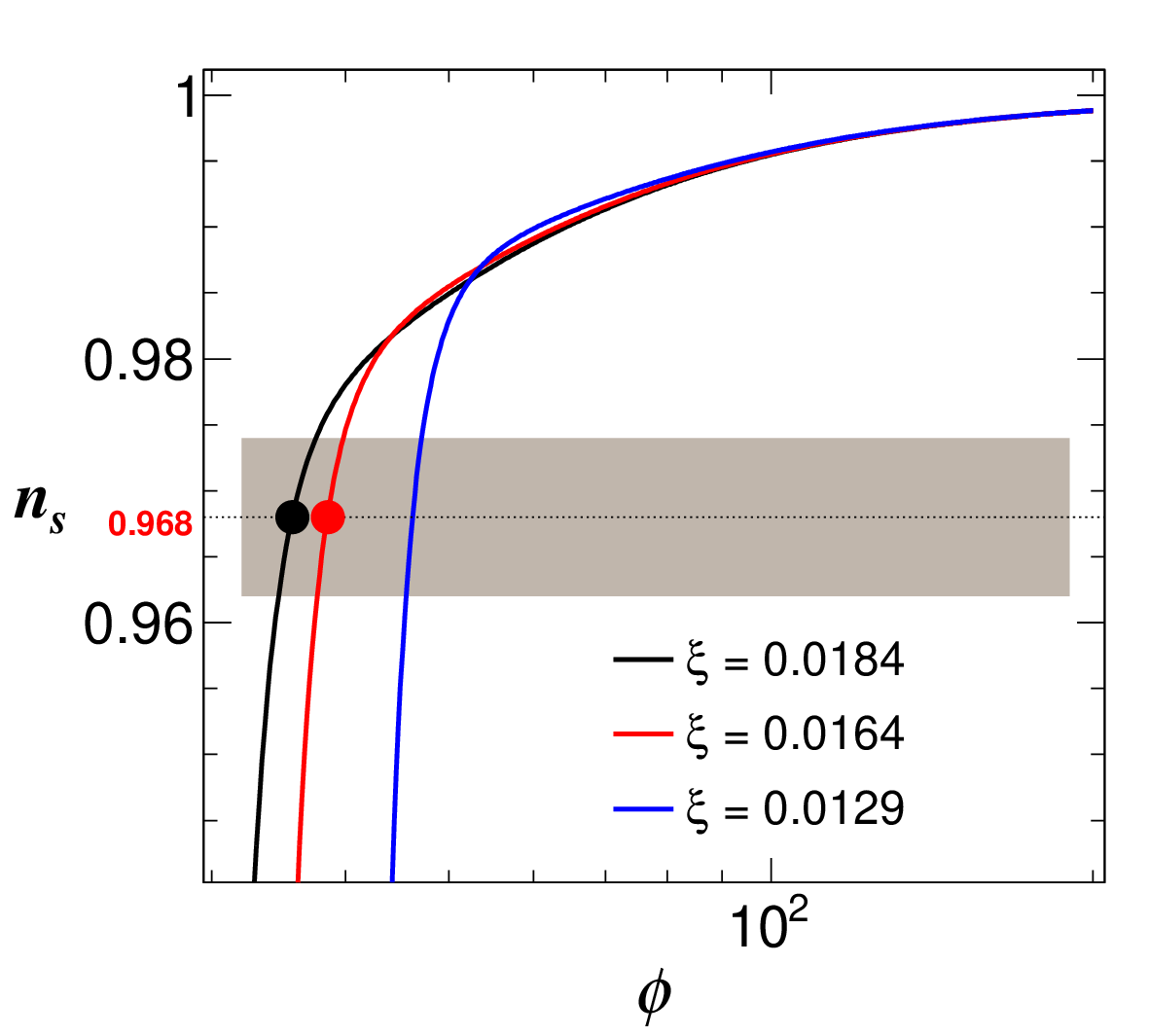}
\hspace{0.5cm}
\includegraphics[scale=0.26]{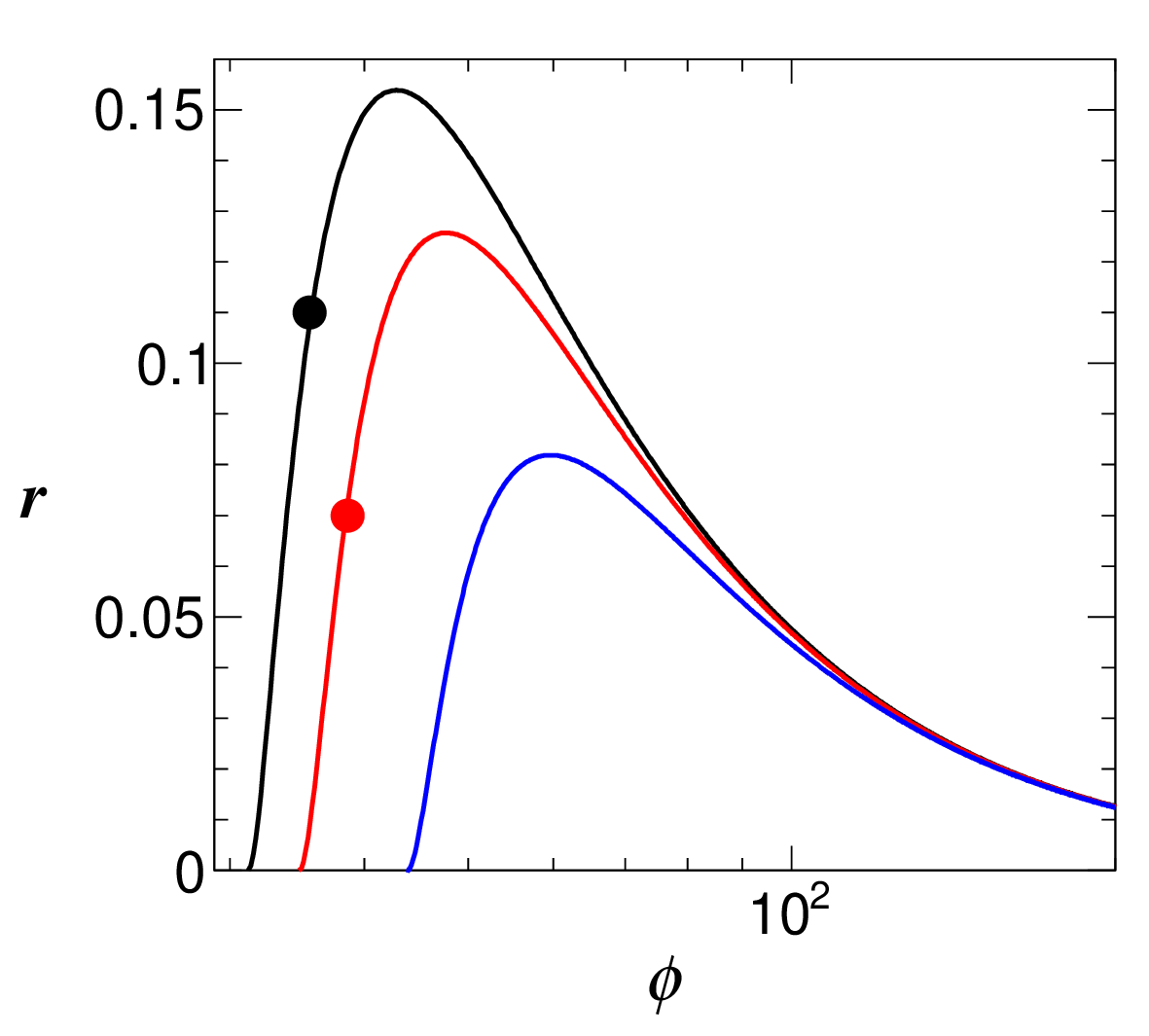}}
\vspace{0.5cm}
\centerline{
\includegraphics[scale=0.26]{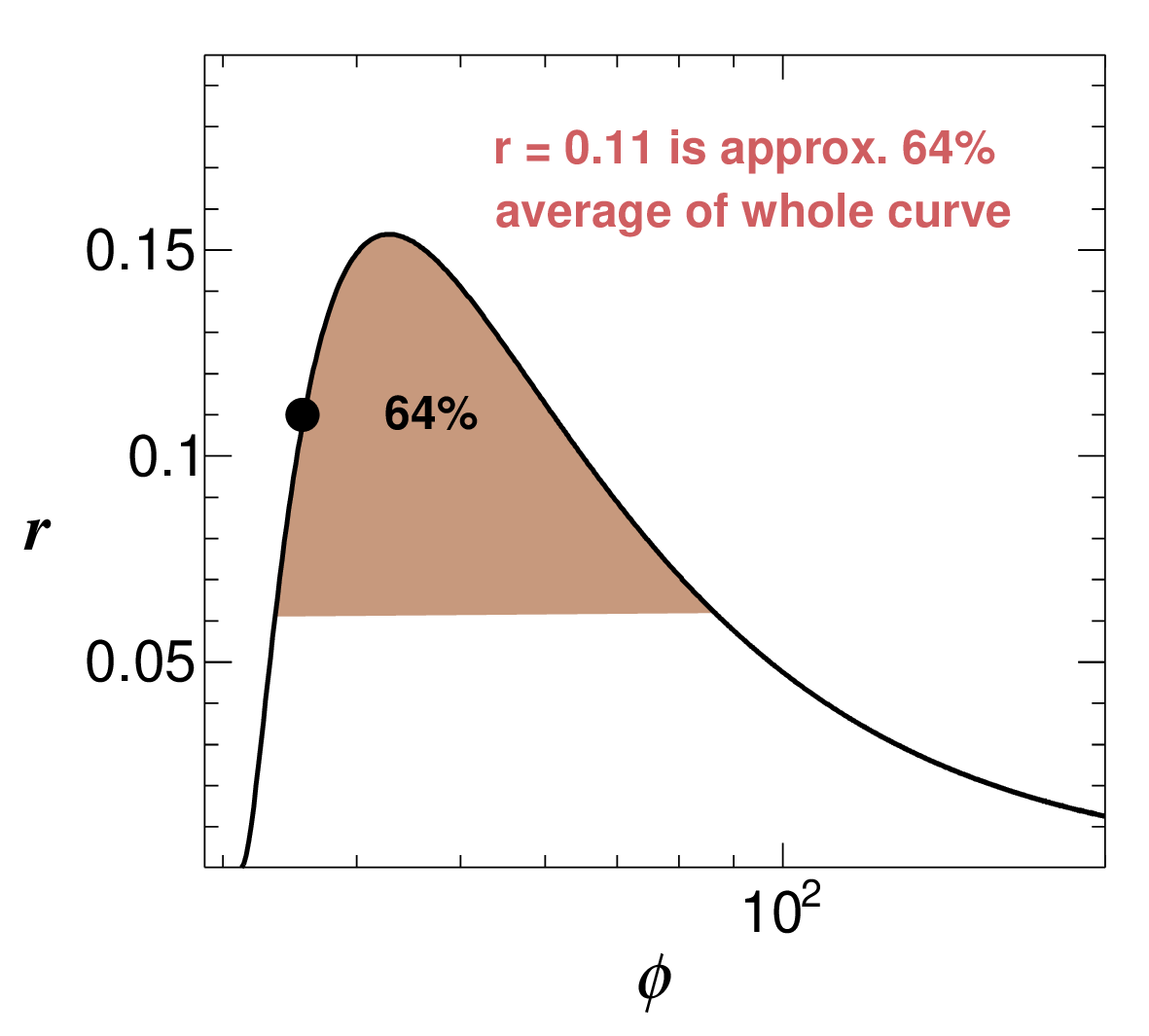}
\hspace{0.5cm}
\includegraphics[scale=0.26]{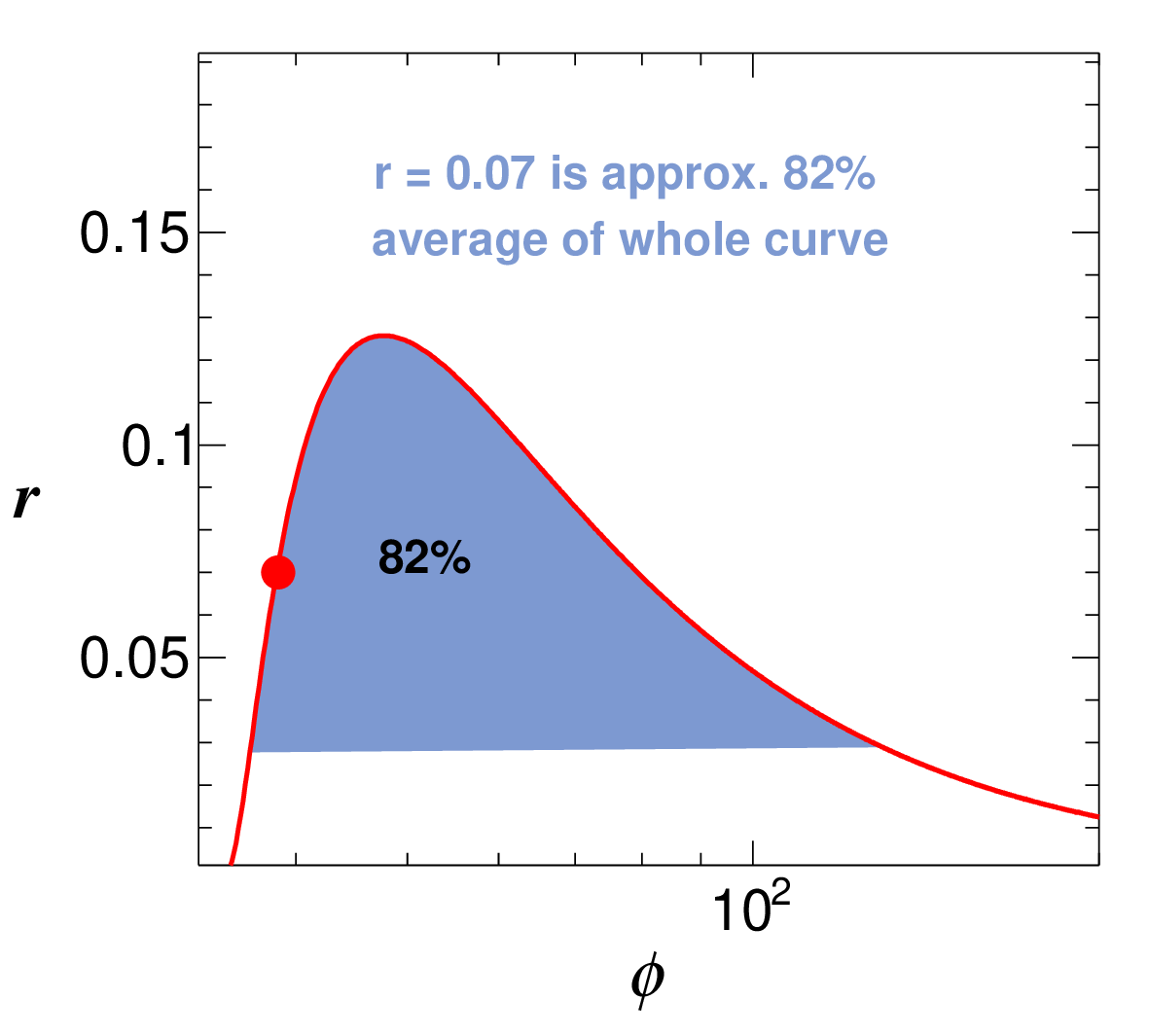}}
\caption{Variation of the tensor-to-scalar ratio $r$ 
in terms of the scalar spectral index $n_s$ (top left plot), variation of 
$n_s$ with respect to the inflaton field $\phi$ (top middle plot) and 
variation of $r$ with respect to $\phi$ (top right plot) for different values 
of non-minimal coupling parameter $\xi$ constrained by the Planck's \cite{PLANCK,PLANCK2} and BICEP2/Keck's \cite{BICEP2Keck} data. In the plots, black 
coloured solid circle indicates the Planck's data and red coloured solid 
circle indicates the BICEP2/Keck's data. In the $\phi-n_s$ and $\phi-r$ plots, the upper bounds of 
$r$ of Planck's and BICEP2/Keck's data correspond to $\phi \approx 35.7$ and
$\approx 38.5$ respectively. Contours in the $n_s-r$ plot indicate the
$95\%$ and $68\%$ c.l. obtained from Planck's data. The shaded area in the
$n_s-r$ and $\phi-n_s$ plots indicates the range of Planck's $n_s$ data. The 
bottom panel's plots represent 
the average values of $r$ over certain ranges of distributions of $r$ w.r.t. 
$\phi$ that are approximately equal to different observed bounds on $r$.}
\label{fig5}
\end{figure*}

For visualization of variation of the tensor-to-scalar ratio $r$ and the scalar
spectral index $n_s$ with respect to the inflaton field $\phi$ as well as the
variation of $r$ with respect to $n_s$, we may do the numerical calculations  
by substituting the equations (\ref{eq41}) and (\ref{eq42}) in equations 
(\ref{eq32}) and substituting equation (\ref{eq41}) in the equation 
(\ref{eq37}) for different values of non-minimal coupling parameter $\xi$. 
The values of the parameter $\xi$ are decided on the basis of bounds of data 
of $r$ and $n_s$ as mentioned above. The results of these 
calculations are shown in the Fig.\ref{fig5}. It is seen that as the 
constraint set by the Planck's data on $n_s = 0.968 \pm 0.006$, according to 
the upper bounds of $r$ given by the Planck and BICEP2/Keck experiments, the 
actual value of $r$ should lies below the values of the same for $\xi < 0.0184$
and $\xi < 0.0164$ respectively. Moreover, the BICEP2/Keck's data on upper 
bound of $r$ lies well within the $95\%$ and $68\%$ confidence level (c.l.) 
contours obtained from Planck's data. We found that under the same constraint,
the lower bound of $95\%$ c.l. contour is obtained for the value of 
$\xi = 0.0129$ (see the top left plot of the Fig.\ref{fig5}). This indicates 
that the most probable value of $\xi$ should lie within $0.0164^{+0.0020}_{-0.0035}$.

The values of $\phi$ corresponding to data points in the $\phi-n_s$ plot
(top middle plot of the Fig.\ref{fig5}) are guided to draw the same points in
the $\phi-r$ plot (top right plot of the Fig.\ref{fig5}). These values 
of $\phi$ are $\approx 35.7$ and $\approx 38.5$ corresponding to upper 
bounds of Planck's ($r < 0.11$) and BICEP2/Keck's ($r < 0.07$) data 
respectively. From these two values of $\phi$, we see that Planck's and 
BICEP2/Keck's bounds on $r$ are closed in the inflationary phases. 

It is also clear from the $\phi-r$ plot that the value of $r$ is maximum 
($r_{max}$) at a particular value of $\phi$ ($\phi_{rmax}$). However, both 
data points lie at initial phases than $\phi_{rmax}$, such that smaller the 
value of data point with smaller $\xi$, the phase of $\phi$ would be slightly 
more initial than corresponding $\phi_{rmax}$. Thus data points in $\phi-r$ 
plot may be considered as the 
average value of $r$ over a certain range of $\phi$ including $\phi_{rmax}$ 
for the concerned value of $\xi$. The upper bounds of BICEP2/Keck's and 
Planck's data on $r$ are found to be the average values of $r$ over 
$\approx 82\%$ and $\approx 64\%$ distributions of $r$ with respect to $\phi$ 
for $\xi = 0.0164$ and $\xi = 0.0184$ respectively (see the bottom plots of 
the Fig.\ref{fig5}). Therefore during the inflation, the evolution of $\phi$ 
from $\phi_H$ to $\phi_e$ (at the end of inflation) should always contains the 
value $\phi_{rmax}$, where $r$ reaches its maximum value $r_{max}$ before 
starts to decrease. Moreover, the average values of $r$ over whole 
distributions for $\xi = 0.0164$ and $\xi = 0.0184$ are
$\approx 0.048$ and $\approx 0.053$ respectively. Both these values of $r$ lie 
within the upper bounds set by the Planck and BICEP2/Keck collaboration.  

It should be noted that upto certain initial value of the inflaton field 
$\phi$, the inflation is not well defined (giving random and large values
$r$ and $n_s$) due to the reason discussed in the previous section related to 
the masses of the scalar fields. From the potential (\ref{eq25a}) this required
initial value of $\phi$ can be found as 
$|\phi| \equiv \phi_c = \sqrt{\frac{3+ 16\xi}{10\xi^2}}$.  Thus the initial or 
the minimum values of $\phi$ from which the inflation is found to be well 
defined for $\xi = 0.0184$, $0.0164$ and $0.0129$ are $\approx 31.2$, 
$\approx 34.8$ and $\approx 43.9$ respectively. From this initial value of 
$\phi$ depending on the
value of $\xi$, the potential (\ref{eq25a}) becomes positive and sufficiently 
flat, and consequently the mass squared of all fields ($S$, $S'$, $\theta$ and
$\phi$) remain positive favouring for the slow-roll approximation to the
inflation dynamics. Again, we take that value of $\phi$ as end value 
at which the value of $n_s$ becomes equal to $1$, and both $r$ and $n_s$ become 
independent of the parameter $\xi$. We found this end value of $\phi\sim 200$.
So, our range of $\phi$ lies between the initial value (for a particular $\xi$) 
and the end value.
   
\subsection{Number of e-foldings}
\label{nef}
To see what are the values of e-foldings that are related with the values of
inflaton field $\phi$ corresponding to upper bounds of tensor-to-scalar ratio 
$r$ set by different experiments as mentioned above, we have derived the 
expression for the number of e-foldings \cite{Bassett} given by the potential 
(\ref{eq25a}) as
\begin{eqnarray}
N = \frac{\phi^2-\phi_c^2}{16} + a\,\ln\!\frac{\phi}{\phi_c}+
b\,\ln\!\Big(\frac{5\xi\phi^2-2}{5\xi\phi_c^2-2}\Big)+
\alpha\,\ln\!\Big(\frac{10\xi^2\phi^2 - 16\xi -3}{10\xi^2\phi_c^2 - 16\xi -3}\Big),
\label{eq45}
\end{eqnarray}
where 
\begin{align}
\notag
a = &\; (3+ 16\xi)^{-1},\\\notag
b = &\; 3\big(25\xi(1 + 4\xi)\big)^{-1},\\
\notag
\alpha = &\; \frac{3 (1 + 12 \xi)^2 (3 + 20 \xi)}{800 \xi^2 (3 + 28\xi + 64 \xi^2)}
\end{align}
and as discussed in the previous subsection $\phi_c$ is the critical value of 
$\phi$ below which (specially) the field $\theta$ becomes unstable due to 
tachyonic instability ($m_\theta^2<0$). As mentioned there, these critical
values of $\phi$ are $\approx 31.2$, $\approx 34.8$ and $\approx 43.9$ for 
$\xi = 0.0184$, $0.0164$ and $0.0129$ respectively. The equation 
(\ref{eq45}) also shows that for a physically acceptable value of $N$, 
$\phi_c$ should be greater than $\sqrt{\frac{3 + 16\xi}{10\xi^2}}$. This is 
obvious because at $\phi = \sqrt{\frac{3 + 16\xi}{10\xi^2}}$ the potential 
becomes just positive as mentioned earlier.  
\begin{figure}[hbt]
\centerline{
\includegraphics[scale=0.32]{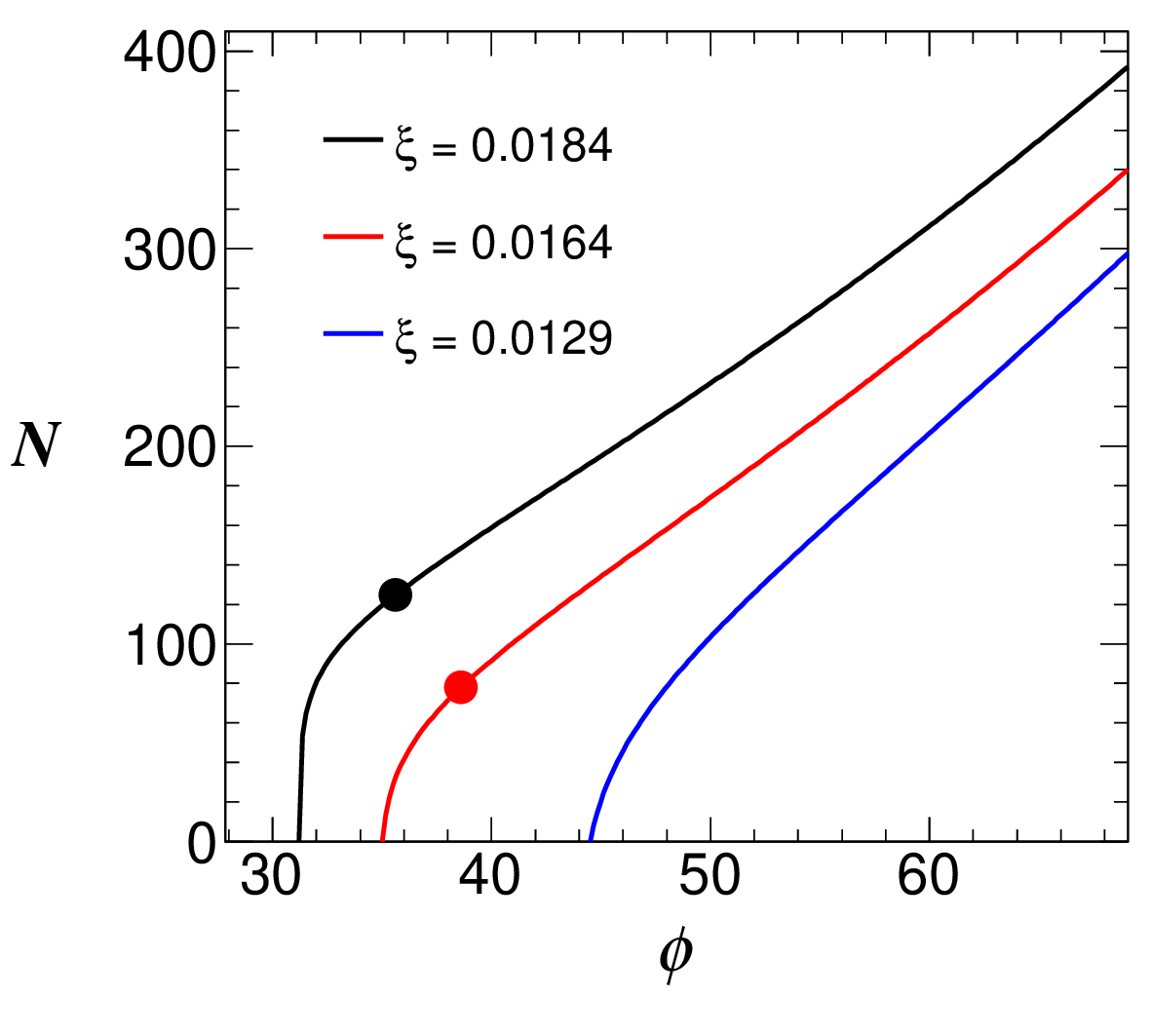}}
\caption{Variation of the number of e-foldings ($N$) with respect to the 
inflaton field $\phi$ for different values of the non-minimal coupling
parameter $\xi$ that are used on the basis of bounds of tensor-to-scalar ratio 
$r$ set by the Planck \cite{PLANCK,PLANCK2} and
BICEP2/Keck \cite{BICEP2Keck} collaborations. In the plot, the black and
red coloured solid circles represent respectively the results corresponding to 
Planck's and 
BICEP2/Keck's data on upper bound of $r$. The Planck's and BICEP2/Keck's data 
give $N \approx 125$ with $\phi \approx 35.7$ and $N \approx 78$ with 
$\phi \approx 38.6$ respectively.}
\label{fig6}
\end{figure} 

The numerical calculations using the equation (\ref{eq45}) give that the upper 
bounds of $r$ from the BICEP2/Keck and the Planck experiments correspond to 
$\approx 78$ and $\approx 125$ e-foldings with $\phi = 38.5$ and 
$\phi = 35.7$ respectively (see the Fig.\ref{fig6}). It 
is clear that the smaller value of $\xi$ or $r$ gives smaller value of number 
of e-foldings for all values of $\phi$. Yet, it is difficult 
to infer something from the prediction of number of e-foldings by the potential using this analysis. However, it is certain that the model could predict a 
sufficient number of e-foldings ($> 50-60$) for upper bounds of $r$ set by the 
BICEP2/Keck and the Planck experiments. The number of e-foldings $
N\ge 50-60$ is required for any model of inflation to be valid \cite{Bassett}. 
Our model's prediction in terms of the number of e-foldings is discussed in the
subsection \ref{secef} below. 
 
\subsection{Relations between $r$, $n_{tk}$ and $\phi$}        
\begin{figure*}[hbt]
\centerline{
\includegraphics[scale=0.32]{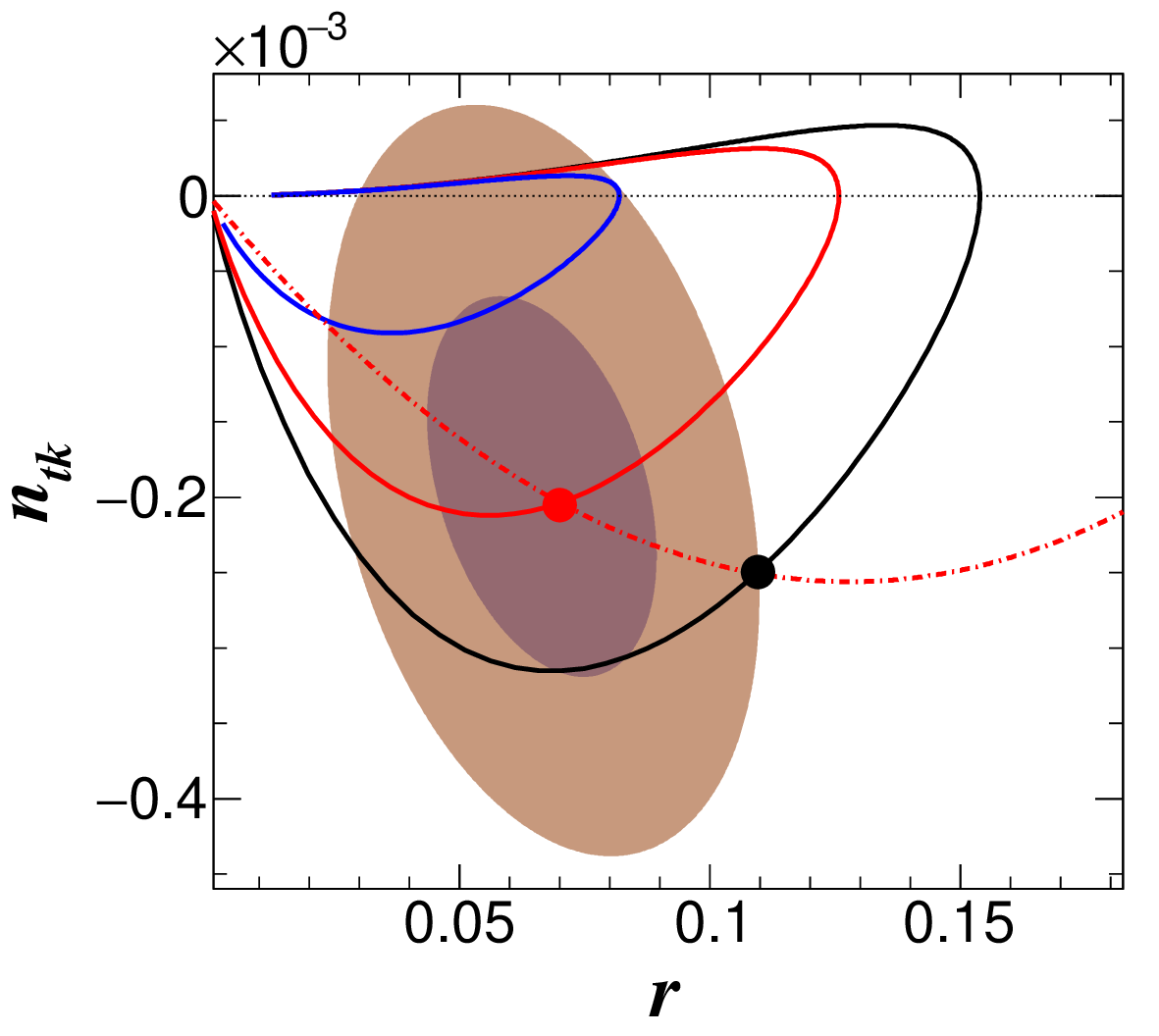}
\hspace{0.5cm}
\includegraphics[scale=0.32]{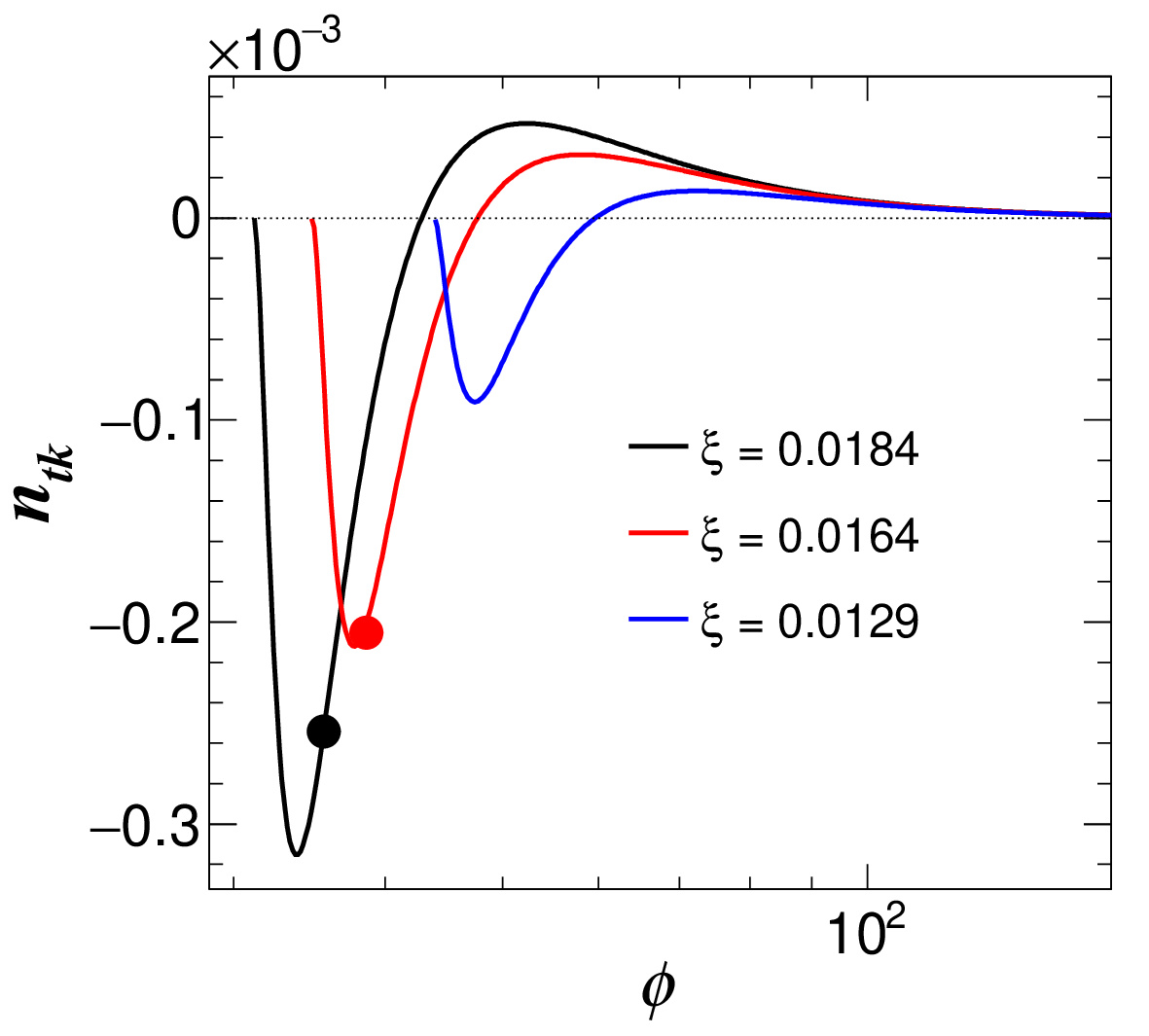}}
\caption{Variation of the running of the tensor spectral index $n_{tk}$ as a 
function of $r$ and of the field $\phi$ for different values of non-minimal
coupling parameter $\xi$ used on the basis of experimental upper bounds on 
$r$. In the $r-n_{tk}$ plot, the dash-dot-dash curve represents the model 
independent values of $n_{tk}$ obtained from the equation (\ref{eq40}) for 
values of $r$ from $0$ to $0.17$. In the same plot, contours represent $95\%$ 
and $68\%$ c.l. obtained on the basis of Planck's data. In both plots, the 
black and red solid circles represent the results for the upper bounds of 
the Planck's and the BICEP2/Keck's data on $r$ respectively.}
\label{fig7}
\end{figure*}

Numerical calculations are done on the equation (\ref{eq40}) using 
equations (\ref{eq41}) and (\ref{eq42}) via equations (\ref{eq32}) and 
(\ref{eq37}) to see the variation of the running of the tensor spectral index 
$n_{tk}$ as a function of the tensor-to-scalar ratio $r$ and the field 
$\phi$ for different values of the parameter $\xi$ corresponding to 
experimental upper bounds on $r$ as mentioned above. The model 
independent numerical calculation is also done using equation (\ref{eq40}) for 
values of $r$ from $0$ to $0.17$. The results of all these calculations are 
shown in the Fig.\ref{fig7}. The model independent $r-n_{tk}$ curve is found 
to passages through points on model based $r-n_{tk}$ curves for different 
$\xi$ (see the left plot of Fig.\ref{fig7}). These points give the 
values of $n_{tk}$ corresponding to experimental upper bounds on $r$ 
(refer to previous section). This suggests that the values of 
$n_{tk}$ for $\xi<0.0164$ and $\xi<0.0184$ lie within the upper bounds of 
BICEP2/Keck's and Planck's data respectively. From this result we may say that 
the slow-roll approximation imposed on our model is appropriate.  
Furthermore, the value of $n_{tk}$ for the upper bound of 
BICEP2/Keck's data is found to lies well within the $95\%$ and $68\%$ c.l. 
contours drawn on the basis Planck's data. Hence, the most probable 
value of the non-minimal coupling parameter $\xi$ should lie within 
$0.0164^{+0.0020}_{-0.0035}$. This is the same inference as we have drawn above.
      
$n_{tk}$ is appeared to has oscillatory nature from negative to positive 
value with increasing value of $\phi$ (see the right plot of 
the Fig.\ref{fig7}). But the oscillation is asymmetric, as it is seen that the 
amplitude of the oscillatory curve on negative side is very large in 
comparison to its
positive counterpart. Thus on average, each curve will give negative value of
$n_{tk}$ for a particular value of $\xi$. Comparison with the top right plot 
of the Fig.\ref{fig5} it is clear that at $\phi_{rmax}$, $n_{tk}$ reaches 
to its zero value. This is because, the $n_{tk}$ represents the second 
harmonics of the tensor power spectrum, whereas $r$ represents its principal 
harmonics. $n_{tk}$ attains its minimum negative value at a particular value 
of $\phi$ during its initial stage of evolution in the inflation phase. Almost 
at or near to this stage, the values of $n_{tk}$ for $\xi<0.0164$ and 
$\xi<0.0184$ lie within the upper bounds of BICEP2/Keck's and Planck's data 
respectively. The values of $\phi$ corresponding to both these experimental 
limits of $r$ are same as mentioned earlier.  
 
\subsection{Relations between $r$, $n_{sk}$ and $\phi$}
\begin{figure*}[hbt]
\centerline{
\includegraphics[scale=0.26]{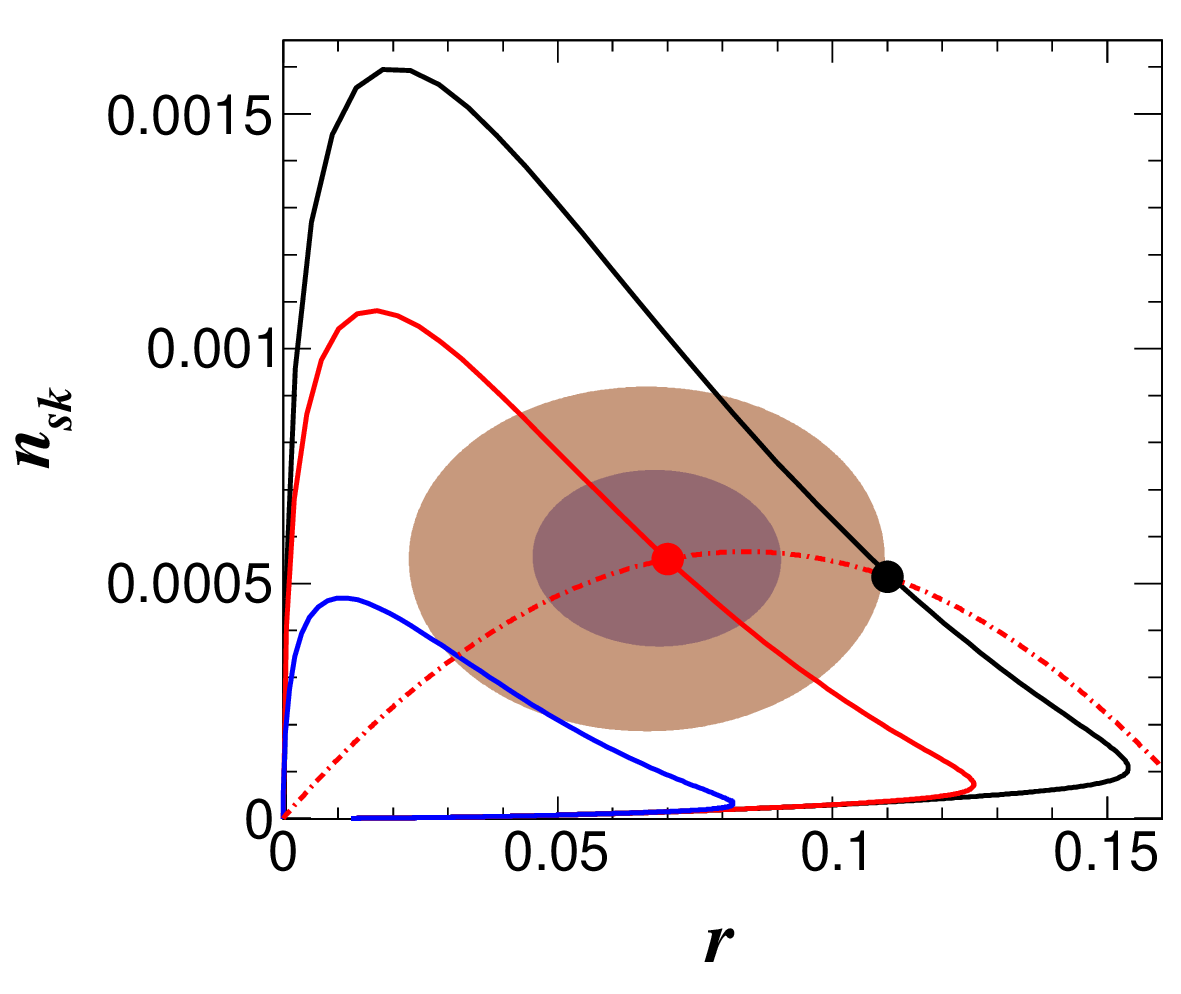}
\hspace{0.5cm}
\includegraphics[scale=0.26]{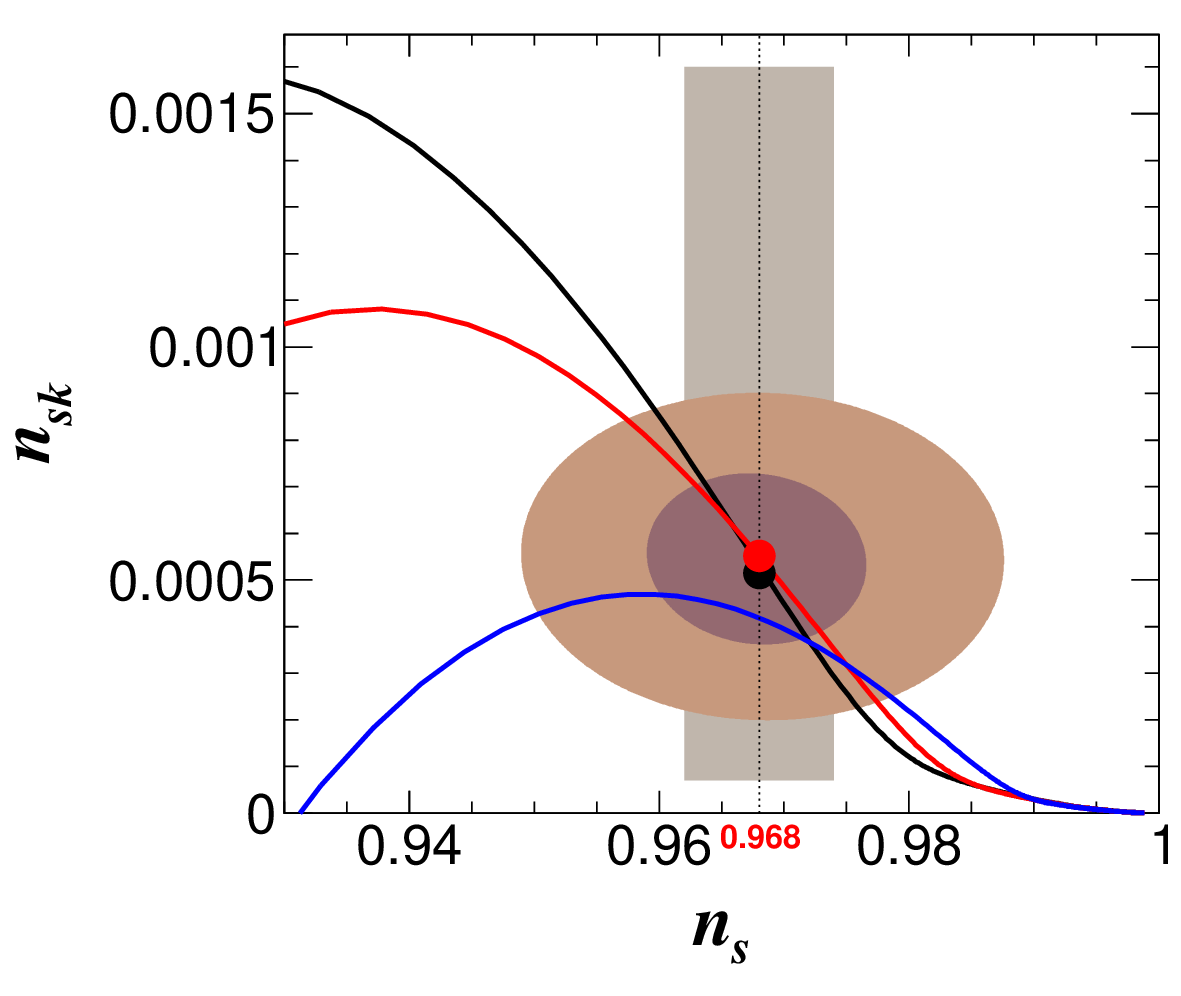}
\hspace{0.5cm}
\includegraphics[scale=0.26]{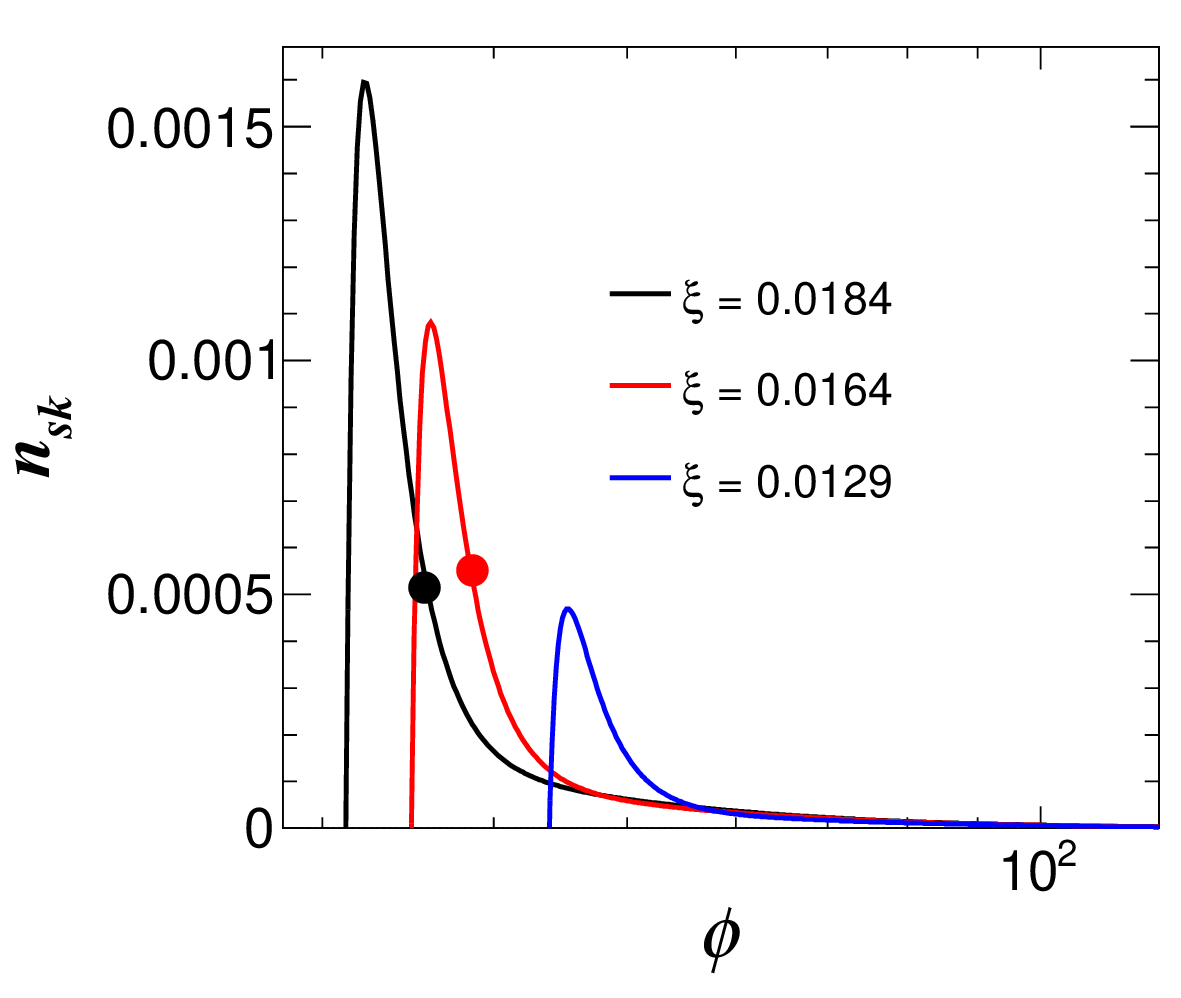}}
\caption{Variation of the running of the scalar spectral index $n_{sk}$ as a
function of $r$ (left plot), as a function of $n_s$ (middle plot) and
as a function of the field $\phi$ (right plot) for different values of
non-minimal coupling parameter $\xi$ as in the previous cases. In the 
$r - n_{sk}$ plot, the dash-dot-dash line represents the model independent
values of $n_{sk}$ obtained from the equation (\ref{eq46}) for values of $r$ 
from $0$ to $0.17$. Contours in the $r - n_{sk}$ and $n_s-n_{sk}$ plots 
indicate the $95\%$ and $68\%$ c.l. obtained from Planck's data. The shaded 
area in the $n_s-n_{sk}$ plot indicates range of Planck's $n_s$ data. 
In all plots, the black and red solid circles represent the $n_{sk}$ values 
for upper bounds of the Planck's and the BICEP2/Keck's data on $r$ 
respectively.}
\label{fig8}
\end{figure*}

For the same purpose as in the previous cases, numerical calculation was
done on $n_{sk}$ by using the generic equation (\ref{eq34}) via equations 
(\ref{eq41}), (\ref{eq42}) and (\ref{eq43}) for different values of $\xi$ 
as in the earlier cases, the results of which are shown in the 
Fig.\ref{fig8}.   

We found that the upper bounds of BICEP2/Keck's and Planck's data give that 
$n_{sk}$ should take values within $0< n_{sk} < 5.52\times 10^{-4}$ and 
$0 < n_{sk} < 5.15\times 10^{-4}$ respectively. From these results we used the
upper bounds of $n_{sk}$ corresponding to the upper bounds of $r$ to establish 
a numerical relation among $n_{sk}$, $r$ and $\delta_{ns}$ as given by
\begin{equation}
n_{sk} = \frac{r}{16}(6.74\,\delta_{ns} - 1.28\,r).
\label{eq46}
\end{equation} 

It is observed that as in the case of $n_{tk}$, the model independent values 
of $n_{sk}$ for $r = 0$ to $0.17$ obtained from the equation (\ref{eq46}), 
passage through different points on the model based $r-n_{sk}$ curves 
depending on the values of $\xi$ (see the left plot of Fig.\ref{fig8}).
These points correspond to the results for upper limits of observed value of 
$r$ as already 
mentioned. Thus same inference can be made from $r-n_{sk}$ plot together with 
$n_s-n_{sk}$ plot as in the case of previous $r-n_{tk}$ plot of the 
Fig.\ref{fig7} as well as in the case of $n_s-r$ plot of the Fig.\ref{fig5}. 
That is, our model could incorporate both experimental upper limits of $r$ as 
well as the experimental value of $n_s$ within a specific range of values of 
the 
non-minimal coupling parameter $\xi$. The value of $n_{sk}$ obtained for
the upper bound of BICEP2/Keck's data is found to lie well within the $68\%$ 
c.l. contour, drawn as mentioned earlier, in the $r-n_{sk}$ plot. However in 
the $n_s-n_{sk}$ plot, the value of $n_{sk}$ obtained for the upper bound of 
Planck's data is also found to lie within the $68\%$ c.l. contour. Hence, we 
may infer from here also that the most probable value of $\xi$ should lie 
within $0.0164^{+0.0020}_{-0.0035}$.        

As in the case of $\phi-r$ plot of the Fig.\ref{fig5}, the value of $n_{sk}$ 
becomes maximum ($n_{skmax}$) at a particular value of $\phi$ ($\phi_{nskm}$).
But this happens in a earlier phase of $\phi$ for this case than $\phi-r$ plot
and accordingly data points are shifted away from $n_{skmax}$ to retain their 
position over the specified values of $\phi$ as mentioned earlier. Hence, 
there is a reasonable (and should be observable) phase
difference between the values of $r$ and $n_{sk}$. It is mainly due to the 
$\delta_{ns}$ factor that remains effective in the equation of $n_{sk}$ as
clear from the numerical equation (\ref{eq46}). Taking the 
reference of $\phi-r$ plot, we have found that the values $n_{sk}$ for upper 
bounds of BICEP2/Keck's and Planck's data correspond to $\approx 61\%$ and 
$\approx 78\%$ of total values of $\phi-n_{sk}$ curves for $\xi = 0.0164$ and
$\xi = 0.0184$ respectively. Thus $\phi_{nskm}$ should be contained in the 
process of evolution of $\phi$ during the inflation as in the case of 
$\phi_{rmax}$.\\
\indent Again, comparing with the $\phi-n_s$ plot of the the Fig.\ref{fig5}, 
it can be seen that where the value of $n_s$ remains as constant 
(maximum $= 1$), there the values of $n_{sk}$ is zero. Moreover, with higher 
values of $\xi$ as the values of $n_s$ increase very fast at slightly earlier 
phases of $\phi$ than that with the lower values of $\xi$, so $n_{sk}$ also 
increases with similar fashion, but with a prominent manifestation. These 
corresponding behaviours of $n_s$ and 
$n_{sk}$ with respect to $\phi$ are due to the fact that $n_{sk}$
is the principal harmonics of the scalar power spectrum and is the running of 
$n_s$. 
     
\subsection{Relations between $r$, $n_{skk}$ and $\phi$}
\begin{figure*}[hbt]
\centerline{
\includegraphics[scale=0.26]{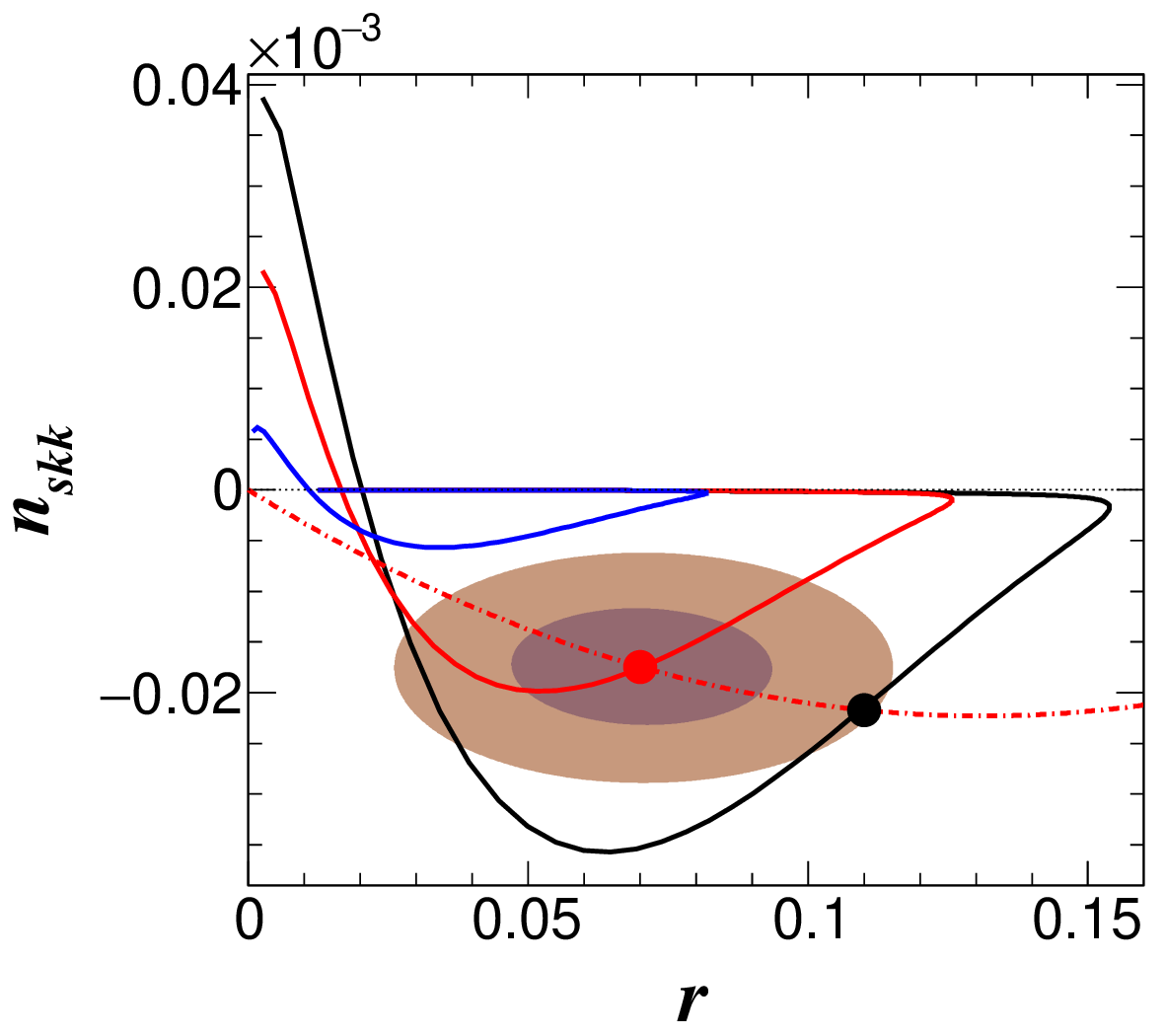}
\hspace{0.2cm}
\includegraphics[scale=0.26]{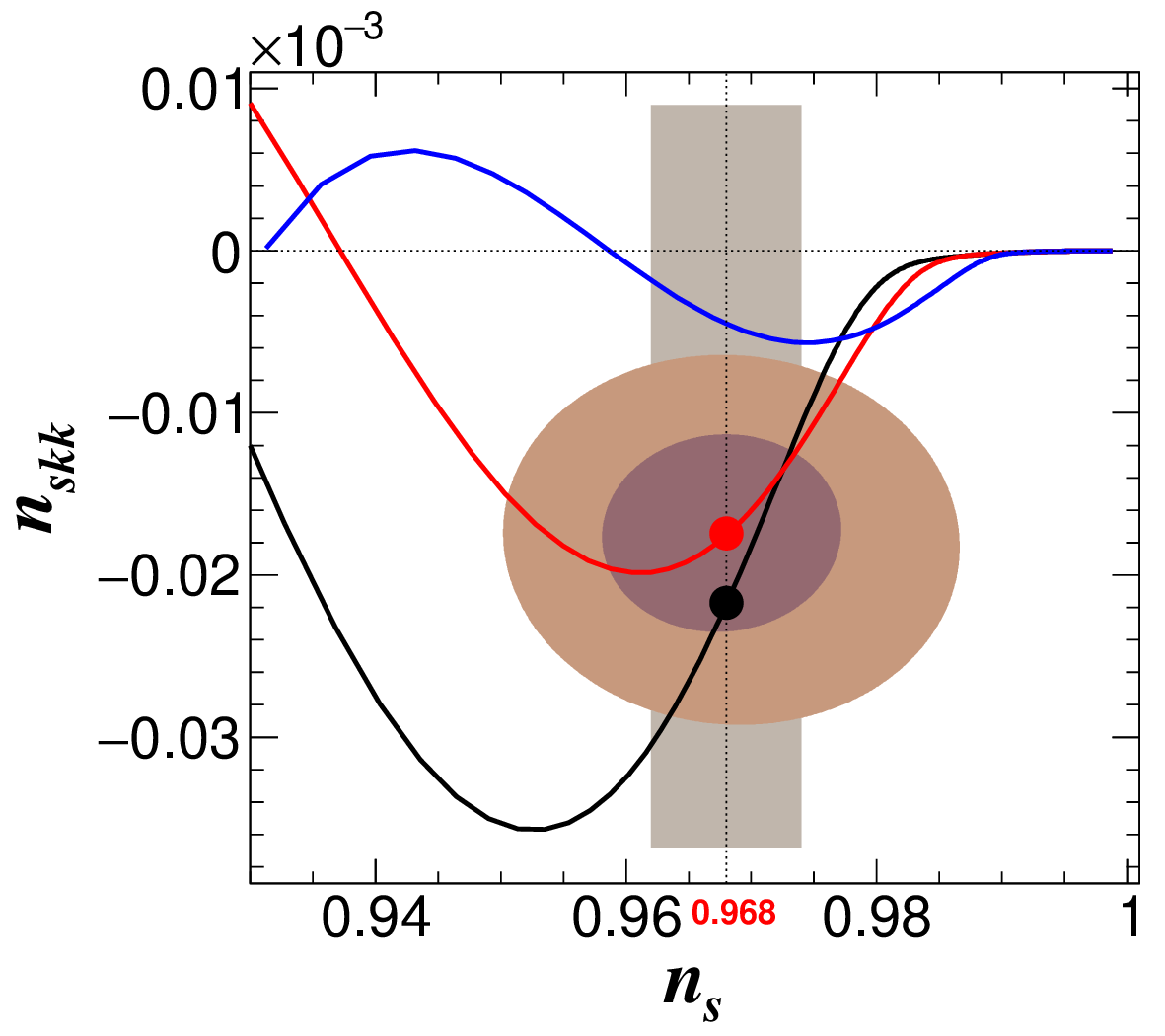}
\hspace{0.2cm}
\includegraphics[scale=0.26]{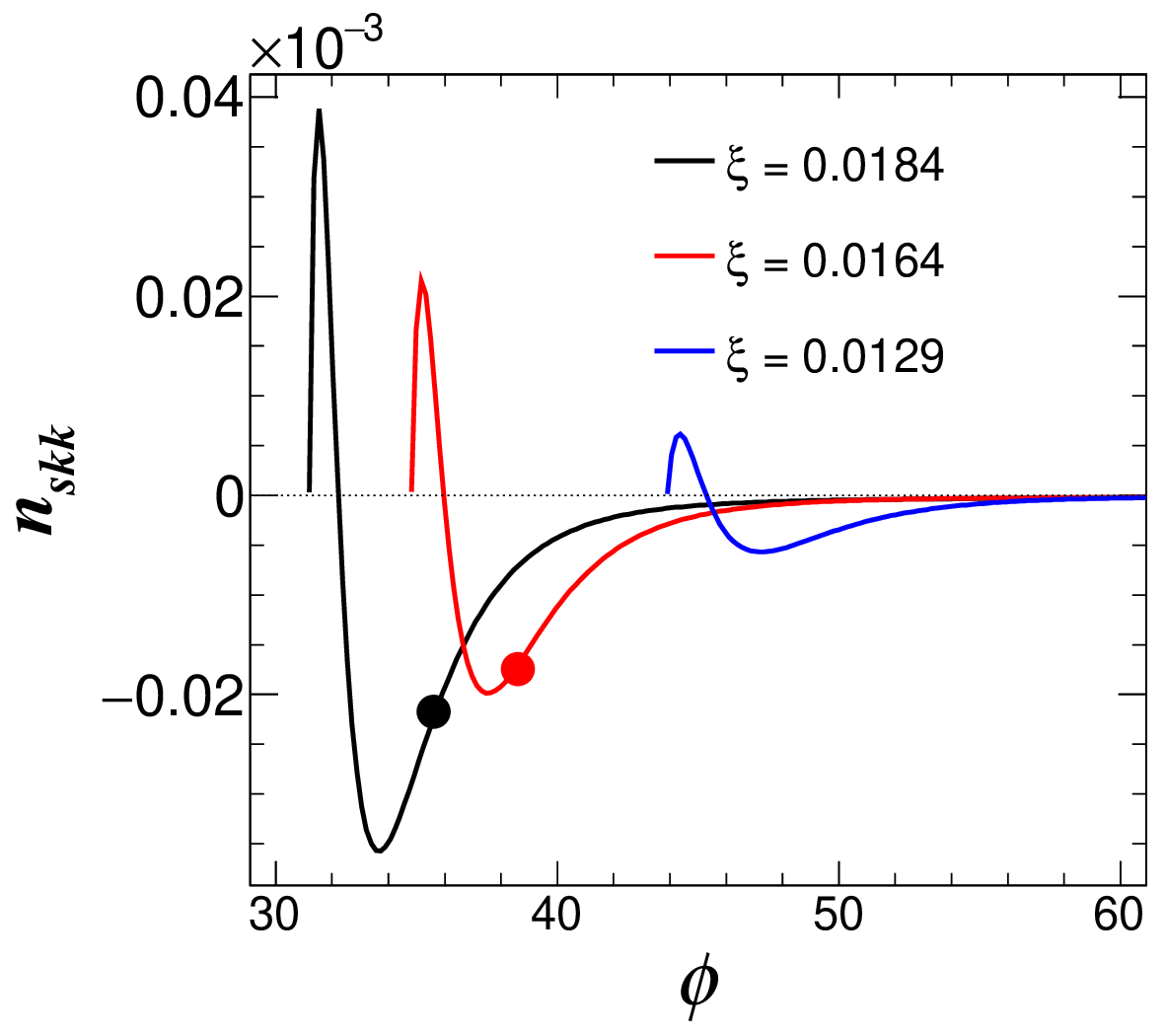}}
\caption{Variation of the running of the running of scalar spectral index 
$n_{skk}$ as a function of $r$ (left plot), as function of $n_s$ 
(middle plot) and as a function of the field $\phi$ (right plot) for 
different values of non-minimal coupling parameter $\xi$ used as in the
previous cases.  In the $r - n_{skk}$ plot, the dash-dot-dash line represents 
the model independent values of $n_{skk}$ obtained from the equation (\ref{eq47}) for values of $r$ from $0$ to $0.17$. Contours in the $r - n_{skk}$ and 
$n_s-n_{skk}$ plots indicate the $95\%$ and $68\%$ c.l. obtained from Planck's 
data. The shaded area in the $n_s-n_{skk}$ plot indicates the range of Planck's 
$n_s$ data. In all plots, the black and red coloured solid circles represent 
the values of $n_{skk}$ for the upper bounds of the Planck's and the 
BICEP2/Keck's data on $r$ respectively.}
\label{fig9}
\end{figure*}

Finally, to study the behaviours of $n_{skk}$ with respect to $r$, $n_s$ and 
$\phi$ we have to do the numerical calculations using the generic expression 
(\ref{eq35}) via equations (\ref{eq41}), (\ref{eq42}), (\ref{eq43}) and 
(\ref{eq44}). The results of these calculations are shown in the 
Fig.\ref{fig9} for different values $\xi$ as in the earlier cases. It is found 
that as per the upper bounds of the BICEP2/Keck's and Planck's data of
$r$, the actual value of $n_{skk}$ should lie within the ranges of 
$-1.744\times10^{-5}<n_{skk}<0$ and $-2.171\times10^{-5}<n_{skk}<0$ 
respectively. 

Similar to the case of $n_{sk}$, we derived a model independent relation for
the $n_{skk}$ also in terms of $r$ and $\delta_{ns}$ from the numerical data
in the form:
\begin{equation}
n_{skk} = -\frac{r}{256}(84.96\,\delta_{ns} - 10.36\,r)\delta_{ns}.
\label{eq47}
\end{equation} 
Same inference can be made from this equation as that is made from the equation
(\ref{eq46}). Moreover, this equation indicates that the probable value of 
$n_{skk}$ should be negative for any value of $r$ other than zero or 
$>8.2\,\delta_{ns}$. Here the later limit of $r$ sets a constraint that the
value of $r$ can not be greater than $8.2\,\delta_{ns}$. This is in agreement
with the constraint set by the model independent equation (\ref{eq40}) that
for the value of $n_{tk}$ should remain as negative the value of $r$ should be
less than $8\,\delta_{ns}$. Again, comparatively a more squeezed constraint on
$r$ is set by the equation (\ref{eq46}) is that for the value of $n_{sk}$ to
be remain positive the value of $r$ should be less than $5.62\,\delta_{ns}$.
These constraints on $r$ definitely rule out the reported BICEP2 experiment 
\cite{BICEP2} data on $r$.     

From the $r-n_{skk}$ and $n_s-n_{skk}$ plots a similar conclusion can be drawn 
on the value of non-minimal coupling parameter $\xi$, i.e. on the
model prediction as in the previous cases.
     
Moreover, the values of $n_{skk}$ oscillate in between positive
to negative values with the increasing values of $n_s$ and $\phi$ as 
$n_{skk}$ represents the second harmonics of scalar power spectrum. 
Hence at $\phi_{nskm}$, the value of $n_{skk}$ is found to be zero (refer to 
the Fig.\ref{fig8} and Fig.\ref{fig9}). Due to this fact, points corresponding 
to the data on 
$\phi-n_{skk}$ plot are shifted towards minimum negative side. 
  
Thus, our model could predict all conventional observables of the slow-roll 
approximation of inflation corresponding to the experimental bounds
of the tensor-to-scalar ratio $r$ as well as to the Planck's precise 
experimental value of scalar spectral index $n_s$ as mentioned above. The 
corresponding predicted values of the inflaton field $\phi$ are found to be 
same for all observables.  

Furthermore, it is worthy to mention that although the equation (\ref{eq40}) 
is based on the model independent derivation, but the equations (\ref{eq46}) 
and (\ref{eq47}) are derived from numerical results obtained from our model. 
Thus equations (\ref{eq46}) and (\ref{eq47}) can be used explicitly to test our
model from future cosmological observations. 

\subsection{Observables in terms of number of e-foldings}
\label{secef}
In order to compare our model's predictions precisely with other models as well
as with observations, it is necessary to calculate the observables of slow-roll
parameters in terms of number of e-foldings. For this purpose, we have to solve 
the equation (\ref{eq45}) of the number of e-foldings ($N$) to obtain
the inflaton field $\phi$ as a function of $N$. As this equation is very 
complex to find such a solution, we may simplify it by neglecting its second
and 3rd terms. This simplification is justified by the fact that the
contributions
of these two terms to the value of $N$ are found to be very negligible in
comparison to other two terms from the numerical calculation for the range of 
values of $\xi$ we have considered. The second term contribute only 
$\sim 0.045\%$ and the third term only $\sim 2\%$ on average to the sum of the
contributions made by the first and fourth terms. Thus with this 
simplification, the 
field $\phi$ can be obtained from the equation (\ref{eq45}) approximately
as  
\begin{equation}
\phi^2 \approx \frac{a^{-1} +\beta\exp{\left[\frac{16N + \phi^2_c}{16\alpha}- \frac{a^{-1}}{160\xi^2 \alpha}\right]}}{10\xi^2},
\label{eq48}
\end{equation}   
where $a$ and $\alpha$ represent the same expression as given in the 
subsection \ref{nef}  and 
$\beta = 10\xi^2\phi^2_c-16\xi-3$. As mentioned earlier, for 
physically acceptable value of $\phi$ (or $N$) 
$\phi_c > \sqrt{\frac{3 + 16\xi}{10\xi^2}}$.

Now, the values of $r$, $n_s$, $n_{tk}$,
$n_{sk}$ and $n_{skk}$ can be obtained in terms of the number of e-foldings 
$N$ for a given value of $\xi$ by substituting the equation (\ref{eq48}) into 
the the respective equations (\ref{eq37}), (\ref{eq32}), (\ref{eq33}), 
(\ref{eq34}) and (\ref{eq35}) via equations (\ref{eq41}), (\ref{eq42}),
(\ref{eq43}) and (\ref{eq44}).   

Presently, it is almost universally accepted that the number of e-foldings 
before the
end of inflation were $50-60$. So in our calculation we have taken three values
of number of e-foldings, viz., $N=50,\; 60\; \mbox{and}\; 70$ by widening the 
e-folding window in the upper side by one decade value. Moreover, from 
above analysis it is clear that the most probable value of non-minimal 
coupling parameter $\xi$ should be laid within $0.0164^{+0.0020}_{-0.0035}$. 
Hence, under 
the constraints of $N = 50 - 70$ and the Planck's data of $n_s$, for this 
calculation also we have taken three values of $\xi = 0.0129$, $0.0134$ and 
$0.0139$. Also
taking into account of values of the $\phi$ for the experimental upper limits
of $r$ as mentioned in the earlier occasions, we set the values of 
$\phi_c = (\sqrt{\frac{3 + 16\xi}{10\xi^2}}) + 0.25$ in this particular 
numerical calculation. 
Table \ref{tab1} shows the estimated values of different observables 
obtained from our model for the said values of $N$ and $\xi$. 


\begin{center}
\begin{table*}[ht]
\caption{\label{tab1} Estimated values of inflationary observables for the
potential (\ref{eq25a}) with non-minimal coupling parameter $\xi$ = $0.0129$, 
$0.0134$, $0.0139$ and number e-foldings $N$ = $50$, $60$, $70$. Estimation of
values of $r$, $n_s$, $n_{tk}$, $n_{sk}$ and $n_{skk}$ are made by using 
the equation (\ref{eq48}) into the the respective equations (\ref{eq37}), 
(\ref{eq32}), (\ref{eq33}), (\ref{eq34}) and (\ref{eq35}) via equations 
(\ref{eq41}), (\ref{eq42}), (\ref{eq43}) and (\ref{eq44}).}
\begin{center}
\begin{tabular}{ccccccc}\\[-5.0pt]
\hline \hline
$\xi$&$N$&$r$&$n_s$&$n_{tk}$&$n_{sk}$&$n_{skk}$\\\hline
&$50$&$0.0118$&$0.9596$&$-5.71\times10^{-5}$&$4.69\times10^{-4}$&$-5.31\times10^{-7}$\\
$0.0129$&$60$&$0.0190$&$0.9659$&$-7.55\times10^{-5}$&$4.39\times10^{-4}$&$-3.67\times10^{-6}$\\
&$70$&$0.0290$&$0.9719$&$-8.85\times10^{-5}$&$3.67\times10^{-4}$&$-5.49\times10^{-6}$\\\hline
&$50$&$0.0153$&$0.9594$&$-7.38\times10^{-5}$&$5.29\times10^{-4}$&$-2.25\times10^{-6}$\\
$0.0134$&$60$&$0.0249$&$0.9665$&$-9.45\times10^{-5}$&$4.69\times10^{-4}$&$-5.79\times10^{-6}$\\
&$70$&$0.0378$&$0.9732$&$-1.04\times10^{-4}$&$3.64\times10^{-4}$&$-6.93\times10^{-6}$\\\hline
&$50$&$0.0203$&$0.9600$&$-9.51\times10^{-5}$&$5.80\times10^{-4}$&$-4.86\times10^{-6}$\\
$0.0139$&$60$&$0.0331$&$0.9679$&$-1.15\times10^{-4}$&$4.78\times10^{-4}$&$-8.13\times10^{-6}$\\
&$70$&$0.0494$&$0.9750$&$-1.16\times10^{-4}$&$3.35\times10^{-4}$&$-7.78\times10^{-6}$\\
\hline \hline
\end{tabular}
\end{center}
\end{table*}
\end{center}

\begin{figure*}[hbt]
\centerline{
\includegraphics[scale=0.26]{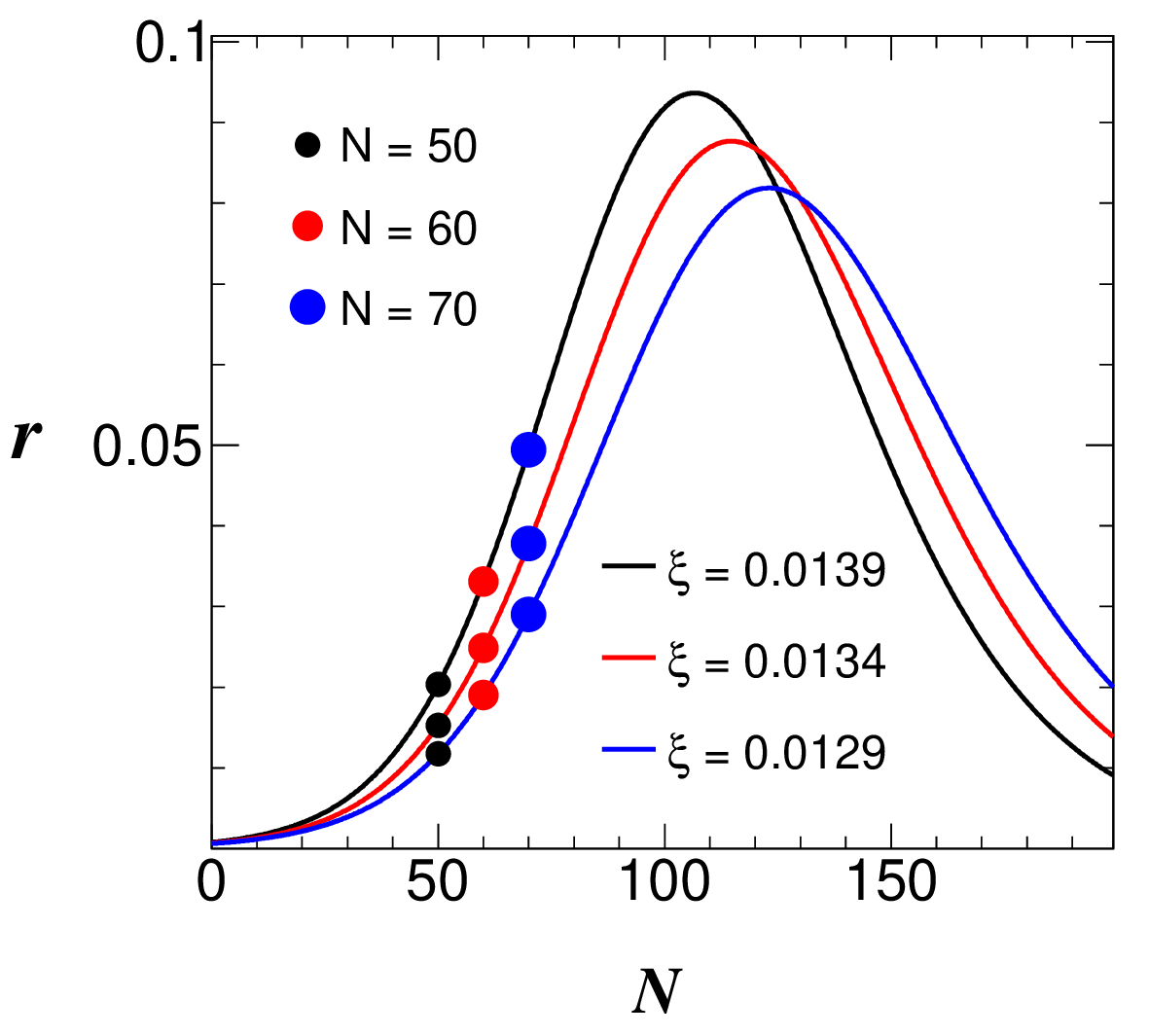}
\hspace{0.2cm}
\includegraphics[scale=0.26]{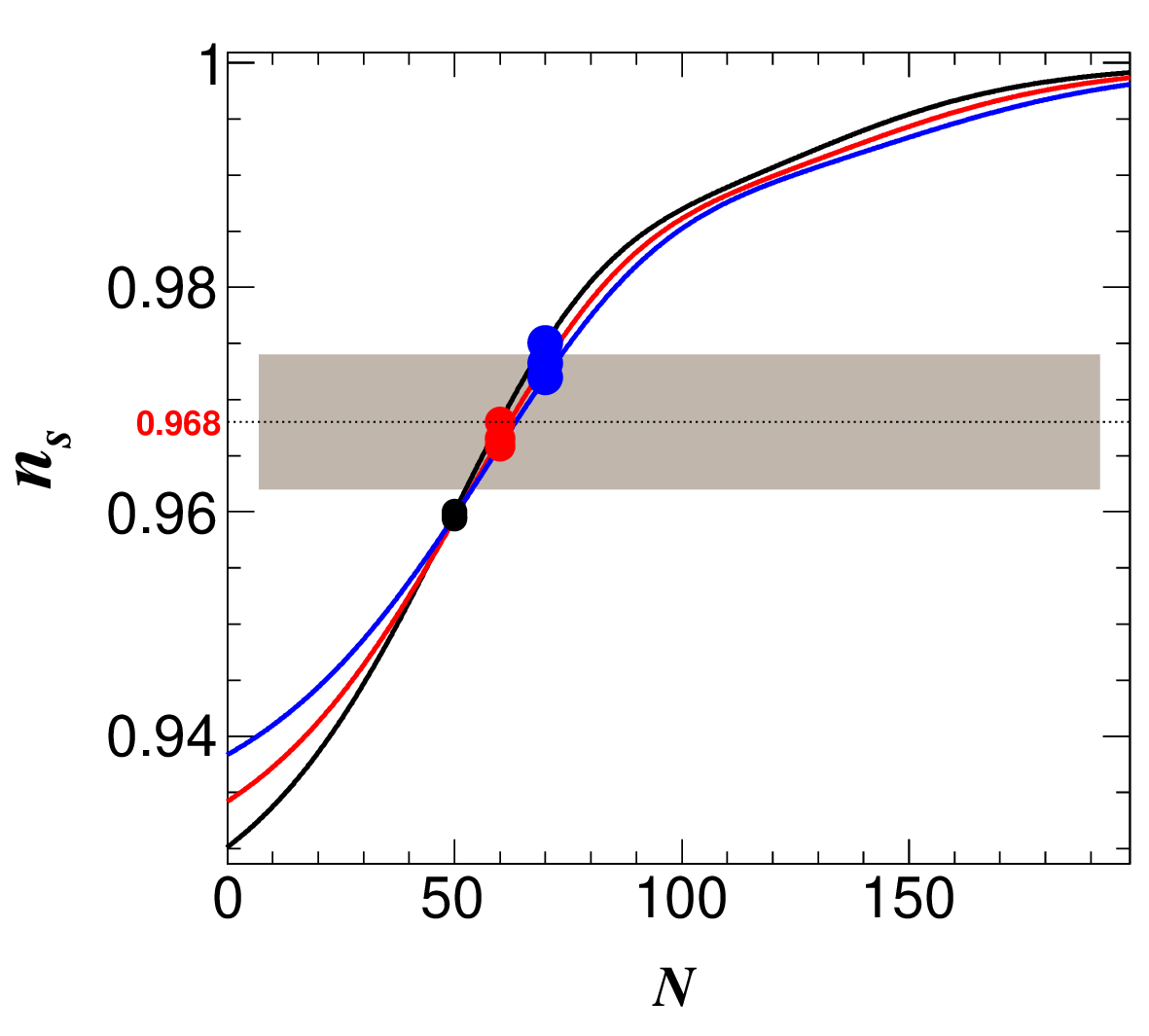}
\hspace{0.2cm}
\includegraphics[scale=0.26]{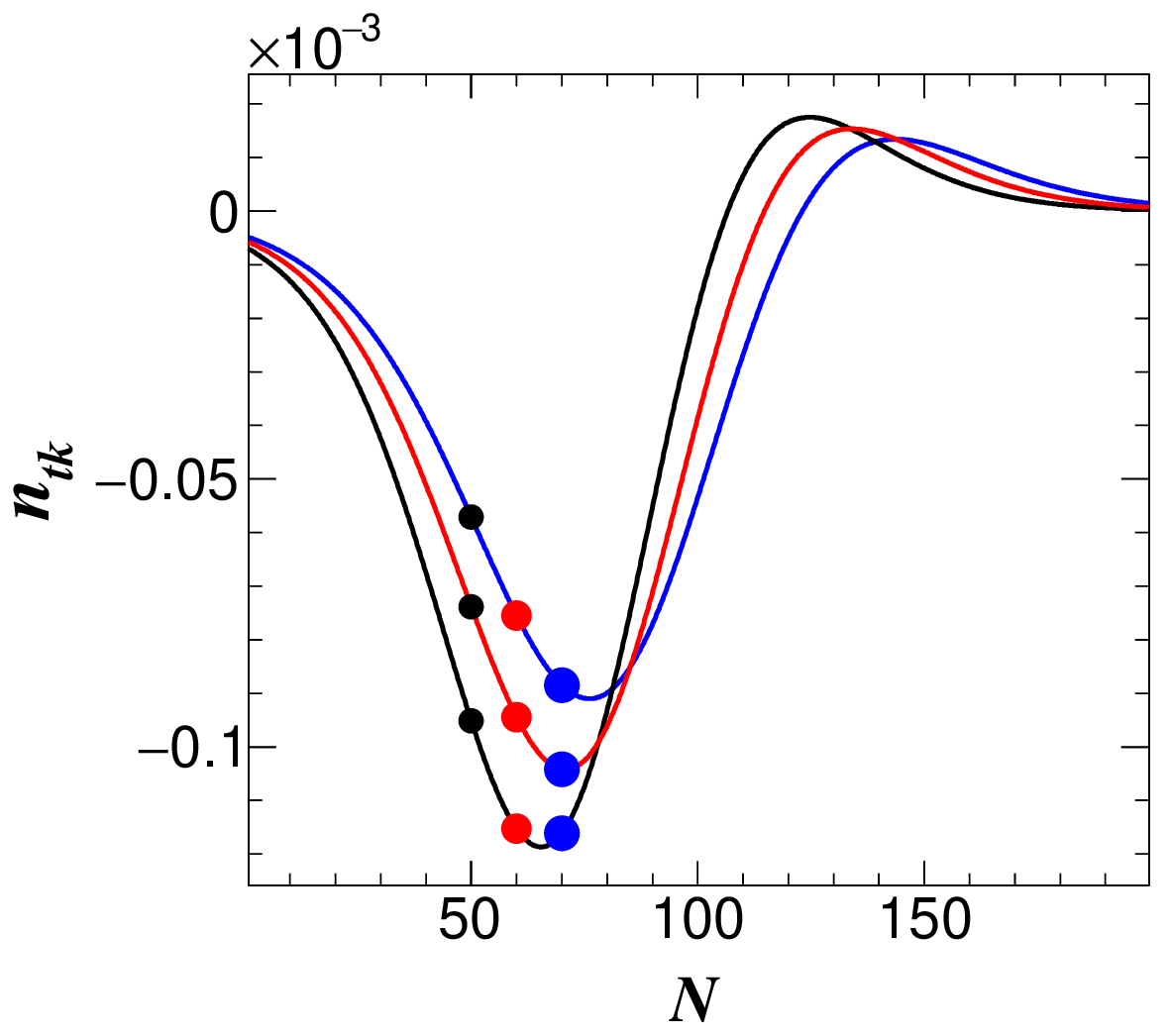}}
\vspace{0.2cm}
\centerline{
\includegraphics[scale=0.26]{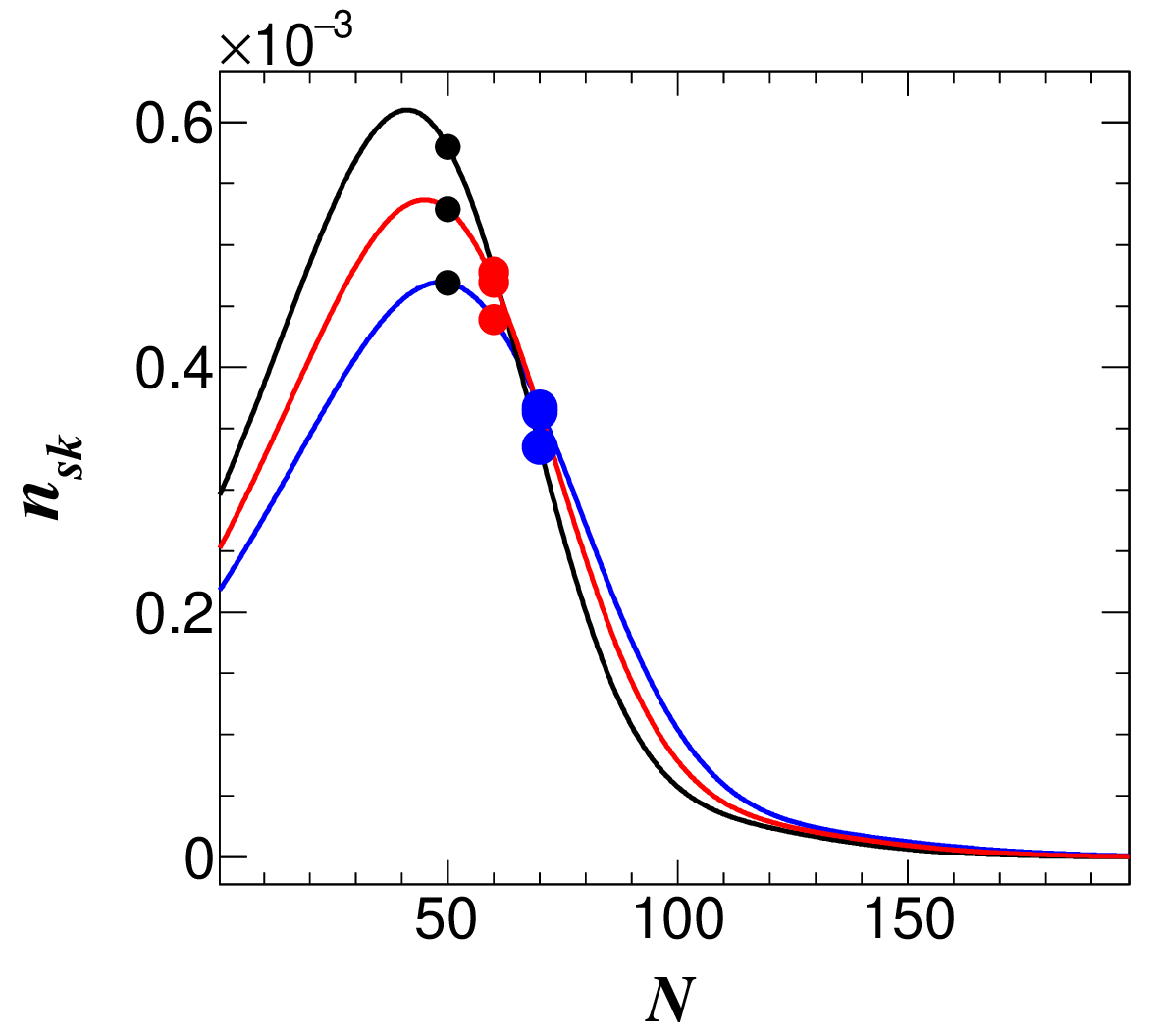}
\hspace{0.2cm}
\includegraphics[scale=0.26]{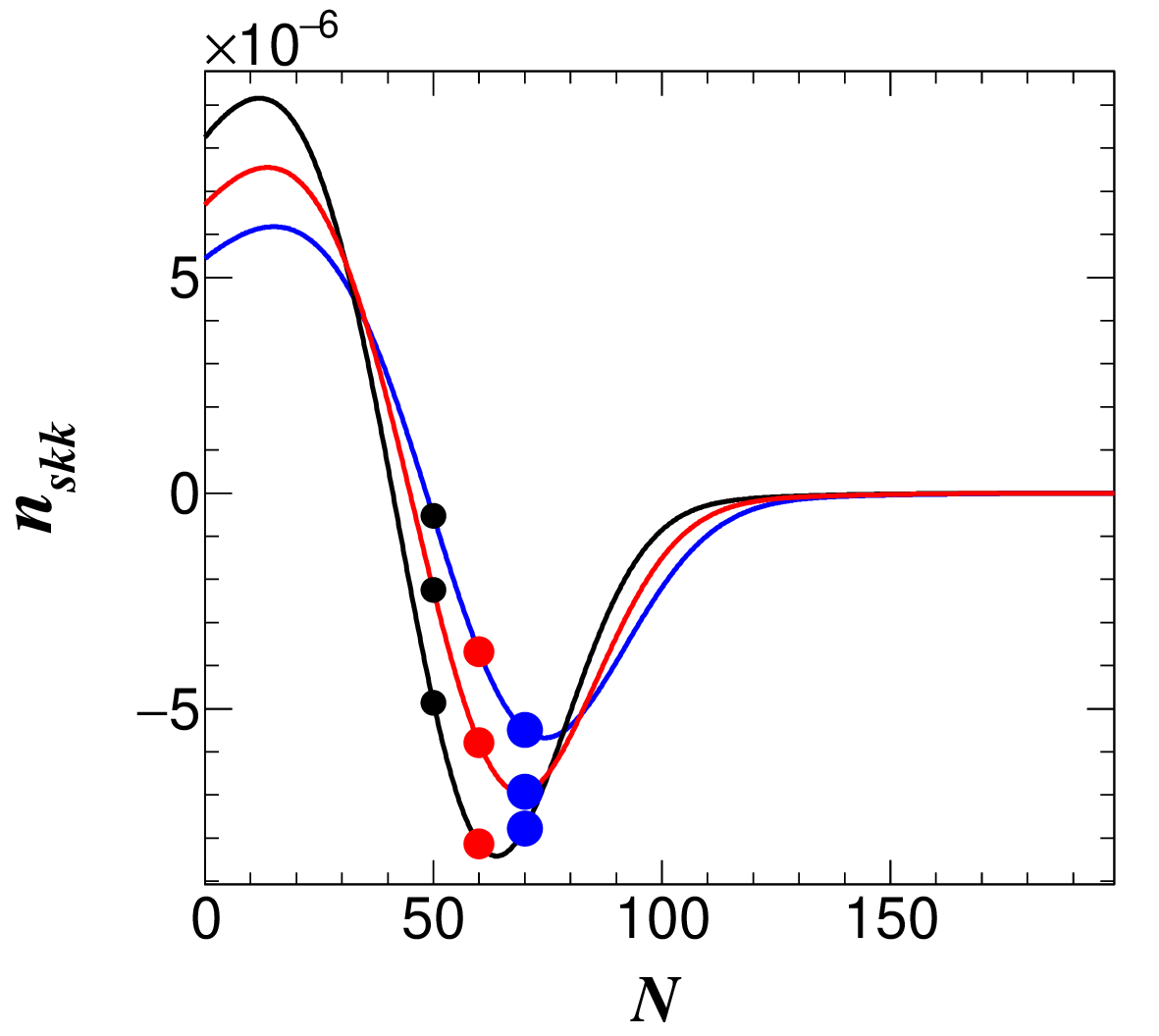}}
\caption{Variation of estimated values of different observables with respect to 
number of e-foldings $N$ for the values of non-minimal coupling parameter
$\xi = 0.0129$, $0.0134$ and $0.0139$. The shaded area in the $N-n_{s}$ plot 
indicates range of Planck's $n_s$ data.}
\label{fig10}
\end{figure*}

It is seen from the table \ref{tab1} that, the values of the 
tensor-to-scalar ratio $r$ estimated from our model for the number of 
e-foldings $N$ = $50-70$ with non-minimal 
coupling parameter $\xi$ = $0.0129$, $0.0134$ and $0.0139$ lie well below the 
upper limits set by Planck ($r<0.11$ at $95\%$ c.l.) \cite{PLANCK,PLANCK2} and 
BICEP2/Keck ($r<0.07$ at $95\%$ c.l.) \cite{BICEP2Keck} collaborations. 
Similarly, the estimated values of the scalar spectral index $n_s$ from our 
model for these parameters are seen
to be very close to the latest value of $n_s$ ($0.968\pm0.006$ at $68\%$ c.l.) 
as measured by the Planck collaboration \cite{PLANCK2} (in fact for $N$ = $60$ 
and $70$ almost all values of $n_s$ are laying within the range of data). 
Particularly, the value of $n_s$ for $N$ = $60$ with $\xi$ = $0.0139$ is in 
agreement with the mean value of the Planck's data (see the Fig.\ref{fig10} 
also). Within $N=50-70$ both these 
observables $r$ and $n_s$ increase in their magnitudes with increasing value of $N$. The general behaviours of all observables with respect to $N$ and $\xi$
are shown in the  Fig.\ref{fig10}. 

      
\section{Summary and conclusions}
We have found that the idea of supersymmetric hybrid inflation with non-minimal 
coupling to gravity is very interesting and useful, because the simplest 
version of the model based on it can accommodate the upper limits for the 
tensor-to-scalar ratio $r$ set by the Planck and BICEP2/Keck experiments for 
the values of non-minimal coupling parameter $\xi = 0.0184$ and $0.0164$
respectively. The values of $r$ for $\xi < 0.0164$ and $\xi < 0.0184$, 
restricted by the  Planck's data on $n_s$, fall within the range of upper 
bounds of the BICEP2/Keck's and the Planck's data respectively. Within these 
ranges of data as well as the said ranges of parameter $\xi$, the expected 
values of the running of the tensor spectral index $n_{tk}$, scalar spectral 
index $n_{sk}$ and its running $n_{skk}$ are found to belong. The values 
of the inflaton field $\phi$ corresponding to all these observables are found 
to be $\approx 38.5$, and $\approx 35.7$ respectively for upper bound of 
BICEP2/Keck's data and upper bound of Planck's data. The number of e-foldings 
corresponding to these values of $\phi$ are found to be $>50$ as expected. 
The $95\%$ and $68\%$ confidence level predictions based on the Planck's data 
for all the said observables indicate that the most probable value of $\xi$ 
for our model should lie within $0.0164^{+0.0020}_{-0.0035}$. 
Implying that the coupling is very week. More precisely, from the calculations
using the number of e-foldings, the values of $r$ estimated from our model for 
$\xi = 0.0134\pm0.0005$ and for $50-70$ number of e-foldings are laying well 
below the upper limits set by the Planck and BICEP2/Keck collaborations. 
Similarly, for the said parameters, the values of $n_s$ estimated from our 
model are in good agreement with its latest Planck data, specially the 
agreement is excellent for $N$ = $60$ with $\xi = 0.0139$. Moreover, these 
estimated values of 
$r$ and $n_s$ are also found to be in agreement with the results of the Ref. \cite{Pallis1}. Therefore, the most probable value of the non-minimal coupling 
parameter $\xi$ is found to be $\sim 0.0134\pm0.0005$.
   
Thus, this type of model can be used to predict a wide range of observational 
data on the cosmic inflation. So, it would be interesting to see the 
predictions of observables of the cosmic inflation by other versions of such 
model in future. In this context, it should be noted that the running of the 
scalar spectral index $n_{sk}$ can be considered as a probe of both 
inflation itself, and of the overall evolution of the very early universe as 
its magnitude is relatively consistent in comparison to tensor amplitude 
\cite{Peter}. It is potentially detectable by future large scale observations 
through measurements of clustering of high-redshift galaxy surveys together 
with CMB data  or the 21 cm forest signal observations \cite{Peter}. 
In the event of such measurements, our model of non-minimal coupling of 
gravity with SUSY hybrid inflation will play a significant role.               

The constrained equation for the running of the tensor spectral index $n_{tk}$ 
(equation (\ref{eq40})) is derived independent of any model. However, such 
equations for the running of the scalar spectral index $n_{sk}$ and its
running $n_{skk}$ (equations (\ref{eq46}) and (\ref{eq47}) respectively) are 
derived numerically based on the structure of our model.
All these equations can be used as 
constraint equations of slow-roll approximation. Specially, the 
equation (\ref{eq40}) can be used to test slow-roll approximation itself, 
and equations (\ref{eq46}) and (\ref{eq47}) can be used to test our model from
the data of future cosmological observations.        

Lastly, it is necessary to mention that to avoid the negative potential 
of the inflaton field of our model within the effective range of inflation, we 
introduced a second order non-minimal coupling parameter $\zeta$ along with 
the coupling parameter $\xi$ into the superconformal factor $\Omega$. For
simplified formulations and calculations we consider a very effective
relation between $\zeta$ and $\xi$ as $\zeta = 3\xi^2$. It is already clear that
this relationship works very well for our model. However, there may exists 
other such possible relations also.    

\section*{Acknowledgments}

The author is thankful to the Inter-University Centre for
Astronomy and Astrophysics (IUCAA), Pune for hospitality during his visits as
a Visiting Associate of the institute.


\begin{thebibliography}{99}
\bibitem{Starobinsky}
A. A. Starobinsky, \emph{A new type of isotropic cosmological models without 
singularity}, \emph{Phys. Lett. B} {\bf 91} (1980) 99. 

\bibitem{Kazanas}
D. Kazanas, \emph{Dynamics of the universe and spontaneous symmetry breaking},  \emph{Astrophys. J. Lett.} {\bf 241} (1980) L59.

\bibitem{Guth}
A. H. Guth, \emph{Inflationary universe: A possible solution to the horizon and 
flatness problems}, \emph{Phys. Rev. D} {\bf 23} (1981) 347. 

\bibitem{Sato1}
K. Sato, \emph{First-order phase transition of a vacuum and the expansion of 
the Universe}, \emph{Mon. Not. R. Astron. Soc.} {\bf 195} (1981) 467. 

\bibitem{Sato2}
K. Sato, \emph{Cosmological baryon-number domain structure and the first order 
phase transition of a vacuum}, \emph{Phys. Lett. B} {\bf 99} (1981) 66.

\bibitem{Bassett}
B. A. Bassett, S. Tsujikawa and D. Wands, \emph{Inflation dynamics and 
reheating}, \emph{Rev. Mod. Phys.} {\bf 78} (2000) 537 [arXiv:astro-ph/0507632].

\bibitem{Martin}
J. Martin, C. Ringeval and V. Vennin, \emph{Encyclopedia Inflationaris}, 
\emph{Phys. Dark Univ.} {\bf 5-6} (2014) 75 [ArXiv:1303.3787].

\bibitem{COBE}
C. L. Bennett et al., \emph{4-Year COBE DMR Cosmic Microwave Background 
Observations: Maps and Basic Results}, \emph{Astrophys. J.} {\bf 464} (1996) 
L1 [arXiv:astro-ph/9601067].
 
\bibitem{PLANCK}
P. A. R. Ade et al. (Planck Collaboration), \emph{Planck 2013 results. XVI. 
Cosmological parameters}, \emph{A\&A} {\bf 571} (2014) A16 
[arXiv:1303.5076].

\bibitem{PLANCK2}
P. A. R. Ade et al. (Planck Collaboration), \emph{Planck 2015 results. XIII. 
Cosmological parameters}, \emph{A\&A} {\bf 594} (2016) A13 [arXiv:1502.01589].

\bibitem{BICEP2}
P. A. R. Ade et al. (BICEP2 Collaboration), \emph{BICEP2 I: Detection Of 
B-mode Polarization at Degree Angular Scales}, \emph{Phys. Rev. Lett.} 
{\bf 112} (2014) 241101 [arXiv:1403.3985]. 

\bibitem{BICEP2Keck}
P. A. R. Ade et al. (Keck Array and BICEP2 Collaborations), \emph{BICEP2 / 
Keck Array VI: Improved Constraints On Cosmology and Foregrounds When Adding 
95 GHz Data From Keck Array}, \emph{Phys. Rev. Lett.} {\bf 116} (2016) 031302 
[arXiv:1510.09217].

\bibitem{WMAP}
G. Hinshaw et al. [WMAP Collaboration], \emph{Nine-Year Wilkinson Microwave 
Anisotropy Probe (WMAP) Observations: Cosmological Parameter Results}, 
\emph{Astrophys. J. Suppl.} {\bf 208} (2013) 19 [arXiv:1212.5226].

\bibitem{Linde0}
A. Linde, \emph{Chaotic inflation}, \emph{Phys. Lett. B} {\bf 129} (1983) 177.

\bibitem{Linde1}
A. D. Linde, \emph{Hybrid Inflation}, \emph{Phys. Rev. D} {\bf 49} (1994) 748
[arXiv:astro-ph/9307002]. 

\bibitem{Copeland}
E. J. Copeland, A. R. Liddle, D. H. Lyth, E. D. Stewart and D. Wands, 
\emph{False vacuum inflation with Einstein gravity}, \emph{Phys. Rev.} 
D {\bf 49} (1994) 6410 [arXiv:astro-ph/9401011].

\bibitem{Kawasaki}
M. Kawasaki, M. Yamaguchi and T. Yanagida, \emph{Natural Chaotic Inflation in 
Supergravity}, \emph{Phys. Rev. Lett.} {\bf 85} (2000) 3572 
[arXiv:hep-ph/0004243].

\bibitem{Feng} 
J. L. Feng, \emph{Naturalness and the Status of Supersymmetry}, \emph{Annual 
Rev. Nucl. Part. Sci.} {\bf 63} (2013) 351 [arXiv:1302.6587].

\bibitem{Mendes}
L. E. Mendes and A. R. Liddle,\emph{Initial conditions for hybrid inflation},
\emph{Phys. Rev.} {\bf D 62} (2000) 103511 [arXiv:astro-ph/0006020]. 

\bibitem{Lyth}
D. H. Lyth and A. Riotto, \emph{Particle Physics Models of Inflation and the 
Cosmological Density Perturbation}, \emph{Phys. Rep.} {\bf 314} (1999) 1 
[arXiv:hep-ph/9807278].

\bibitem{Carmi}
D. Carmi, A. Falkowski, E. Kuflik, T. Volansky, J. Zupan, \emph{Higgs After 
the Discovery: A Status Report}, \emph{JHEP} {\bf 10} (2012) 196 
[arXiv:1207.1718].  

\bibitem{Ferrara0}
S. Ferrara, R. Kallosh, A. Linde, A. Marrani and A. Van Proeyen, 
\emph{Superconformal Symmetry, NMSSM, and Inflation},
\emph{Phys. Rev.} {\bf D83} (2011) 025008 [arXiv:1008.2942].

\bibitem{Ferrara}
S. Ferrara, R. Kallosh, A. Linde, A. Marrani and A. Van Proeyen,
\emph{Jordan Frame Supergravity and Inflation in NMSSM}, \emph{Phys. Rev.
D} {\bf 82} (2010) 045003 [arXiv:1004.0712].

\bibitem{Lee}
H. M. Lee, \emph{Chaotic inflation in Jordan frame supergravity}, 
\emph{JCAP} {\bf 08} (2010) 003 [arXiv:1005.2735].

\bibitem{Salopek}
D. S. Salopek, J. R. Bond and J. M. Bardeen, \emph{Designing density fluctuation spectra in inflation}, \emph {Phys. Rev. D} {\bf 40} (1989) 1753.

\bibitem{Futamase}
T. Futamase and K.-i. Maeda, \emph{Chaotic inflationary scenario of the 
Universe with a nonminimally coupled "inflaton" field}, \emph{Phys. Rev. D} 
{\bf 39} (1989) 399.

\bibitem{Fakir}
R. Fakir and W.G. Unruh, \emph{Improvement on cosmological chaotic inflation 
through nonminimal coupling}, \emph{Phys. Rev. D} {\bf 41} (1990) 1783.

\bibitem{Mukaigawa} 
S. Mukaigawa, T. Muta, S. D. Odintsov, \emph{Finite Grand Unified Theories and 
Inflation}, \emph{Int. J. Mod. Phys.} {\bf A 13} (1998) 2739 
[arXiv:hep-ph/9709299]. 

\bibitem{Bezrukov}
F. L. Bezrukov and M. Shaposhnikov, \emph{The Standard Model Higgs boson as 
the inflaton}, \emph{Phys. Lett. B} {\bf 659} 
(2008) 703 [arXiv:0710.3755].

\bibitem{Barvinsky}
A. O. Barvinsky, A. Y. Kamenshchik and A. A. Starobinsky, \emph{Inflation 
scenario via the Standard Model Higgs boson and LHC}, \emph{JCAP} 
{\bf 11} (2008) 021 [arXiv:0809.2104]. 

\bibitem{Shaposhnikov}
F. Bezrukov and M. Shaposhnikov, \emph{Standard Model Higgs boson mass from 
inflation: two loop analysis}, \emph{JHEP} {\bf 07}
(2009) 089 [arXiv:0904.1537].

\bibitem{Gorbunov}
F. Bezrukov, D. Gorbunov and M. Shaposhnikov, \emph{On initial conditions for 
the Hot Big Bang}, \emph{JCAP} {\bf 06} (2009) 029 [arXiv:0812.3622]. 

\bibitem{Bellido}
J. G.-Bellido, D. G. Figueroa and J. Rubio, \emph{Preheating in the Standard 
Model with the Higgs-Inflaton coupled to gravity}, \emph{Phys. Rev. D} 
{\bf 79} (2009) 063531 [arXiv:0812.4624].

\bibitem{Simone}
A. D. Simone, M. P. Hertzberg and F. Wilczek,  \emph{Running Inflation in the 
Standard Model}, \emph{Phys. Lett. B} {\bf 678} (2009) 1 [arXiv:0812.4946]. 

\bibitem{Barvinsky1}
A. O. Barvinsky, A. Y. Kamenshchik, C. Kiefer, A. A.
Starobinsky and C. Steinwachs, \emph{Asymptotic freedom in inflationary 
cosmology with a non-minimally coupled Higgs field}, \emph{JCAP} {\bf 12} 
(2009) 003 [arXiv:0904.1698].

\bibitem{Kamenshchik}
A. O. Barvinsky, A. Y. Kamenshchik, C. Kiefer, A. A.
Starobinsky and C. Steinwachs, \emph{Higgs boson, renormalization group, and 
naturalness in cosmology}, \emph{Euro. Phys. J. C} {\bf 72} 
(2012) 2219 [arXiv:0910.1041]. 

\bibitem{Magnin}
F. Bezrukov, A. Magnin, M. Shaposhnikov and S. Sibiryakov, \emph{Higgs 
inflation: consistency and generalisations}, \emph{JHEP} {\bf 01} (2011) 016 
[arXiv:1008.5157]. 

\bibitem{Choudhury}
S. Choudhury, T. Chakraborty and S. Pal, \emph{Higgs inflation from new 
K\"ahler potential}, \emph{Nucl. Phys. B} {\bf 880} (2014), 155 
[arXiv:1305.0981]. 

\bibitem{Noorbala}
A. Linde, M. Noorbala and A. Westphal, \emph{Observational consequences of 
chaotic inflation with nonminimal coupling to gravity}, \emph{JCAP} {\bf 03} 
(2011) 013 [ArXiv:1101.2652]. 

\bibitem{Elizalde}
E. Elizalde, S. D. Odintsov, E. O. Pozdeeva and S. Yu. Vernov, 
\emph{Renormalization-group inflationary scalar electrodynamics and SU(5) 
scenarios confronted with Planck2013 and BICEP2 results}, \emph{Phys. Rev. D} 
{\bf 90} (2014) 084001 [arXiv:1408.1285]. 

\bibitem{Inagaki}
T. Inagaki, R. Nakanishi and S. D. Odintsov, \emph{Inflationary Parameters in 
Renormalization Group Improved \boldmath $\varphi^4$ Theory}, \emph{Astrophys. 
Space Sci.} {\bf 354} (2014) 2108 [arXiv:1408.1270]. 

\bibitem{Odintsov}
T. Inagaki, R. Nakanishi and S. D. Odintsov, \emph{Non-Minimal Two-Loop 
Inflation}, \emph{Phys. Lett. B} {\bf 745} (2015) 105 [arXiv:1502.06301].

\bibitem{Myrzakulov}
R. Myrzakulov, S. Odintsov and L. Sebastiani, \emph{Inflationary universe from higher derivative quantum gravity coupled with scalar electrodynamics}, 
\emph{Nucl. Phys. B} {\bf 907} (2016) 646 [arXiv:1604.06088]. 

\bibitem{Elizalde1}
E. Elizalde, S. D. Odintsov, E. O. Pozdeeva and S. Y. Vernov, 
\emph{Cosmological attractor inflation from the RG-improved Higgs sector of f
inite gauge theory}, \emph{JCAP} {\bf 02} (2016) 025 [arXiv:1509.08817].

\bibitem{Kallosh}
R. Kallosh and A. Linde, \emph{New models of chaotic inflation in
supergravity}, \emph{JCAP} {\bf 11} (2010) 011 [ArXiv:1008.3375].

\bibitem{David}
D. I. Kaiser and A. T. Todhunter, \emph{Primordial Perturbations from 
Multifield Inflation with Nonminimal Couplings}, \emph{Phys. Rev. D} 
{\bf 81} (2010) 124037 [arXiv:1004.3805]. 

\bibitem{Kaiser}
D. I. Kaiser, E. A. Mazenc and E. I. Sfakianakis, \emph{Primordial Bispectrum 
from Multifield Inflation with Nonminimal Couplings}, \emph{Phys. Rev. D} 
{\bf 87} (2013) 064004 [arXiv:1210.7487]. 

\bibitem{Greenwood}
R. N. Greenwood, D. I. Kaiser and E. I. Sfakianakis, \emph{Multifield Dynamics 
of Higgs Inflation}, \emph{Phy. Rev. D} {\bf 87} (2013) 064021 
[arXiv:1210.8190]. 

\bibitem{Sfakianakis}
D. I. Kaiser and E. I. Sfakianakis, \emph{Multifield Inflation after Planck: 
The Case for Nonminimal Couplings}, \emph{Phys. Rev. Lett.} {\bf 112} 
(2014) 011302 [arXiv:1304.0363]; 

\bibitem{Schutz}
K. Schutz, E. I. Sfakianakis and D. I. Kaiser, \emph{Multifield Inflation 
after Planck: Isocurvature Modes from Nonminimal Couplings}, 
\emph{Phys. Rev. D} {\bf 89} (2014) 064044 [arXiv:1310.8285]. 

\bibitem{Pallis1}
G.~Lazarides and C.~Pallis,
  \emph{Shift Symmetry and Higgs Inflation in Supergravity with Observable Gravitational Waves}, \emph{JHEP} {\bf 1511} (2015) 114 [arXiv:1508.06682]; 

\bibitem{Pallis2}
C.~Pallis,
  \emph{Kinetically modified nonminimal Higgs inflation in supergravity},
  \emph{Phys.\ Rev.\ D} {\bf 92} (2015) 121305 [arXiv:1511.01456].

\bibitem{Mazumdar}
A. Mazumdar and J. Rocher, \emph{Particle physics models of inflation and 
curvaton scenarios}, \emph{Phys. Rep.} {\bf 497} (2011) 85 
[arXiv:1001.0993].

\bibitem{Dvali}
G. Dvali, Q. Shafi and R. Schaefer, \emph{Large Scale Structure and 
Supersymmetric Inflation without Fine Tuning}, \emph{Phys. Rev. Lett.} {\bf 73} 
(1994) 1886 [arXiv:hep-ph/9406319].

\bibitem{Orani}
S. Orani and A. Rajantie, \emph{Supersymmetric hybrid inflation with a light 
scalar}, \emph{Phys. Rev. D} {\bf 88} (2013) 043508.

\bibitem{Koushik}
K. Das, V. Domcke and K. Dutta, \emph{Supergravity Contributions to Inflation
in models with non-minimal coupling to gravity},
\emph{JCAP} {\bf 03} (2017) 036 [arXiv:1612.07075].

\bibitem{Kallosh1}
R. Kallosh, L. Kofman, A. D. Linde and A. Van Proeyen, \emph{Superconformal 
Symmetry, Supergravity and Cosmology}, \emph{Class. 
Quant. Grav.} {\bf 17} (2000) 4269 [Erratum-ibid {\bf 21} (2004) 5017] 
[arXiv:hep-th/0006179].

\bibitem{Cremmer}
E. Cremmer, B. Julia and J. Scherk, S. Ferrara, L. Girardello and P. van 
Nieuwenhuizen, \emph{Spontaneous symmetry breaking and Higgs effect in 
supergravity without cosmological constant}, \emph{Nucl. Phys. B} {\bf 147} 
(1979) 105.

\bibitem{Proeyen}
E. Cremmer, S. Ferrara, L. Girardello and A. Van Proeyen, \emph{Yang-Mills 
theories with local supersymmetry: Lagrangian, transformation laws and 
superHiggs Effect}, \emph{Nucl. Phys. B} {\bf 212} (1983) 413. 

\bibitem{Matthew}
M. Civiletti, C. Pallis  and Q. Shafi,\emph{Upper Bound on the 
Tensor-to-Scalar Ratio in GUT-Scale Supersymmetric Hybrid Inflation}, 
\emph{Phys. Lett. B} {\bf 733} (2014) 276 [arXiv:1402.6254].

\bibitem{Tatsuo}
T. Kobayashi and O. Seto, \emph{Beginning of Universe through large field 
hybrid inflation}, \emph{Mod. Phys. Lett. A} {\bf 30} (2015) 1550106 
[arXiv:1404.3102].

\bibitem{Felder2002}
G. Felder, A. Frolov, L. Kofman and A. Linde, \emph{Cosmology with negative 
potentials}, \emph{Phys. Rev. D} {\bf 66} (2002) 023507 [arXiv:hep-th/0202017]


\bibitem{Mariana}
M. C.-Gonz\'alez, G. German, A. H.-Aguilar, J. C. Hidalgo, R. Sussman, 
\emph{Testing Hybrid Natural Inflation with 
BICEP2}, \emph{Phys. Lett. B} {\bf 734} (2014) 345 [arXiv:1404.1122].

\bibitem{Peter}
P. Adshead, R. Easther, J. Pritchard and A. Loeb, \emph{Inflation and the 
Scale Dependent Spectral Index: Prospects and Strategies}, \emph{JCAP} 
{\bf 1102} (2011) 021 [arXiv:1007.3748].

\end{thebibliography}
\end{document}